\documentclass{emulateapj}
\pdfoutput=1

\usepackage{amsmath}
\usepackage{gensymb}
\usepackage{chngcntr}

\shorttitle{Planck cold clumps in the $\lambda$ Orionis complex: PGCC G192.32-11.88 } \shortauthors{Liu et al.}

\begin{document}

\title{Planck cold clumps in the $\lambda$ Orionis complex: I. Discovery of an extremely young Class 0 protostellar object and a proto-brown dwarf candidate in a bright rimmed clump PGCC G192.32-11.88}
\author{Tie Liu\altaffilmark{1}, Qizhou Zhang\altaffilmark{2}, Kee-Tae Kim\altaffilmark{1}, Yuefang Wu\altaffilmark{3}, Chang Won Lee\altaffilmark{1,4}, Jeong-Eun Lee\altaffilmark{5}, Ken'ichi Tatematsu\altaffilmark{6}, Minho Choi \altaffilmark{1}, Mika Juvela\altaffilmark{7}, Mark Thompson\altaffilmark{8}, Paul F. Goldsmith\altaffilmark{9}, Sheng-yuan Liu\altaffilmark{10}, Hirano Naomi\altaffilmark{10}, Patrick Koch\altaffilmark{10}, Christian Henkel\altaffilmark{11}, Patricio Sanhueza\altaffilmark{6}, JinHua He\altaffilmark{12}, Alana Rivera-Ingraham\altaffilmark{13}, Ke Wang\altaffilmark{14}, Maria R. Cunningham\altaffilmark{15}, Ya-Wen Tang\altaffilmark{10}, Shih-Ping Lai\altaffilmark{16}, Jinghua Yuan\altaffilmark{17}, Di Li\altaffilmark{17,24}, Gary Fuller\altaffilmark{18}, Miju Kang\altaffilmark{1}, Quang Nguyen Luong\altaffilmark{6}, Hauyu Baobab Liu\altaffilmark{10}, Isabelle Ristorcelli\altaffilmark{19}, Ji Yang\altaffilmark{20}, Ye Xu\altaffilmark{20}, Tomoya Hirota\altaffilmark{6}, Diego Mardones\altaffilmark{21}, Sheng-Li Qin\altaffilmark{22}, Huei-Ru Chen\altaffilmark{10,16}, Woojin Kwon\altaffilmark{1}, FanYi Meng\altaffilmark{23}, Huawei Zhang\altaffilmark{3},  Mi-Ryang Kim\altaffilmark{1}, Hee-Weon Yi\altaffilmark{5}    }
\altaffiltext{1}{Korea Astronomy and Space Science Institute, 776 Daedeokdae-ro, Yuseong-gu, Daejeon 34055, Korea; liutiepku@gmail.com}
\altaffiltext{2}{Harvard-Smithsonian Center for Astrophysics, 60 Garden Street, Cambridge, MA 02138, USA}
\altaffiltext{3}{Department of Astronomy, Peking University, 100871, Beijing China}
\altaffiltext{4}{University of Science \& Technology, 217 Gajungro, Yuseong-gu, 305-333 Daejeon, Korea}
\altaffiltext{5}{School of Space Research, Kyung Hee University, Yongin-Si, Gyeonggi-Do 446-701, Korea}
\altaffiltext{6}{National Astronomical Observatory of Japan, 2-21-1 Osawa, Mitaka, Tokyo 181-8588, Japan}
\altaffiltext{7}{Department of physics, University of Helsinki, FI-00014 Helsinki, Finland}
\altaffiltext{8}{Centre for Astrophysics Research, University of Hertfordshire, College Lane, Hatfield, AL10 9AB, UK}
\altaffiltext{9}{Jet Propulsion Laboratory, California Institute of Technology, 4800 Oak Grove Drive, Pasadena, CA 91109, USA 0000-0002-6622-8396}
\altaffiltext{10}{Academia Sinica, Institute of Astronomy and Astrophysics, P.O. Box 23-141, Taipei 106, Taiwan}
\altaffiltext{11}{Max-Planck-Institut f{\"u}r Radioastronomie, Auf dem H{\"u}gel  69, D-53121, Bonn, Germany; Astronomy Department, Abdulaziz University, PO Box 80203, 21589, Jeddah, Saudi Arabia}
\altaffiltext{12}{Key Laboratory for the Structure and Evolution of Celestial Objects, Yunnan observatories, Chinese Academy of Sciences,
P.O. Box 110, Kunming, 650011, Yunnan Province, PR China.}
\altaffiltext{13}{ESA/ESAC, 28691 Villanueva de la Ca\~{n}ada, Madrid, Spain}
\altaffiltext{14}{European Southern Observatory, Karl-Schwarzschild-Str.2, D-85748 Garching bei M\"{u}nchen, Germany}
\altaffiltext{15}{School of Physics, University of New South Wales, Sydney, NSW 2052, Australia}
\altaffiltext{16}{Institute of Astronomy and Department of Physics, National Tsing Hua University, Hsinchu, Taiwan}
\altaffiltext{17}{National Astronomical Observatories, Chinese Academy of Sciences, Beijing, 100012, China}
\altaffiltext{18}{Jodrell Bank Centre for Astrophysics, Alan Turing Building, School of Physics and Astronomy, University of Manchester, Manchester M13 9PL, UK}
\altaffiltext{19}{IRAP, CNRS (UMR5277), Universit\'{e} Paul Sabatier, 9 avenue du Colonel Roche, BP 44346, 31028, Toulouse Cedex 4, France}
\altaffiltext{20}{Purple Mountain Observatory, Chinese Academy of Sciences, China, Nanjing 210008}
\altaffiltext{21}{Departamento de Astronom\'{\i}a, Universidad de Chile, Casilla 36-D, Santiago, Chile}
\altaffiltext{22}{Department of Astronomy, Yunnan University, and Key Laboratory of Astroparticle Physics of Yunnan Province, Kunming, 650091, China}
\altaffiltext{23}{Physikalisches Institut, Universit\"{a}t zu K\"{o}ln, Z\"{u}lpicher Str. 77, D-50937 K\"{o}ln, Germany}
\altaffiltext{24}{Key Lab of Radio Astronomy, Chinese Academy of Sciences, China}

\begin{abstract}
We are performing a series of observations with ground-based telescopes toward Planck Galactic cold clumps (PGCCs) in the $\lambda$ Orionis complex in order to systematically investigate the effects of stellar feedback. In the particular case of PGCC G192.32-11.88, we discovered an extremely young Class 0 protostellar object (G192N) and a proto-brown dwarf candidate (G192S). G192N and G192S are located in a gravitationally bound bright-rimmed clump. The velocity and temperature gradients seen in line emission of CO isotopologues indicate that PGCC G192.32-11.88 is externally heated and compressed. G192N probably has the lowest bolometric luminosity ($\sim0.8$ L$_{\sun}$) and accretion rate (6.3$\times10^{-7}$ M$_{\sun}$~yr$^{-1}$) when compared with other young Class 0 sources (e.g. PACS Bright Red sources (PBRs)) in the Orion complex. It has slightly larger internal luminosity ($0.21\pm0.01$ L$_{\sun}$) and outflow velocity ($\sim$14 km~s$^{-1}$) than the predictions of first hydrostatic cores (FHSCs). G192N might be among the youngest Class 0 sources, which are slightly more evolved than a FHSC. Considering its low internal luminosity ($0.08\pm0.01$ L$_{\odot}$) and accretion rate (2.8$\times10^{-8}$ M$_{\sun}$~yr$^{-1}$), G192S is an ideal proto-brown dwarf candidate. The star formation efficiency ($\sim$0.3\%-0.4\%) and core formation efficiency ($\sim$1\%) in PGCC G192.32-11.88 are significantly smaller than in other giant molecular clouds or filaments, indicating that the star formation therein is greatly suppressed owing to stellar feedback.
\end{abstract}


\section{Introduction}

Stellar feedback from massive stars can strongly influence their surrounding interstellar medium and regulate star formation through outflows, photoionizing radiation, energetic winds, supernova explosions. Such stellar feedback may have both ``negative" and ``positive" effects on molecular clouds (Dale et al. 2015). In the negative aspect, expanding H{\sc ii} regions or supernova explosions are energetic enough to sweep-up the surrounding molecular gas and destroy giant molecular clouds (GMCs). Stellar feedback might also be responsible for the low star formation efficiencies (SFE) or low star formation rates (SFR) in GMCs (Solomon et al. 1979; Dale et al. 2012,
2013). However, whether the low SFE and SFR in GMCs are caused by stellar feedback is still under debate. For example, most nearby GMCs uniformly have
low star formation efficiencies (Evans et al. 2009). While it has  been stated that there are no obvious signs of stellar feedback such as bubbles (Krumholz et al.  2014, pp. 243-266), more recent studies suggest that bubbles are prevalent around even low mass YSOs when inspected closely (Arce et al. 2011;
Li et al. 2015).

The positive aspect of stellar feedback is usually referred to as triggered star formation. The shock front (SF) that emerges as bubbles around
massive stars expands can compress the ISM, triggering star
formation in very dense layers (Chen et al. 2007). Thompson et
al. (2012) discovered an enhanced surface density of young stellar objects (YSOs) projected against the rim of infrared bubbles, suggesting that the YSOs were triggered by the expansion of the bubbles. They also estimated that the fraction of massive
stars in the Milky Way formed by triggering process could be between 14\% and 30\% (Thompson et
al. 2012). Theoretically, the ionization front (IF) generated from
an H{\sc ii} region can generate a shock front (SF) in molecular clouds and gather molecular gas to form a dense
shell between the IF and the SF (Elmegreen \& Lada 1977; Whitworth et al.
1994). The gas and dust in the collected layer collapse to form stars when they reach the critical
density (Chen et al. 2007; Ogura 2010; Whitworth et al. 1994). This so-called ``collect and
collapse" process can self propagate and lead to sequential star formation (Elmegreen \& Lada 1977; Whitworth et al.
1994). Star formation induced through the ``collect and
collapse" process has been revealed in the borders of several
H{\sc ii} regions evidenced by fragmented shells or sequential star formation (Deharveng et al. 2003; Deharveng, Zavagno, \& Caplan
2005; Zavagno et al. 2006, 2007; Deharveng et al. 2008;
Pomares, M. et al. 2009; Petriella et al. 2010; Brand
et al. 2011; Liu et al. 2012a, 2015).  Another model for triggered star formation is the so-called ``radiation-driven implosion" (RDI) (Bertoldi \& McKee 1990; Lefloch \&
Lazareff. 1994; Ogura 2010), in which the surface layer of
a pre-existing cloud is ionized by the UV photons from nearby massive stars and then
the cloud is compressed and collapsed due to the shock generated
from the ionization front. The difference between the ``collect and
collapse" and the ``radiation-driven implosion" processes is whether the molecular cloud is collected as the IF advances or is pre-existing before the bubble expands. Bright-rimmed clouds are often treated as a hint of triggered star formation due to the RDI process because their rims are the glowing ionized boundary layers on the irradiated faces of the clouds. Candidates
for star formation by RDI have been suggested in many bright rimmed clouds (Sugitani et al. 1991; Sugitani \& Ogura
1994; Urquhart et al. 2004; Thompson et al. 2004; Lee
et al. 2005; Urquhart et al. 2006, 2007; Morgan et al.
2010; Liu et al. 2012a).

However, recent numerical simulations indicate that most signs in previous observational works, such as the ages and geometrical distribution of stars relative to the feedback
source or feedback-driven structure (e.g. shells, pillar structures), are not substantially helpful in distinguishing triggered star formation from spontaneous star formation (Dale et al. 2015). Triggered star formation can either increase the SFR, the SFE or the total numbers of stars formed (Dale et al. 2015). Therefore, to trace triggered star formation we need to carefully investigate how molecular clouds interact with feedback sources (H{\sc ii} regions or Supernova remnants) and how the SFR and SFE change in molecular clouds due to stellar feedback.

The $\lambda$ Orionis complex contains one of the most prominent OB associations in Orion. As shown in the left panel of Figure 1, it is a giant H{\sc ii} region with a ring-like structure of dust and gas of $8\arcdeg-10\arcdeg$ in diameter (Maddalena \& Morris 1987; Zhang et al. 1989; Dolan
\& Mathieu 2002). The central $\lambda$ Ori cluster contains a (O8 III) massive star with a massive companion that became a supernova (Maddalena \& Morris 1987; Cunha
\& Smith 1996; Dolan \& Mathieu 2002), removing nearby molecular gas at the center of the cluster about 1 Myr ago and leading to the formation of the ring of dust and gas (Mathieu et al. 2008). As one of the nearest large H{\sc ii} region, the $\lambda$ Orionis complex is an ideal laboratory for studies of stellar feedback. In this region, the age sequence of classical T Tauri stars (CTTSs) and the high star formation efficiency of some bright-rimmed clouds (e.g. B35 \& LDN 1634) provide strong evidence of triggered star formation (Lee et al. 2005). However, the properties (e.g. substructures, gravitational stability) of molecular clouds in the $\lambda$ Orionis complex are far from well understood due to a lack of high angular resolution observations in both molecular lines and (sub)millimeter continuum.

\begin{figure*}
\centering
\includegraphics[angle=-90,scale=0.5]{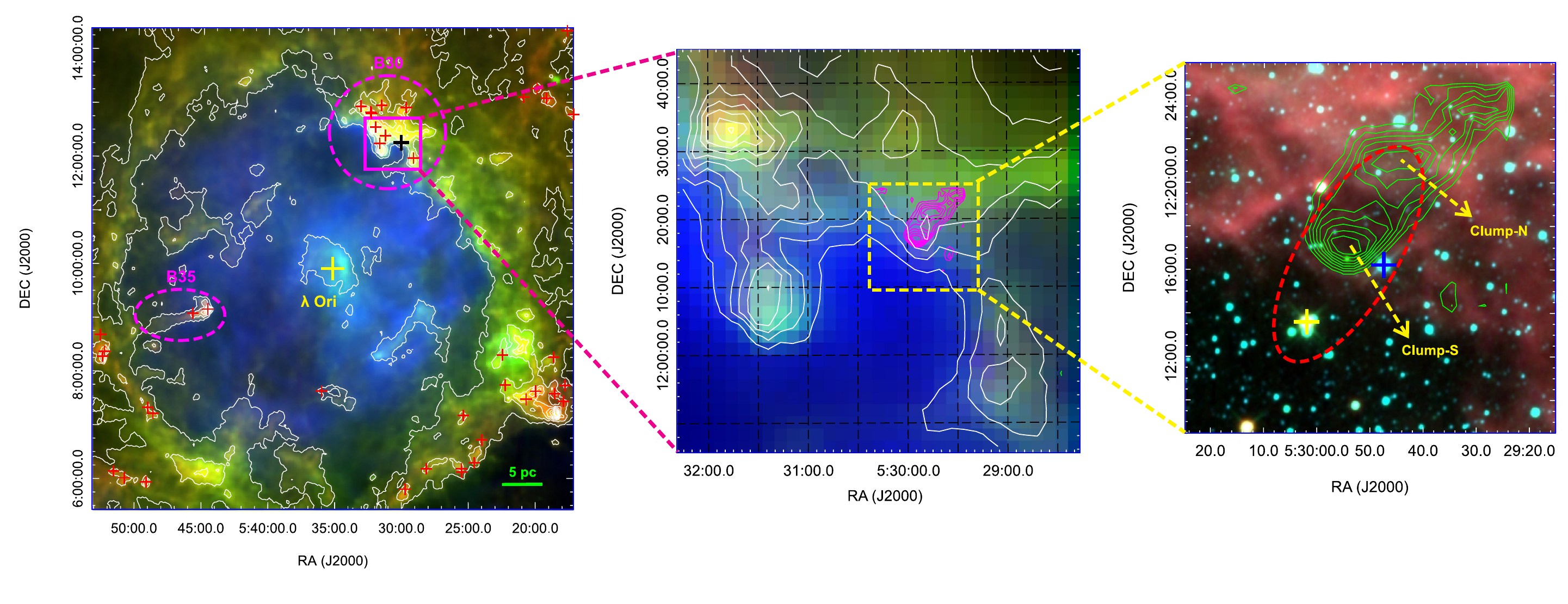}
\caption{Left panel: Three-color composite image (Red: Planck 857 GHz; Green: IRAS 100 $\mu$m; Blue: H$\alpha$) of $\lambda$ Orionis complex. The white contours represent the flux density of Planck 857 GHz continuum emission. The contours are from 10\% to 90\% in steps of 10\% of peak value 133.1 MJy/sr. The two bright-rimmed clouds B30 and B35 are marked with pink ellipses. The ``$\lambda$ Ori" OB binary or multiple star system is marked with a yellow cross. PGCC G192.32-11.88 is marked with a black cross. All Planck cold clumps in this region with a clump-averaged column density larger than $5\times10^{20}$ cm$^{-2}$ are shown as red crosses. Middle panel: The environment of PGCC G192.32-11.88. The color image and white contours are the same as in the left panel. H$_{2}$ column density calculated from $^{13}$CO (1-0) emission (see section 3.1.3) of PGCC G192.32-11.88 is shown in pink contours. The contour levels are from 20\% to 90\% in steps of 10\% of the peak value 1.22$\times10^{22}$ cm$^{-2}$.  Right panel: H$_{2}$ column density calculated from CO isotopologues is shown in green contours overlayed on WISE three color composite image (3.4 $\micron$ in blue, 4.6 $\micron$ in green, 22 $\micron$ in red). The contours are the same as in the middle panel. The dashed ellipse marks the shape of PGCC G192.32-11.88. The B type star HD 36104 is marked with a blue cross. The nearby multiple star system which contains four Variable Stars (V\*GX Ori, V\* V370 Ori, V\* V443 Ori and V\*V444 Ori) of Orion Type is marked with a yellow cross. The two main gas clumps ``Clump-N" and ``Clump-S" are marked with yellow dashed arrows. }
\end{figure*}

Planck is the third generation mission to measure the anisotropy of the cosmic microwave background (CMB) and observed the sky in nine frequency bands (the 30, 44, 70, 100, 143, 217, 353, 545, and 857 GHz bands). The high frequency channels of Planck cover the peak thermal emission frequencies of dust colder than 14 K, indicating that Planck could probe the coldest parts of the ISM (Planck Collaboration et al. 2011, 2015).
The Planck team has cataloged 13188 Planck galactic cold clumps (PGCCs) (Planck Collaboration et al. 2015). Having low dust temperatures of 6-20 K, Planck cold clumps show the smallest line widths but modest column densities when compared to other kinds of star forming clouds (Planck Collaboration et al. 2011, 2015; Wu et
al. 2012; Liu, Wu \& Zhang 2013; Liu et al. 2014). A large fraction of PGCCs seem to be quiescent, not affected by on-going star forming activities, suggesting that they might trace very initial conditions of star formation (Wu et al. 2012; Liu, Wu \& Zhang 2013; Liu
et al. 2014; T´oth et al. 2014). Planck cold clumps are highly structured as indicated in Herschel follow-up observations (Juvela et al. 2012) and mapping observations in CO isotopologues with the PMO 13.7-m telescope (Liu,
Wu \& Zhang 2012b; Meng, Wu \& Liu 2013; Liu et
al. 2014). There are about 180 PGCCs located in the $\lambda$ Orionis complex. As shown in the left panel of Figure 1, most of the densest PGCCs with a clump-averaged column density larger than $5\times10^{20}$ cm$^{-2}$ are located in the rim of the H{\sc ii} region. Therefore, the studies of PGCCs in $\lambda$ Orionis complex will provide us information on how the stellar feedback of the H{\sc ii} region influences the initial conditions of star formation. To improve our understandings of star formation in $\lambda$ Orionis complex, we are conducting a series of observations toward Planck cold clumps with several ground-based telescopes (see section 2 for details).

Here we report the first results from our observations toward a particular source PGCC G192.32-11.88, which is located in the B30 cloud of the $\lambda$ Orionis complex. It has a size of $\sim11\arcmin.2\times4\arcmin.5$, with the long axis parallel to the Galactic plane (Planck Collaboration et al. 2015). The cold component has a clump-averaged temperature of $17.3\pm6.0$ K (Planck Collaboration et al. 2015), while the warm background has a dust temperature of $28.1\pm$4.4 K (Planck Collaboration et al. 2015). As shown in the middle panel of Figure 1, PGCC G192.32-11.88 as a bright-rimmed clump is facing the H{\sc ii} region and is surrounded by H$_{\alpha}$ emission, indicating that it is shaped by the photo-erosion of the ionizing radiation. It contains two main gas clumps ``Clump-N" and ``Clump-S" as revealed in $^{13}$CO (1-0) emission in the right panel of Figure 1. The B type star HD 36104 is at the south-west edge of ``Clump-S". The distance of $\lambda$ Orionis complex is in between 380$\pm$30 pc, as derived by Hipparcos, and 420$\pm$42 pc, as derived from Pan-STARRS1 photometry (Schlafly et al. 2014). However, different clouds in the $\lambda$ Orionis complex might have various distances from $\sim$380 to $\sim$470 pc (Schlafly et al. 2014). The distance of the B type star HD 36104 close to PGCC G192.32-11.88 is 364$\pm$32 pc as derived by Hipparcos (Dolan \& Mathieu
2001). In this paper we adopt a distance of 380 pc for PGCC G192.32-11.88. PGCC G192.32-11.88 only has a mass of 3.0$\pm$2.4 M$_{\sun}$ as inferred from the Planck data at a distance of 380 pc.

\section{Observations}

\subsection{PMO 13.7 m telescope observations}

The observations toward PGCC G192.32-11.88 in the
$^{12}$CO (1-0), $^{13}$CO (1-0) and C$^{18}$O (1-0) lines were carried out
with the Purple Mountain Observatory (PMO) 13.7 m radio telescope
in May 2011. The new 9-beam array receiver system in single-sideband
(SSB) mode was used as front end (Shan et al. 2012). FFTS spectrometers were used as back ends, which have a
total bandwidth of 1 GHz and 16384 channels, corresponding to a
velocity resolution of 0.16 km~s$^{-1}$ for the $^{12}$CO (1-0) and 0.17
km~s$^{-1}$ for the $^{13}$CO (1-0) and the C$^{18}$O (1-0). The $^{12}$CO (1-0) was observed
in the upper sideband, while the $^{13}$CO (1-0) and the C$^{18}$O (1-0) were observed simultaneously in the lower
sideband.  The half-power beam width (HPBW) is 56$\arcsec$ and the main beam efficiency is
$\sim0.45$. The pointing accuracy of the telescope was better than
4$\arcsec$. The typical system temperatures (T$_{sys}$) are around 130 K for LSB and 220 K for USB, which vary about 10\% among the 9 beams. The
On-The-Fly (OTF) observing mode was utilised. The antenna
continuously scanned a region of 22$\arcmin\times22\arcmin$ centered
on the Planck cold clumps with a scan speed of 20$\arcsec$~s$^{-1}$.
However, the edges of the OTF maps have a much higher noise level and thus only the central 14$\arcmin\times14\arcmin$
regions are selected to be further analyzed. The typical rms noise level per channel was
0.3 K in T$_{A}^{*}$ for the $^{12}$CO (1-0), and 0.2 K for the $^{13}$CO (1-0) and the C$^{18}$O (1-0).

\subsection{CSO 10 m telescope observations}

Observations of ``Clump-S" in the
$^{12}$CO (2-1), $^{13}$CO (2-1) and C$^{18}$O (2-1) lines were carried out
with the Caltech Submillimeter Observatory (CSO) 10 m radio telescope in January 2014. We used the Sidecab receiver and the FFTS2 spectrometer. The spectral resolution is $\sim$0.35 km~s$^{-1}$. The weather was good during observations with a system temperature of 244 K. The beam size and beam efficiency of the CSO telescope at 230 GHz is $\sim$34$\arcsec$ and 0.76, respectively. The map size is $5\arcmin\times5\arcmin$ with a spacing of 10$\arcsec$. The typical rms noise level per channel is
0.2 K in T$_{A}^{*}$.

Using the GILDAS software package including CLASS and GREG (Guilloteau \& Lucas 2000), the OTF data of CO isotopologues were
converted to 3-D cube data. Baseline subtraction was performed by fitting linear or sinusoidal functions.
Then the data were exported to MIRIAD (Sault et al. 1995) and CASA for further analysis.

\subsection{KVN 21 m telescope observations}

We also observed ``Clump-S" in the J=1-0 transitions of HCO$^{+}$, H$^{13}$CO$^{+}$ and N$_{2}$H$^{+}$ as well as HDCO ($2_{0,2}-1_{0,1}$) and o-H$_{2}$CO ($2_{1,2}-1_{1,1}$) with the Korean VLBI Network (KVN) 21 m telescope at Yonsei station in single dish mode (Kim et al. 2011). The observations were carried out in February 2015. The spectral resolution and system temperature for the J=1-0 of HCO$^{+}$, H$^{13}$CO$^{+}$ and N$_{2}$H$^{+}$ (1-0) is $\sim$0.052 km~s$^{-1}$ and 230 K, respectively. The spectral resolution and system temperature for HDCO ($2_{0,2}-1_{0,1}$) and o-H$_{2}$CO ($2_{1,2}-1_{1,1}$) are $\sim$0.034 km~s$^{-1}$ and 453 K, respectively. We did single-pointing observations with an rms level per channel of T$_{A}^{*}<$0.1 K. The beam size and beam efficiency of the KVN telescope at 86 GHz band are $\sim$32$\arcsec$ and 0.41, respectively. The beam size and beam efficiency of the KVN telescope in the 129 GHz band is $\sim$24$\arcsec$ and 0.40, respectively. The data were reduced using the GILDAS software package.

\subsection{JCMT 15 m telescope observations}

The observations of ``Clump-S" with James Clerk Maxwell Telescope (JCMT) SCUBA-2 were conducted in April 2015. We used the ``CV Daisy" mapping mode, which is suitable for point sources. The 225 GHz opacity during observations ranges from 0.088 to 0.11. Therefore, we only obtained 850 $\micron$ continuum data. The data were reduced using SMURF in STARLINK package. The FOV of SCUBA-2 at 850 $\micron$ is 8$\arcmin$. The mapping area is about $12\arcmin\times$12\arcmin. The beam size of SCUBA-2 at 850 $\micron$ is $\sim$14$\arcsec$. The rms level in the central 3$\arcmin$ area of the map is around 8 mJy~beam$^{-1}$.

\subsection{The SMA observations}

The observations of ``Clump-S" with the Submillimeter Array (SMA)\footnote{The Submillimeter Array is a joint project between the Smithsonian
Astrophysical Observatory and the Academia Sinica Institute of Astronomy and
Astrophysics, and is funded by the Smithsonian Institution and the Academia
Sinica (Ho, Moran, \& Lo 2004).} were carried out in November 2014 in its compact configuration. The phase reference center was R.A.(J2000)~=~05$^{\rm h}$29$^{\rm m}$54.32$^{\rm s}$ and
DEC.(J2000)~=~$12\arcdeg16\arcmin40.50\arcsec$. In observations, QSOs 0530+135 and 0510+180 were observed for gain and phase correction. QSO 3c279 was used for bandpass calibration. Uranus was applied for flux-density calibration. The channel spacing across the 8 GHz spectral band is 0.812~MHz or $\sim$1.1 km s$^{-1}$. The visibility data
were calibrated with the IDL superset MIR package and imaged with MIRIAD and CASA packages. The 1.3 mm continuum data were acquired by averaging all the line-free channels over both the upper and lower spectral bands. The synthesized beam size and 1 $ \sigma$ rms of the continuum emission are $3\arcsec.4\times2\arcsec.3$ (PA=-59$\arcdeg$) and $\sim$1 mJy~beam$^{-1}$, respectively. The 1$ \sigma$ rms for lines is $\sim$50 mJy~beam$^{-1}$ per channel. We detect strong $^{12}$CO (2-1) emission and weak $^{13}$CO (2-1) emission above 3 $\sigma$
signal-to-noise level in the SMA observations. To recover the missing flux in the SMA $^{12}$CO (2-1) and $^{13}$CO (2-1) emission, we combined the SMA data with the CSO data following a procedure outlined in Zhang et al. (1995). The combined SMA+CSO data for $^{12}$CO (2-1) and $^{13}$CO (2-1) emissions were convolved with the same beam size of $5\arcsec.2\times3\arcsec.8$ (PA=-53$\arcdeg$) to derive the $^{13}$CO column density. The 1$ \sigma$ rms for lines is $\sim$300 mJy~beam$^{-1}$ per channel in the SMA+CSO combined data. We use primary beam corrected continuum and line data in further analysis.

We also used Spitzer MIPS and IRAC archival data.

\section{Results of single-dish observations}

\subsection{Overall properties of PGCC G192.32-11.88 inferred from the observations of CO isotopologues}

\subsubsection{Moment maps}

\begin{figure*}
\centering
\includegraphics[angle=-90,scale=0.6]{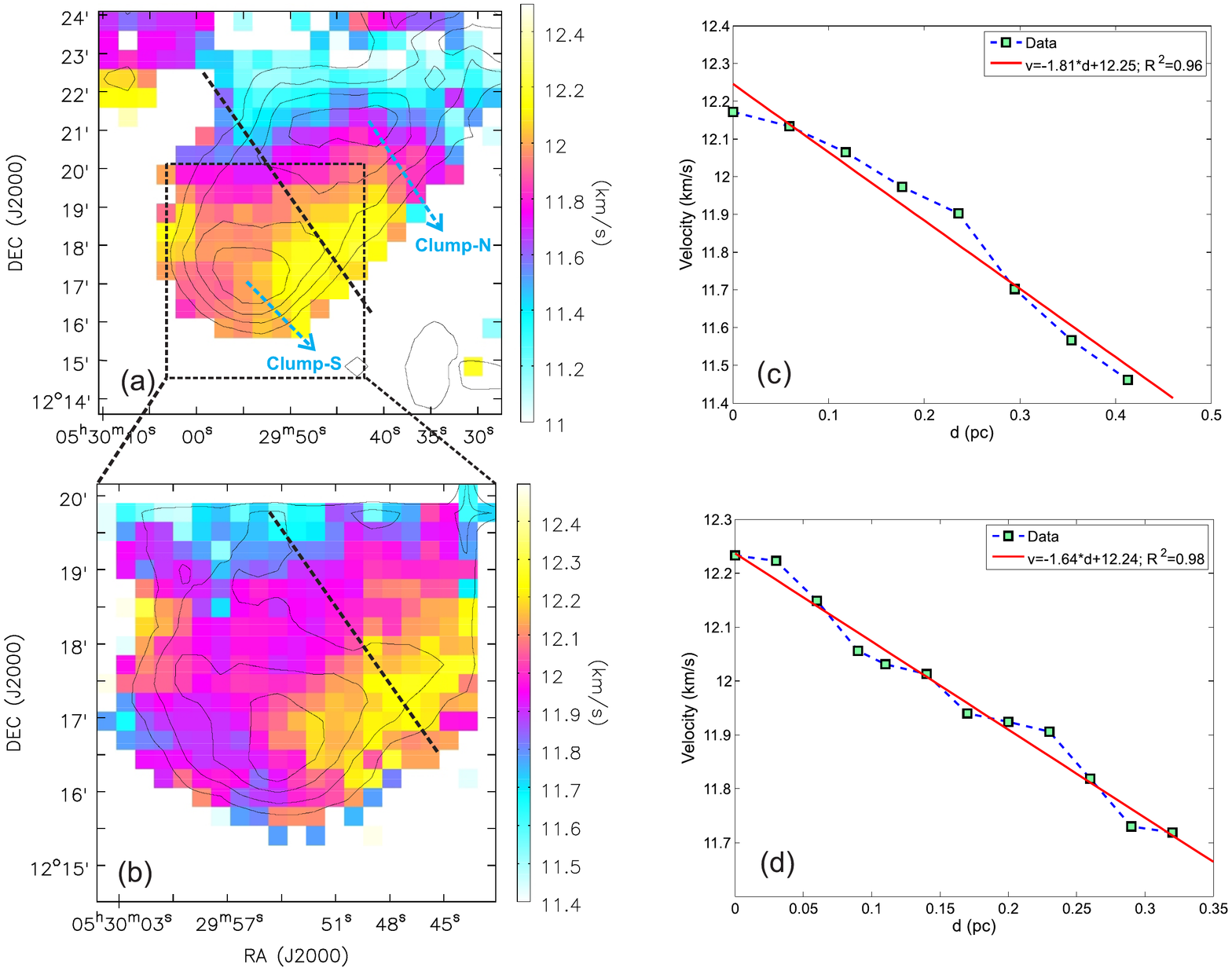}
\caption{The black contours in panels (a) and (b) represent integrated intensity of $^{13}$CO (1-0) and $^{13}$CO (2-1), respectively. The contour levels are from 20\% to 80\% in steps of 20\% of the peak values. The peak values for $^{13}$CO (1-0) and $^{13}$CO (2-1) are 10.8 and 13.0 K~km~s$^{-1}$, respectively. Color images in panels (a) and (b) show the Moment 1 maps of $^{13}$CO (1-0) and $^{13}$CO (2-1), respectively. Panels (c) and (d) show the Position-Velocity diagrams of $^{13}$CO (1-0) and $^{13}$CO (2-1) along the dashed lines in panels (a) and (b), respectively. The distances were calculated from the south-west edge of the clump. The Moment 1 maps are constructed from the data after imposing a cutoff of 5 $\sigma$. }
\end{figure*}

\begin{figure}
\centering
\includegraphics[angle=0,scale=0.4]{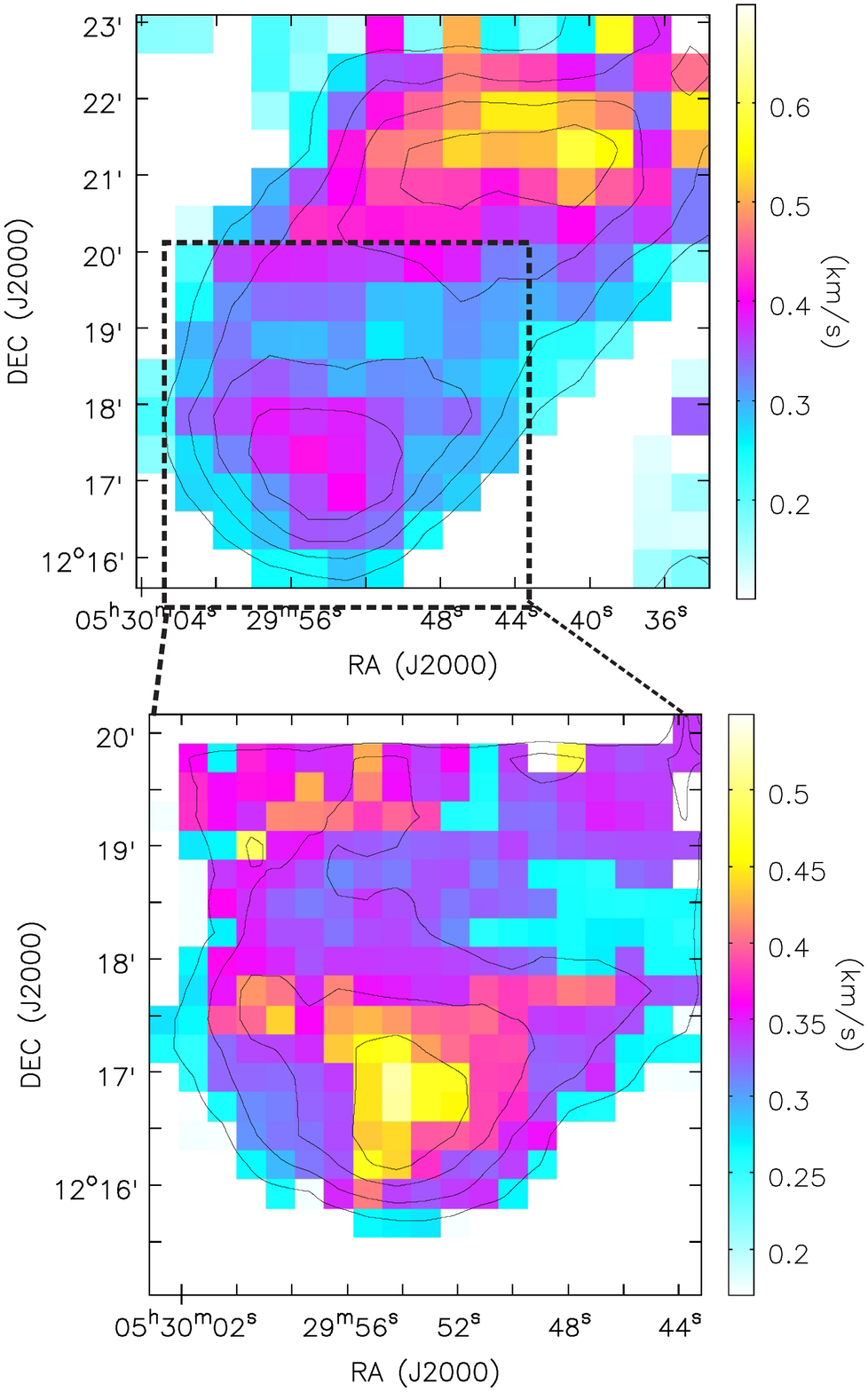}
\caption{Color images show the Moment 2 maps of $^{13}$CO (1-0) and $^{13}$CO (2-1), respectively. The velocity dispersions become larger toward the clump centers, suggesting that the line widths might be broadened by feedback of star formation activity (e.g. outflows) inside the clumps. The black contours in the upper and lower panels represent integrated intensity of $^{13}$CO (1-0) and $^{13}$CO (2-1), respectively. The contour levels are the same as in Figure 2. The Moment 2 maps are constructed from the data after imposing a cutoff of 5 $\sigma$.  }
\end{figure}

The integrated intensity of $^{13}$CO (1-0) is shown as contours overlayed on its Moment 1 color image in panel (a) of Figure 2. Two large clumps (``Clump-N" and ``Clump-S") are revealed in $^{13}$CO (1-0) emission. The integrated intensity of $^{13}$CO (2-1) of ``Clump-S" is shown as contours overlayed on its Moment 1 color image in panel (b) of Figure 2. From the Moment 1 maps, a clear velocity gradient is seen along south-west to north-east direction, indicating that the cloud might be externally compressed. We derived the systemic velocities of $^{13}$CO (1-0) and $^{13}$CO (2-1) from Gaussian fits of the spectra along the dashed lines in panels (a) and (b) and plotted the Position-Velocity diagrams in panels (c) and (d), respectively. The velocities linearly decrease from the south-west to north-east. The direction of velocity gradient is inconsistent with the elongation of the cloud. We notice that the B type star HD 36104 is at the south-west edge of ``Clump-S". Therefore, the velocity gradient may be caused by the joint effect of the radiation pressure from the giant H{\sc ii} region and HD 36104. From linear fits, we found that the velocity gradient of the cloud is 1.6-1.8 km~s$^{-1}$~pc$^{-1}$.

In Figure 3, we present the Moment 2 maps of $^{13}$CO (1-0) and $^{13}$CO (2-1) emission. The velocity dispersions become larger toward the clump centers, suggesting that the line widths might be broadened by feedback of star formation activity (e.g. outflows) inside the clumps.

\subsubsection{Line intensity ratios}

\begin{figure*}
\centering
\includegraphics[angle=-90,scale=0.5]{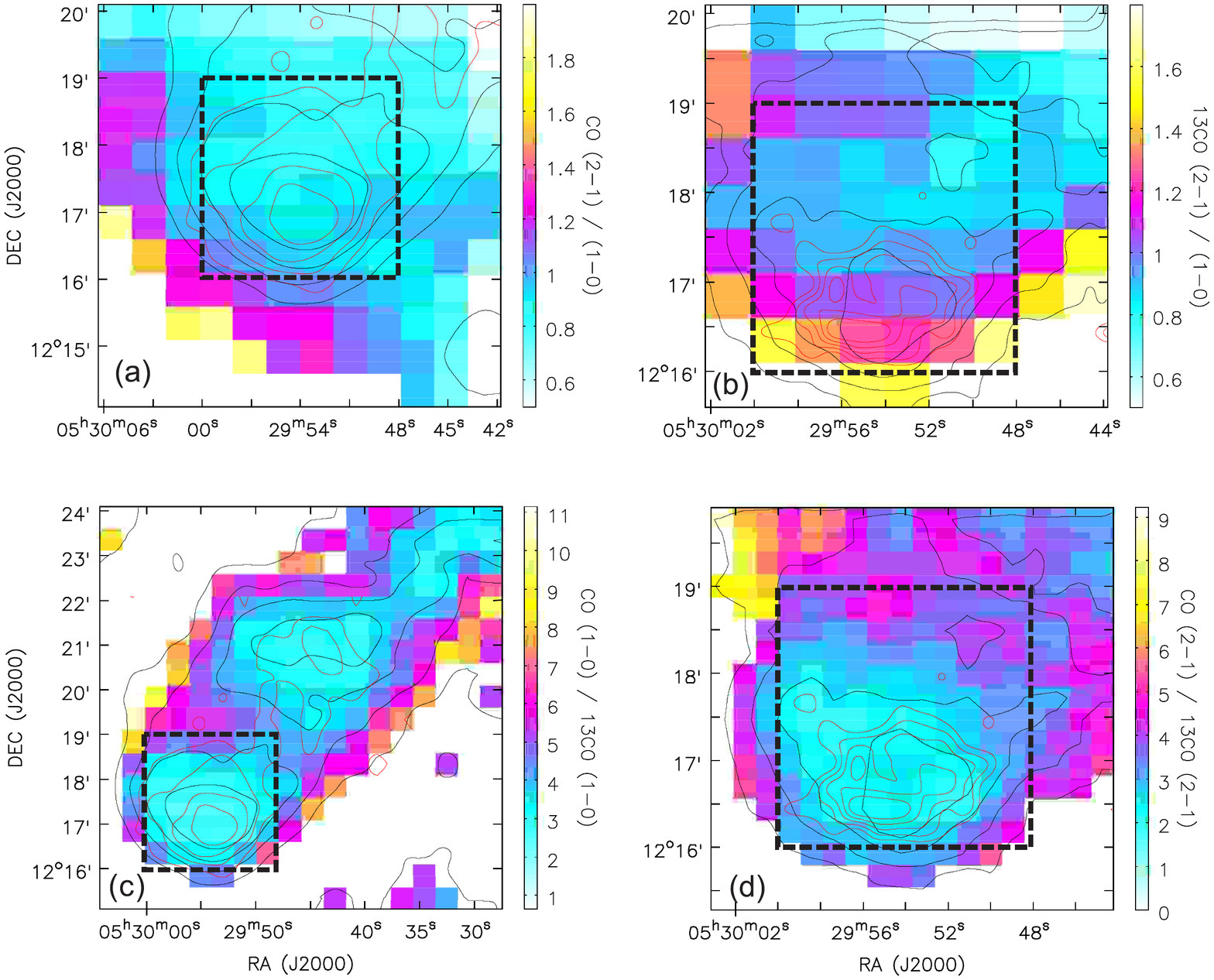}
\caption{Integrated intensity ratios of $^{12}$CO (2-1) to $^{12}$CO (1-0), $^{13}$CO (2-1) to $^{13}$CO (1-0), $^{12}$CO (1-0) to $^{13}$CO (1-0) and $^{12}$CO (2-1) to $^{12}$CO (2-1) are shown in color images in panels (a), (b), (c) and (d), respectively. In panels (a) and (c), the black and red contours represent the integrated intensity of $^{13}$CO (1-0) and C$^{18}$O (1-0), respectively. While in panels (b) and (d), the black and red contours represent the integrated intensity of $^{13}$CO (2-1) and C$^{18}$O (2-1), respectively. The contour levels for $^{13}$CO (1-0), $^{13}$CO (2-1) and C$^{18}$O (1-0)are from 20\% to 80\% in steps of 20\% of their peak values. The peak values of $^{13}$CO (1-0), $^{13}$CO (2-1) and C$^{18}$O (1-0) are 10.8, 13.0 and 1.9 K~km~s$^{-1}$, respectively.  The contour levels for C$^{18}$O (2-1) are from 50\% to 90\% in steps of 10\% of the peak value 3.42 K~km~s$^{-1}$. The ratio maps are constructed from the data after imposing a cutoff of 5 $\sigma$. For easier comparison, a dashed box of the same angular size marking the position and extent of ``Clump-S" was drawn in each panel.}
\end{figure*}

The integrated intensity ratios of $^{12}$CO J=2-1 to J=1-0 (R$_{12CO}$) and $^{13}$CO J=2-1 to J=1-0 (R$_{13CO}$) were calculated by convolving the $^{12}$CO (2-1) and $^{13}$CO (2-1) data with the beam of PMO 13.7 m. The two ratios are shown in color images in Panels (a) and (b) of Figure 4. The ratios become larger from clump center to clump edge. The integrated intensity ratios of $^{12}$CO (1-0) to $^{13}$CO (1-0) (R$^{10}_{12CO/13CO}$) and $^{12}$CO (2-1) to $^{13}$CO (2-1) (R$^{21}_{12CO/13CO}$) were also derived and displayed in panels (c) and (d), respectively. Smaller R$^{10}_{12CO/13CO}$ and R$^{21}_{12CO/13CO}$ ratios indicate higher optical depths for CO emission. The optical depth of CO emission decreases from clump center to clump edge. In Table 1, we present statistical results of these ratios and integrated intensity of $^{13}$CO and C$^{18}$O within 50\% contour (region 1), 20\% contour of C$^{18}$O (1-0) (region 2) and 20\% contour of $^{13}$CO (1-0) (region 3), respectively. These statistical results are not artificial due to low signal-to-noise level because the minimum value of $^{13}$CO (2-1) integrated intensity within its 20\% contour is 1.21 K~km~s$^{-1}$, which is larger than 6 $\sigma$ ($\sigma\sim$0.2 K~km~s$^{-1}$). In general, the average density from region 1 to region 3 decreases. While the mean or maximum values of different ratios generally increase with decreasing average density, which quantitatively indicates that the CO emission at the clump edge is less optically thick and more highly excited than that at the clump center.

\begin{deluxetable*}{ccccc}
\tabletypesize{\scriptsize} \tablecolumns{5} \tablewidth{0pc}
\tablecaption{Statistics of integrated intensities and intensity ratios} \tablehead{
\colhead{Transitions} & \colhead{Mean\tablenotemark{a}} & \colhead{Std\tablenotemark{a}}  & \colhead{Min\tablenotemark{a}} & \colhead{Max\tablenotemark{a}}  } \startdata
&Within 50\% contour of C$^{18}$O (1-0) &\\
\cline{1-5}
CO/$^{13}$CO (1-0)          &  2.80     &  0.31   &   2.49   &   3.28    \\
CO/$^{13}$CO (2-1)          &  2.33     &  0.19   &   1.96   &   2.79    \\
C$^{18}$O (2-1)/(1-0)       &  1.51     &  0.23   &   1.24   &   1.87    \\
CO (2-1)/(1-0)              &  0.91     &  0.04   &   0.85   &   0.96    \\
$^{13}$CO (2-1)/(1-0)       &  1.04     &  0.13   &   0.91   &   1.26    \\
C$^{18}$O (1-0)\tablenotemark{b}             &  1.39     &  0.27   &   1.10   &   1.88    \\
$^{13}$CO (1-0)\tablenotemark{b}             &  9.45     &  1.20   &   7.59   &   10.8    \\
$^{13}$CO (2-1)\tablenotemark{b}             &  10.3     &  1.59   &   7.38   &   13.0    \\
C$^{18}$O (2-1)\tablenotemark{b}             &  2.25     &  0.61   &   1.18   &   3.42    \\
N$_{H_{2}}$\tablenotemark{c}                  &  1.05$\times10^{22}$ & 1.78$\times10^{21}$ & 8.07$\times10^{21}$ & 1.22$\times10^{22}$ \\
T$_{ex}$\tablenotemark{d}                    &  18.84    & 0.79    & 17.66    & 19.89 \\
\cline{1-5}
&Within 20\% contour of C$^{18}$O (1-0) &\\
\cline{1-5}
CO/$^{13}$CO (1-0)          &  3.14      & 0.56   &   2.37   &   4.59    \\
CO/$^{13}$CO (2-1)          &  2.75      & 0.56   &   1.77   &   4.53    \\
C$^{18}$O (2-1)/(1-0)       &  1.79      & 0.52   &   1.00   &   3.30    \\
CO (2-1)/(1-0)              &  0.90      & 0.07   &   0.79   &   1.06    \\
$^{13}$CO (2-1)/(1-0)       &  1.04      & 0.21   &   0.77   &   1.56    \\
C$^{18}$O (1-0)\tablenotemark{b}             &  0.92      & 0.40   &   0.41   &   1.88    \\
$^{13}$CO (1-0)\tablenotemark{b}             &  7.74      & 1.95   &   3.04   &   10.83   \\
$^{13}$CO (2-1)\tablenotemark{b}             &  8.15      & 2.06   &   4.61   &   13.0    \\
C$^{18}$O (2-1)\tablenotemark{b}             &  1.58      & 0.70   &   0.30   &   3.42    \\
\cline{1-5}
&Within 20\% contour of $^{13}$CO (1-0) &\\
\cline{1-5}
CO/$^{13}$CO (1-0)      &  4.14      & 1.40   &   2.37   &   8.14    \\
CO/$^{13}$CO (2-1)      &  3.50      & 1.20   &   1.77   &   10.10   \\
CO (2-1)/(1-0)          &  0.90      & 0.08   &   0.77   &   1.14    \\
$^{13}$CO (2-1)/(1-0)   &  1.11      & 0.29   &   0.72   &   2.27    \\
$^{13}$CO (1-0)\tablenotemark{b}         &  6.23      & 2.35   &   1.79   &    10.83  \\
$^{13}$CO (2-1)\tablenotemark{b}         &  6.62      & 2.41   &   1.21   &    13.0
\enddata
\tablenotetext{a}{We get statistical results from pixel values after beam convolution.}
\tablenotetext{b}{unit in K~km~s$^{-1}$}
\tablenotetext{c}{unit in cm$^{-2}$}
\tablenotetext{d}{unit in K}
\end{deluxetable*}

\subsubsection{Clump properties from LTE analysis}

We first calculated the excitation temperature of $^{12}$CO (1-0) and column density of $^{13}$CO assuming local thermodynamic equilibrium (LTE). The details of the LTE analysis are described in Appendix A.

\begin{figure}
\centering
\includegraphics[angle=90,scale=0.3]{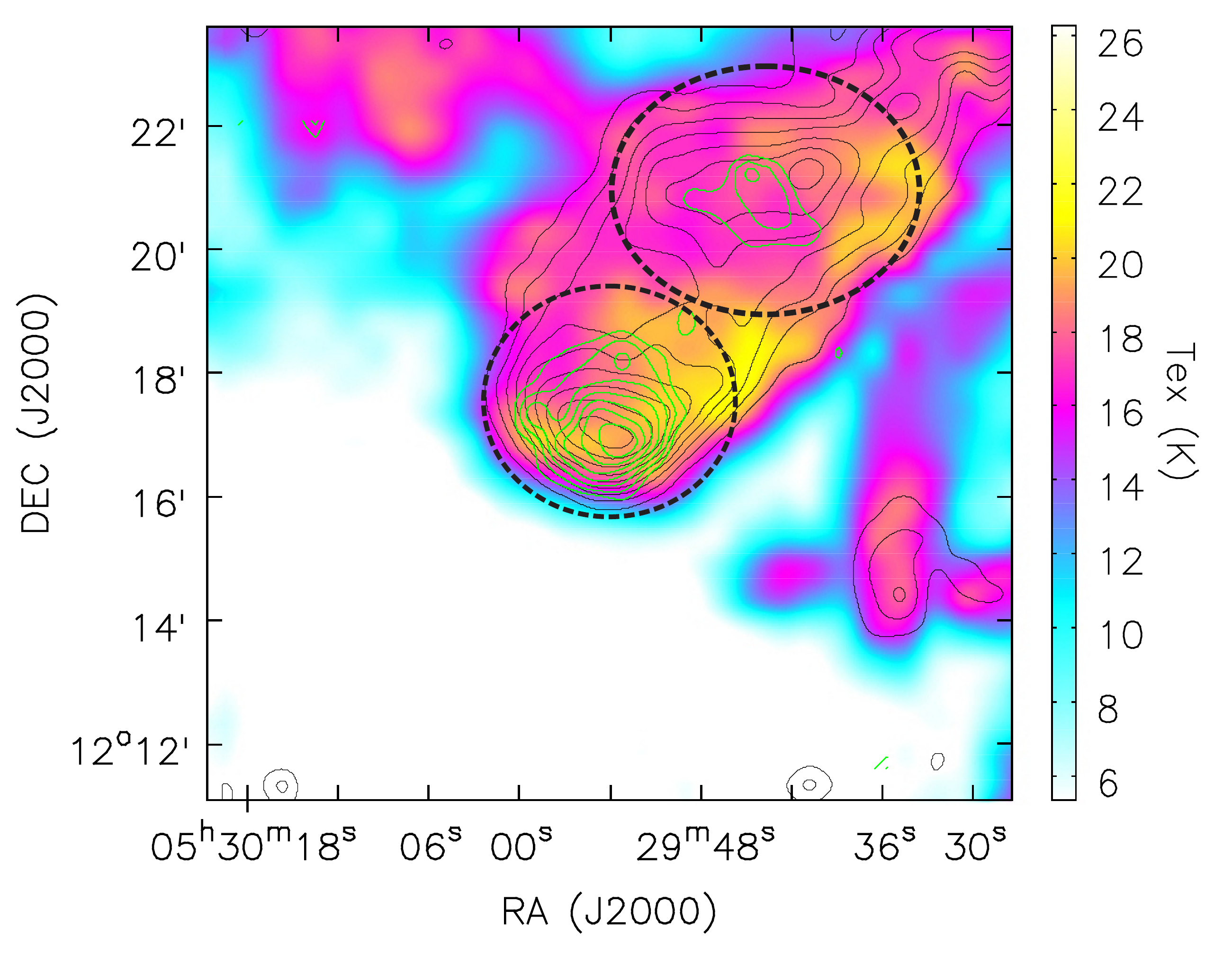}
\caption{H$_{2}$ column density is shown in black contours. The contours are from 20\% to 90\% in steps of 10\% of the peak value 1.22$\times10^{22}$ cm$^{-2}$. Integrated intensity of C$^{18}$O (1-0) is shown in green contours. The contours are from 30\% to 90\% in steps of 10\% of the peak value 1.88 K~km~s$^{-1}$. The excitation temperature of $^{12}$CO (1-0) is shown as a color image. The two dashed ellipses mark the two gas clumps.}
\end{figure}

The mean peak brightness temperature ratio of $^{12}$CO (1-0) to $^{13}$CO (1-0) within 20\% contour of $^{13}$CO (1-0) integrated intensity is $\sim$4, corresponding to an optical depth of 17 for $^{12}$CO (1-0) emission. Thus, $^{12}$CO (1-0) emission is apparently optically thick in PGCC G192.32-11.88. The excitation temperature of $^{12}$CO (1-0) is shown as color image in Figure 5. It can be seen that the excitation temperature (T$_{ex}\sim21$ K) at the western edge is higher than that ($\sim15$K) at the eastern edge of the cloud, indicating that the cloud might be externally heated. The mean excitation temperatures of the ``Clump-S" and ``Clump-N" are 18.1$\pm$0.3 and 18.0$\pm$0.2 K, respectively, which are consistent with the dust temperature (17.3$\pm$6.0 K) of the cold component in Planck observations (Planck Collaboration et al. 2015).

The column density of H$_{2}$ was obtained by adopting typical abundance ratios [H$_{2}$]/[$^{12}$CO]=$10^{4}$ and [$^{12}$CO]/[$^{13}$CO]=60 in the interstellar medium (Deharveng
et al. 2008). The column density of H$_{2}$ is shown as contours in Figure 5. The mean column densities of ``Clump-S" and ``Clump-N" are (6.5$\pm0.4)\times10^{21}$ and (6.3$\pm0.2)\times10^{21}$ cm$^{-2}$, respectively.

The radii of the clumps are defined as $R=\sqrt{a\cdot b}$, where a and b are the full width at half-maximum (FWHM) deconvolved sizes of the major and minor axes, respectively. The radii of ``Clump-S" and ``Clump-N" are 0.33 and 0.47 pc, respectively. The LTE masses of the clumps are estimated as $M_{LTE}=\pi R^{2}\cdot \overline{N}\cdot m_{H}\cdot\mu_{g}$, where $\overline{N}$ is the mean column density, $m_{H}$ is the mass of atomic hydrogen and $\mu_{g}$=2.8 is the mean molecular weight of the gas. The LTE masses of ``Clump-S" and ``Clump-N" are 49$\pm$3 and 99$\pm$3 M$_{\sun}$, respectively. In calculating masses, we only considered the errors in flux measurement. The other uncertainties like distance, abundance and calibration were not taken into account. These systematic uncertainties can have an effect that is much greater than the random noise in the data on mass measurement. For example, the various distance determinations have uncertainties approximately 10\%. If the distance was actually 10\% higher, then the LTE mass determination would be increased by $\sim$20\%. Additionally, as discussed in section 3.3.1, CO gas may be depleted in the central region. Therefore the abundance we adopted here might be larger than the real value. If we adopt an abundance two times smaller than the present value, the corresponding LTE masses will increase by a factor of 2.  To study the gravitationally stability of these two clumps, we also estimated their virial masses. The virial mass considering turbulent support can be derived as $\frac{M_{vir}}{M_{\sun}}=210(\frac{R}{pc})(\frac{\Delta V}{km~s^{-1}})^2$ (Zhang et al. 2015), where $\Delta V$ is the linewidth of C$^{18}$O (1-0). We used C$^{18}$O (1-0) line rather than $^{13}$CO (1-0) line to derive virial masses because that C$^{18}$O (1-0) line emission is more optically thin than $^{13}$CO (1-0) line emission and therefore a better tracer for turbulence. From Gaussian fits toward the spectra at clump centers, the linewidths of C$^{18}$O (1-0) of ``Clump-S" and ``Clump-N" are 0.93$\pm$0.09 and 0.86$\pm$0.11 km~s$^{-1}$, respectively. Thus, the virial masses of ``Clump-S" and ``Clump-N" are 59$\pm$10 and 74$\pm$19 M$_{\sun}$, respectively. The virial masses of the two clumps are consistent with their LTE masses, indicating that these two gas clumps are gravitationally bound.

\subsubsection{CO gas excitation from Non-LTE analysis}

\begin{figure}
\centering
\includegraphics[angle=0,scale=0.4]{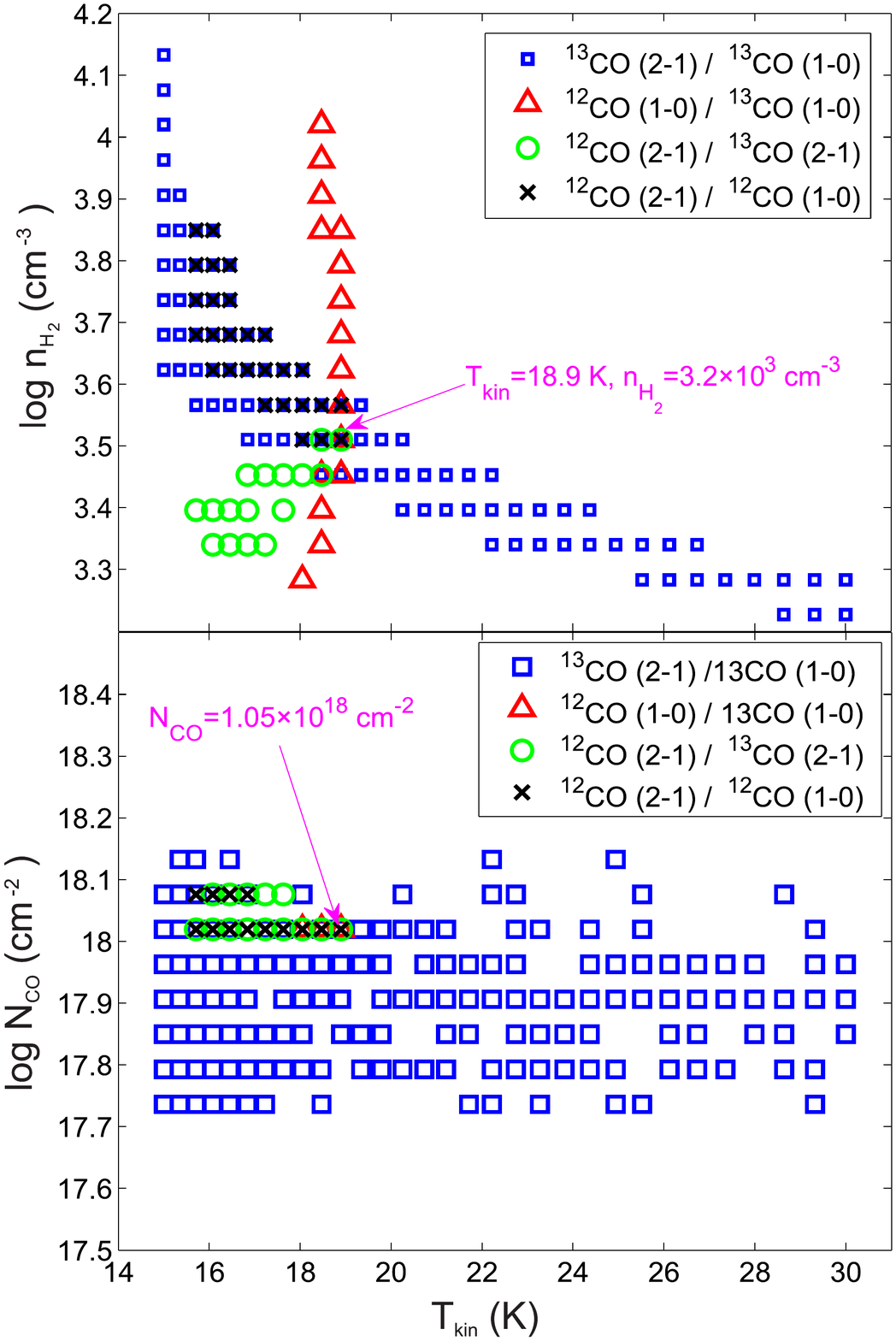}
\caption{RADEX calculation results. Upper panel: n$_{H_{2}}$ vs. T$_{kin}$. Lower panel: N$_{CO}$ vs. T$_{kin}$. The symbols mark the parameter grids that fit the observed line ratios reasonably. }
\end{figure}

In Figure 6, we applied RADEX (Van der Tak et
al. 2007), a one-dimensional non-LTE
radiative transfer code, to investigate the excitation of CO isotopologues within the 50\% contour of C$^{18}$O (1-0) in ``Clump-S" by exploring parameter ranges of  H$_{2}$ volume density in [1$\times10^3$, 5$\times10^4$] cm$^{-3}$, kinetic temperature in [15, 30] K and column density of $^{12}$CO in [1$\times10^{17}$, 5$\times10^{18}$] cm$^{-2}$, respectively. The modelled values which are consistent with observed mean values within 1 $\sigma$ are adopted. For example, the blue boxes in Figure 6 represent the solutions in which modeled R$_{13CO}$ is consistent with the observed value 1.04$\pm$0.13. The best solutions should satisfy the criteria that all the modelled ratios (R$_{12CO}$, R$_{13CO}$, R$^{10}_{12CO/13CO}$, R$^{21}_{12CO/13CO}$) are consistent with the observed values. The kinetic temperature, volume density, CO column density in the overlapped regions are 18.9 K, 3.2$\times$10$^{3}$ cm$^{-3}$ and 1.05$\times10^{18}$ cm$^{-2}$, which are the best solutions. The kinetic temperature (18.9 K) is consistent with the average excitation temperature (18.84$\pm$0.79 K) of $^{12}$CO (1-0) in the LTE analysis as listed in Table 1. Assuming an abundance ratio of [H$_{2}$]/[$^{12}$CO]=$10^{4}$, the H$_{2}$ column density calculated with RADEX is 1.05$\times10^{22}$ cm$^{-2}$, which is the same as the value ((1.05$\pm0.18)\times10^{22}$ cm$^{-2}$) obtained from the LTE analysis, suggesting that the central part of the southern clump satisfies the LTE condition very well. The thickness (h) of ``Clump-S" can be estimated as $h=\frac{\overline{N_{H_{2}}}}{\overline{n_{H_{2}}}}$, where $\overline{N_{H_{2}}}$ and $\overline{n_{H_{2}}}$ are the average column density and volume density, respectively. The thickness of the clump is $\sim$0.95 pc, which is about 1.5 times larger than the diameter (0.66 pc) of the southern clump.

\begin{figure*}
\centering
\includegraphics[angle=-90,scale=0.6]{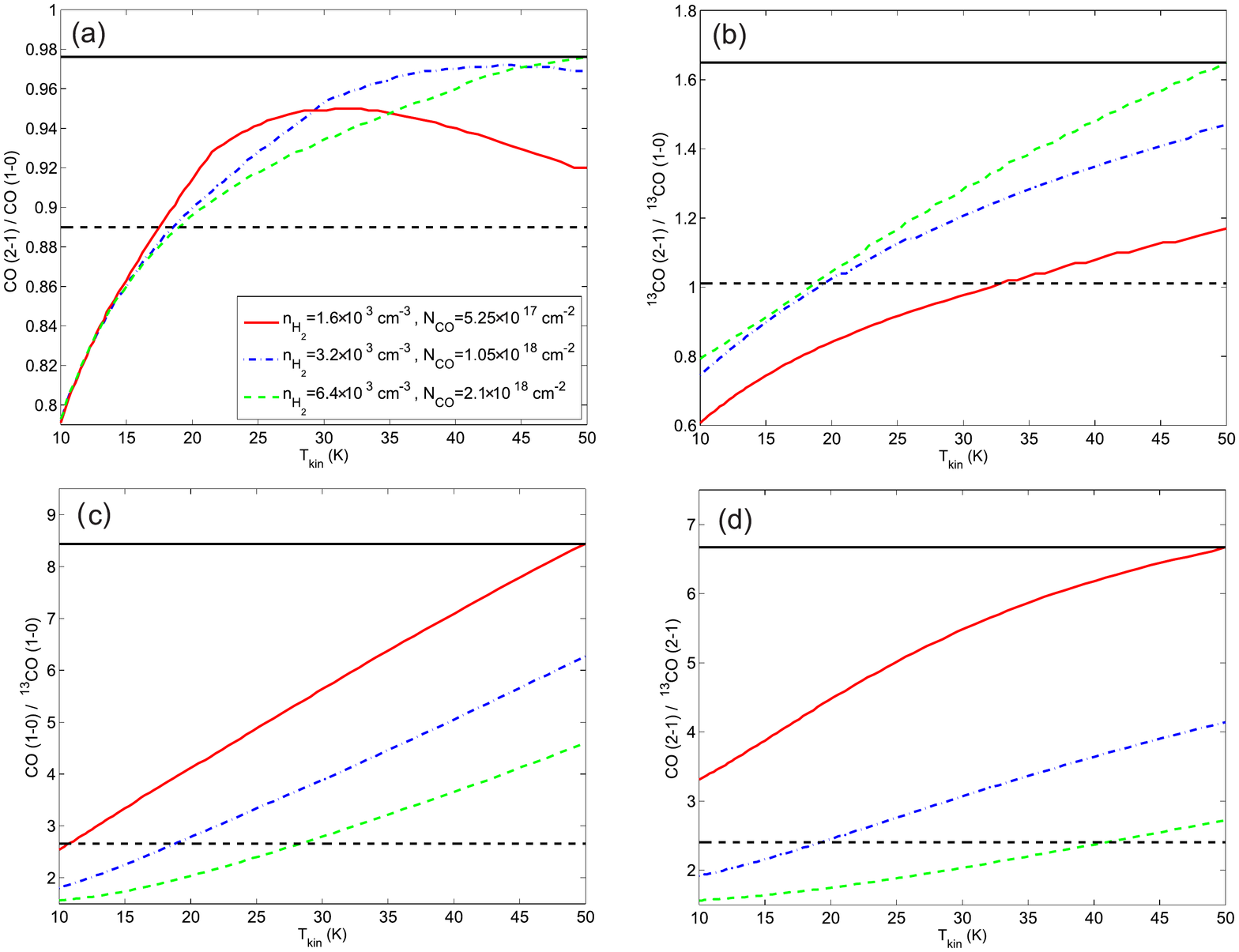}
\caption{RADEX calculation results of intensity ratios vs. T$_{kin}$. The dashed black lines represent the observed intensity ratios at core center. The solid black lines represent the maximum intensity ratios in RADEX calculation. }
\end{figure*}

In Figure 7, we investigate integrated intensity ratios vs. kinetic temperature in three groups of volume density and column density. Roughly speaking, for a given density, integrated intensity ratios increase with kinetic temperature except for R$_{12CO}$ in the lowest density group (n$_{H_{2}}=1.6\times10^{3}$ cm$^{-3}$). For a given kinetic temperature, R$^{10}_{12CO/13CO}$ and R$^{21}_{12CO/13CO}$ decrease with increasing density, indicating that in high density regions CO is more optically thick. As shown in panels (c) and (d), extremely high ratios of R$^{10}_{12CO/13CO}$ and R$^{21}_{12CO/13CO}$ indicate much higher temperature and lower density. As shown in Table 1, the maximum values of R$_{12CO}$, R$_{13CO}$, R$^{10}_{12CO/13CO}$ and R$^{21}_{12CO/13CO}$ within 20\% contour of $^{13}$CO (1-0) are 1.14, 2.27, 8.14, and 10.10, respectively. Such high ratios are larger than the maximum ratios we calculated with RADEX as marked with black solid lines in Figure 7, indicating that the kinetic temperature should be larger than 50 K close to the clump edge. However, the J=1 and 2 levels of CO are only at 5.53 K and 16.60 K above ground level, respectively, indicating that we could not reveal such high temperature (T$_{k}>$ 50 K) conditions accurately with the present data. Observations of higher J transitions of CO are needed to reveal the exact excitation condition at the clump edge. As shown in panels (c) and (d), to achieve extremely high R$^{10}_{12CO/13CO}$ and R$^{21}_{12CO/13CO}$, not only higher temperatures but also lower densities are needed, indicating that the clump might be surrounded by a hot and low density envelope.

\subsection{Dense cores traced by the JCMT/SCUBA-2 850 $\micron$ continuum}

\begin{figure*}
\centering
\includegraphics[angle=90,scale=0.6]{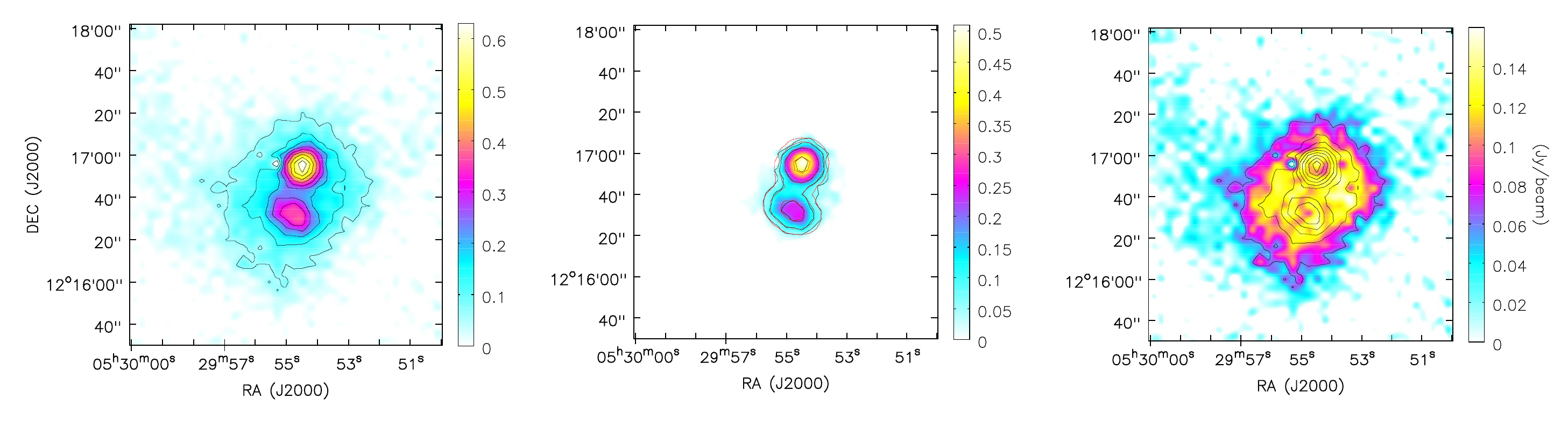}
\caption{Left: JCMT/SCUBA-2 850 $\micron$ continuum emission is shown in contours and color scale. The contours are from 10\% to 90\% in steps of 10\% of the peak value (0.63 Jy~beam$^{-1}$). Middle: JCMT/SCUBA-2 850 $\micron$ continuum emission after subtracting a background value of 0.12 Jy~beam$^{-1}$ is shown in black contours and color scale. The contours are from 20\% to 80\% in steps of 20\% of the peak value (0.51 Jy~beam$^{-1}$). The two Gaussian components of the model emission is shown in red contours. The contours are from 20\% to 80\% in steps of 20\% of the peak value (0.49 Jy~beam$^{-1}$). Right: JCMT/SCUBA-2 850 $\micron$ continuum emission is shown in black contours. JCMT/SCUBA-2 850 $\micron$ continuum emission after subtracting emission of the two Gaussian components of the model emission is shown as a color image. }
\end{figure*}

As shown in the left panel of Figure 8, ``Clump-S" fragments into two dense cores in the 850 $\micron$ continuum from JCMT/SCUBA-2 observations. The southern and northern cores are denoted as G192S and G192N hereafter. The two dense cores are located in a low density halo with a mean flux density of $\sim$0.12 Jy~beam$^{-1}$. After subtracting a local background flux density of 0.12 Jy~beam$^{-1}$, the two dense cores can be fitted with two Gaussian components as shown in the middle panel. The two dense cores were unresolved by JCMT/SCUBA-2. Assuming that the dust emission is optically thin and the dust temperature equals the kinetic temperature (18.9 K) derived from CO isotopologues,
the masses of the two cores can be obtained with the formula
$M=S_{\nu}D^{2}/\kappa_{\nu}B_{\nu}(T_{d})$ (Planck Collaboration et al.
2015), where $S_{\nu}$ is the
flux of the dust emission at 850 $\micron$, D is the distance, and $B_{\nu}(T_d)$ is the Planck
function. Here the ratio of gas to dust is taken as 100. The dust opacity $\kappa_{\nu}=\kappa_{0}(\frac{\nu}{\nu_{0}})^{\beta}$=0.018 cm$^{2}$g$^{-1}$, where $\beta$=1.4 is the dust opacity index obtained from PGCC catalog
and $\kappa_{0}=0.01$ cm$^{2}$g$^{-1}$ is the dust opacity at
230 GHz derived from (Ossenkopf \& Henning 1994). The flux of G192N and G192S from two Gaussian components fit are 0.51$\pm$0.04 and 0.27$\pm$0.03 Jy, corresponding to core masses of 0.43$\pm$0.03 and 0.23$\pm$0.03 M$_{\sun}$, respectively. We only considered flux uncertainties in estimating the mass uncertainties. However, other parameters like dust temperature, dust opacity and distance should also have a great effect on the mass measurement. For example, if we adopt a dust opacity index of 2, the inferred mass should be reduced by $\sim$23\%. In the right panel of Figure 8, we subtracted the two modelled Gaussian components and presented the residual as a color image. The total flux of the residual within the 10\% contour of 850 $\micron$ continuum emission is 1.36$\pm$0.29 Jy, corresponding to a mass of 1.16$\pm$0.25 M$_{\sun}$. The core masses of G192N and G912S without subtracting local background are 0.53$\pm$0.04 and 0.33$\pm$0.03 M$_{\sun}$, respectively. The corresponding beam-averaged column densities of G192N and G192S are (1.1$\pm0.1)\times10^{22}$ and (6.8$\pm0.6)\times10^{21}$ cm$^{-2}$, respectively, which are used to calculate the abundances of molecules in each core as below. The beam-averaged column density of G192N is consistent with the column density estimated from CO isotopologues above. G192S has lower density than G192N. The total mass inferred from JCMT/SCUBA-2 850 $\micron$ continuum, including both dense cores and low density halo, is $1.94\pm0.32$ M$_{\sun}$, which is roughly consistent with the mass (3.0$\pm$2.4) M$_{\sun}$ inferred from the Planck data, indicating that JCMT/SCUBA-2 traced the same cold component as that of Planck. The SCUBA-2 850 $\mu$m continuum band may be contaminated by the $^{12}$CO (3-2) line (Drabek et al. 2012). From RADEX calculation, we estimated that the integrated intensities of $^{12}$CO (3-2) line are $\sim$22 and 21 K~km~s$^{-1}$ for G192N and G192S, respectively. Adopting a molecular line conversion factor of 0.7 mJy~beam$^{-1}$ per K~km~s$^{-1}$ in the grade 3 weather band (Drabek et al. 2012), the contaminations of $^{12}$CO (3-2) line at SCUBA-2 850 $\mu$m continuum band are 15.4 and 14.7 mJy~beam$^{-1}$, corresponding to only $\sim$3\% and $\sim$4\% of the 850 $\micron$ flux for G192N and G192S, respectively. Therefore, line contamination does not greatly affect the mass calculation above.

\subsection{Chemical properties of two dense cores}

\subsubsection{CO depletion}

\begin{figure*}
\centering
\includegraphics[angle=90,scale=0.5]{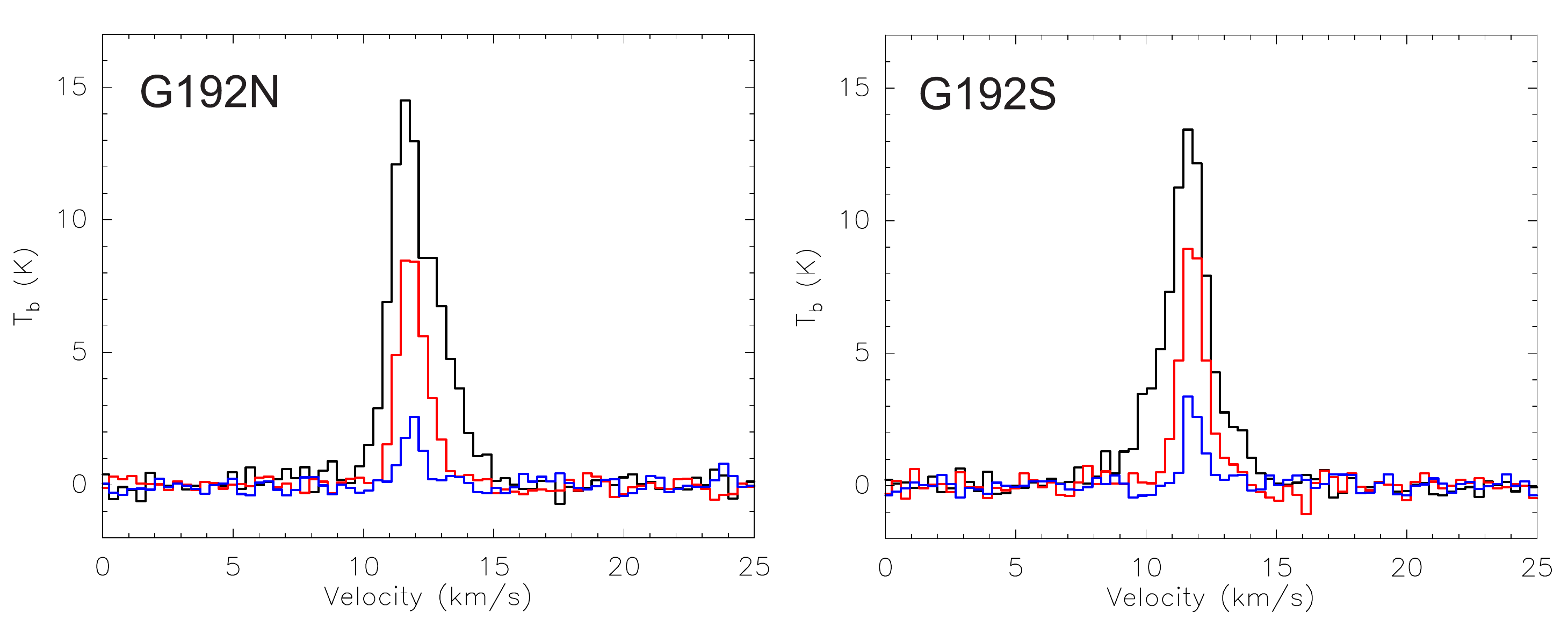}
\caption{CSO spectra toward G192N (left) and G192S (right). Spectra of $^{12}$CO (2-1), $^{13}$CO (2-1) and C$^{18}$O (2-1) are shown in black, red and blue, respectively.}
\end{figure*}

\begin{figure*}
\centering
\includegraphics[angle=90,scale=0.5]{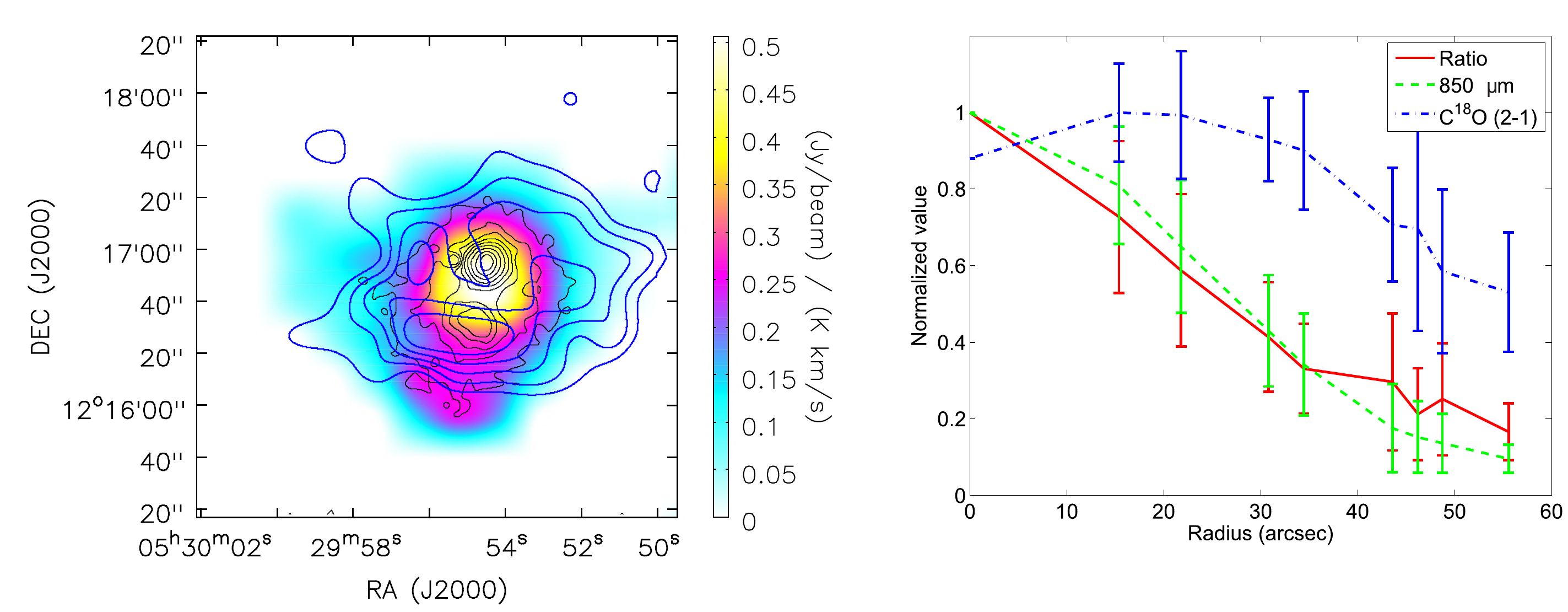}
\caption{Left: The ratio of JCMT/SCUBA-2 850 $\micron$ intensity to integrated intensity of C$^{18}$O (2-1) is shown as color image. JCMT/SCUBA-2 850 $\micron$ continuum emission is shown in black contours. The contours are from 10\% to 90\% in steps of 10\% of the peak value (0.63 Jy~beam$^{-1}$). The blue contours represent integrated intensity of C$^{18}$O (2-1). The contour levels are from 50\% to 90\% in steps of 10\% of the peak value. Right: The circularly averaged JCMT/SCUBA-2 850 $\micron$ intensity (after convolved with CSO beam), integrated intensity of C$^{18}$O (2-1) and their ratio as function of radius centered at the emission peak of 850 $\micron$ continuum emission. }
\end{figure*}

Figure 9 shows the spectra of J=2-1 of CO isotopologues from CSO observations. All the spectra are single-peaked. Weak wing emission can be identified in the $^{12}$CO (2-1) emission. We fitted these spectra with Gaussian profiles and listed the fitted results in Table 2. The brightness temperature ratios of $^{13}$CO (2-1) to C$^{18}$O (2-1) toward G192N and G192S are 3.5 and 2.7, respectively. Assuming a terrestrial abundance ratio
of 5.5 for [$^{13}$CO/C$^{18}$O], the optical depths of $^{13}$CO (2-1) toward G192N and G192S are $\sim1.3$ and $\sim2.3$. While the optical depths of C$^{18}$O (2-1) toward G192N and G192S are $<0.5$. Compared with $^{12}$CO (2-1) and $^{13}$CO (2-1), C$^{18}$O (2-1) emission is more optically thin and can reveal the actual CO column density distribution. In Figure 10, we present the integrated intensity of C$^{18}$O (2-1) emission in blue contours and JCMT/SCUBA-2 850 $\micron$ continuum in black contours. The emission peak of C$^{18}$O (2-1) is clearly displaced from continuum emission peaks, indicating that the CO gas may be depleted toward the clump center. Assuming LTE and optically thin for both 850 $\micron$ continuum and C$^{18}$O (2-1) emission, the ratio of 850 $\micron$ continuum to C$^{18}$O (2-1) could reflect the CO depletion factor in the clump neglecting any temperature change. We convolved the 850 $\micron$ data with the CSO beam and presented the flux ratio of 850 $\micron$ continuum to C$^{18}$O (2-1) as a color image in Figure 10. The flux ratio toward the clump center is larger than the ratio at the clump edge. In the right panel of Figure 10, we investigated the circularly averaged JCMT/SCUBA-2 850 $\micron$ intensity, integrated intensity of C$^{18}$O (2-1) and their ratio as function of radius. JCMT/SCUBA-2 850 $\micron$ intensity drops nearly linearly from center to edge. Integrated intensity of C$^{18}$O (2-1) peaks at $\sim15\arcsec$ from the clump center. The ratio of 850 $\micron$ intensity to integrated intensity of C$^{18}$O (2-1) changes by a factor of five from center to edge, indicating that CO gas is severely depleted in the clump center. The column densities and abundances of C$^{18}$O toward G192N and G192S are presented in Table 3. The C$^{18}$O abundance of G192S is about two times larger than that of G192N, indicating that CO gas in G192N is more depleted than in G192S.

\begin{deluxetable*}{cccccccccccc}
\tabletypesize{\scriptsize} \tablecolumns{12} \tablewidth{0pc}
\tablecaption{Parameters of molecular lines in Single-dish observations} \tablehead{
\colhead{Lines} & & & G192N & & & &&& G192S  &\\
\cline{2-6}\cline{8-12}\\
 & \colhead{Area} & \colhead{V$_{lsr}$} & \colhead{FWHM } & \colhead{T$_{b}$ }  &
\colhead{rms} & & \colhead{Area} & \colhead{V$_{lsr}$} & \colhead{FWHM } & \colhead{T$_{b}$ }  &
\colhead{rms} \\
\colhead{}  & \colhead{(K~km~s$^{-1}$)}    &\colhead{(km~s$^{-1}$)} &
\colhead{(km~s$^{-1}$)} &
\colhead{(K)} &
\colhead{(K)}& & \colhead{(K~km~s$^{-1}$)}    &\colhead{(km~s$^{-1}$)} &
\colhead{(km~s$^{-1}$)} &
\colhead{(K)} &
\colhead{(K)}
  } \startdata
CO (2-1)     &  29.78$\pm$0.46 &    11.88$\pm$0.01 &     2.17$\pm$0.04  & 12.86   & 0.33 &  & 25.99$\pm$0.43 &    11.59$\pm$0.01 &    2.00$\pm$0.04 &   12.19 &  0.31  \\
$^{13}$CO (2-1)   &  12.11$\pm$0.24 &    11.84$\pm$0.01 &     1.30$\pm$0.03  &  8.74   & 0.23 &  & 11.07$\pm$0.33 &    11.78$\pm$0.01 &    1.13$\pm$0.04 &    9.19 &  0.23  \\
C$^{18}$O (2-1)   &   2.43$\pm$0.20 &    11.88$\pm$0.03 &     0.90$\pm$0.09  &  2.53   & 0.23 &  &  2.72$\pm$0.18 &    11.77$\pm$0.02 &    0.74$\pm$0.06 &    3.43 &  0.22  \\
HCO$^{+}$ (1-0)   &   7.15$\pm$0.04 &    12.48$\pm$0.00 &     1.79$\pm$0.01  &  3.75   & 0.07 &  &  5.95$\pm$0.04 &    12.25$\pm$0.01 &    1.37$\pm$0.01 &    4.06 &  0.07  \\
H$^{13}$CO$^{+}$ (1-0) &   1.05$\pm$0.03 &    12.08$\pm$0.01 &     0.80$\pm$0.03  &  1.23   & 0.08 &  &  1.07$\pm$0.02 &    12.10$\pm$0.01 &    0.68$\pm$0.02 &    1.47 &  0.09  \\
H$_{2}$CO ($2_{1,2}-1_{1,1}$)         &   5.05$\pm$0.05 &    12.50$\pm$0.01 &     2.73$\pm$0.03  &  1.73   & 0.09 &  &  3.46$\pm$0.04 &    12.27$\pm$0.01 &    1.60$\pm$0.02 &    2.02 &  0.08  \\
HDCO ($2_{0,2}-1_{0,1}$)      &   0.59$\pm$0.03 &    12.04$\pm$0.02 &     0.94$\pm$0.05  &  0.59   & 0.08 &  &  0.51$\pm$0.02 &    12.15$\pm$0.01 &    0.50$\pm$0.03 &    0.96 &  0.09
\enddata
\end{deluxetable*}

\subsubsection{Molecular abundances from KVN observations}

\begin{figure*}
\centering
\begin{minipage}[c]{0.45\textwidth}
  \centering
  \includegraphics[angle=-90,scale=.3]{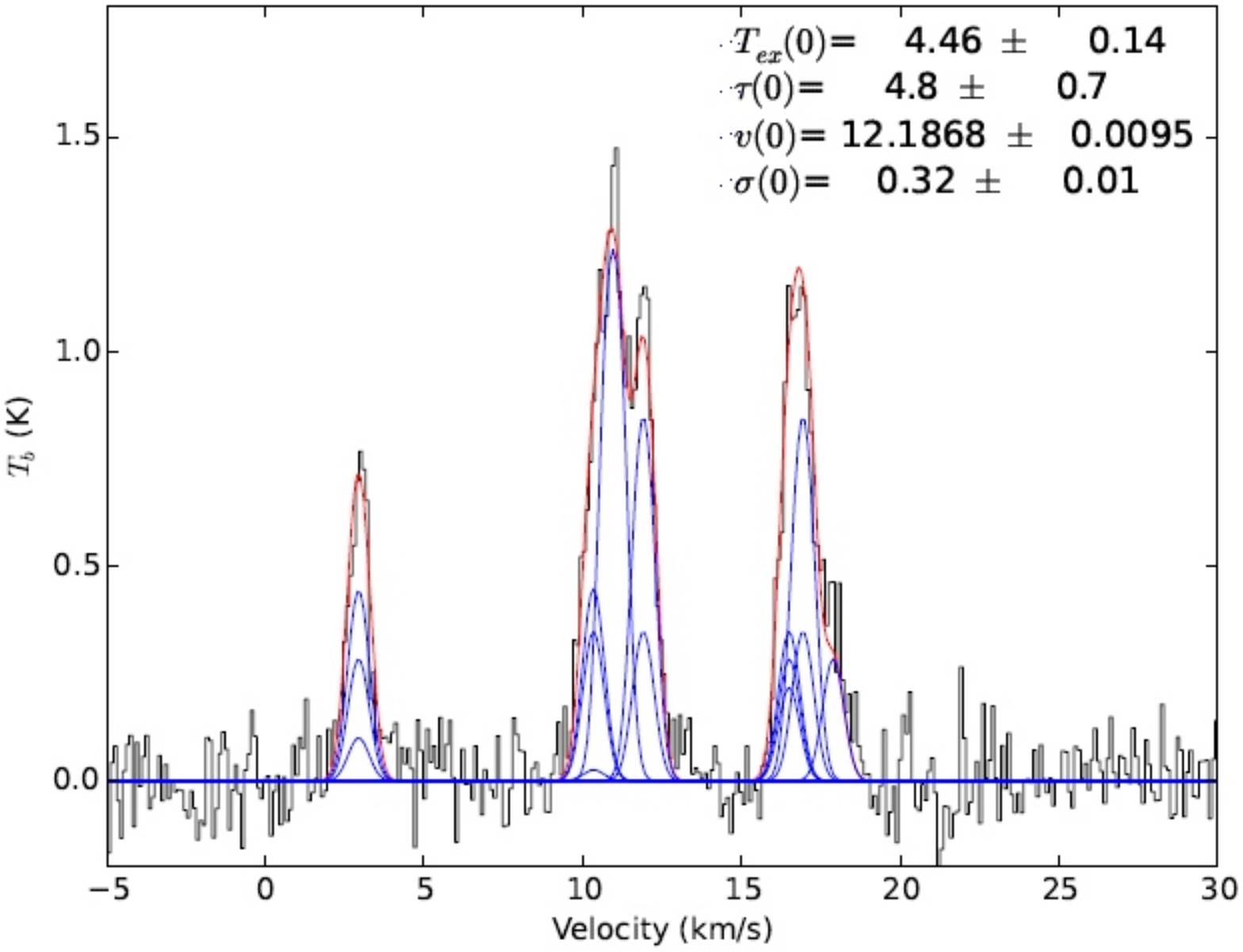}
\end{minipage}
\begin{minipage}[c]{0.45\textwidth}
  \centering
  \includegraphics[angle=-90,scale=.3]{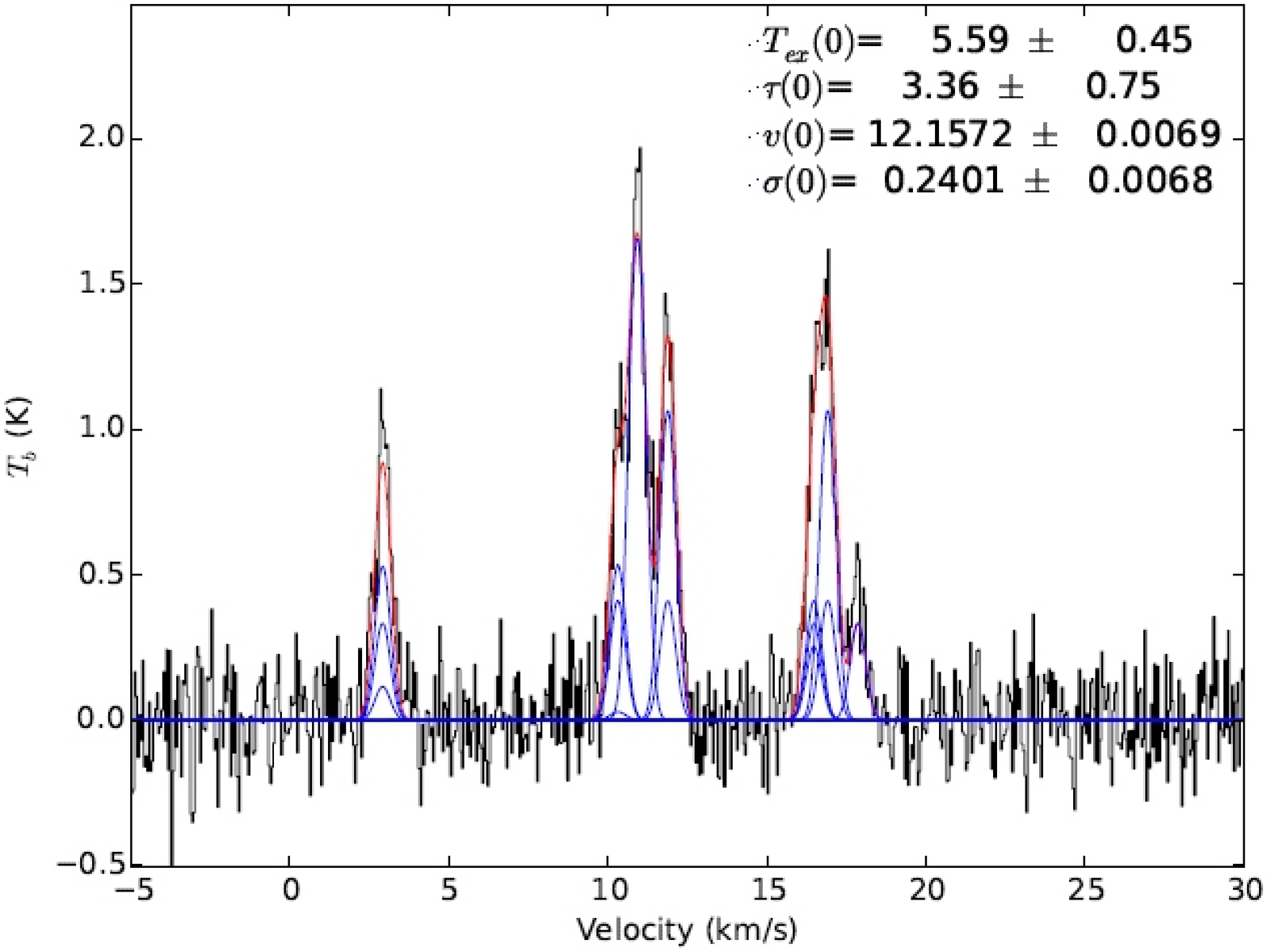}
\end{minipage}
\caption{KVN N$_{2}$H$^{+}$ (1-0) spectra toward G192N (left) and G192S (right). The fitted hyperfine components and their sums are shown in blue and red.}
\end{figure*}

\begin{figure}
\centering
\includegraphics[angle=0,scale=0.4]{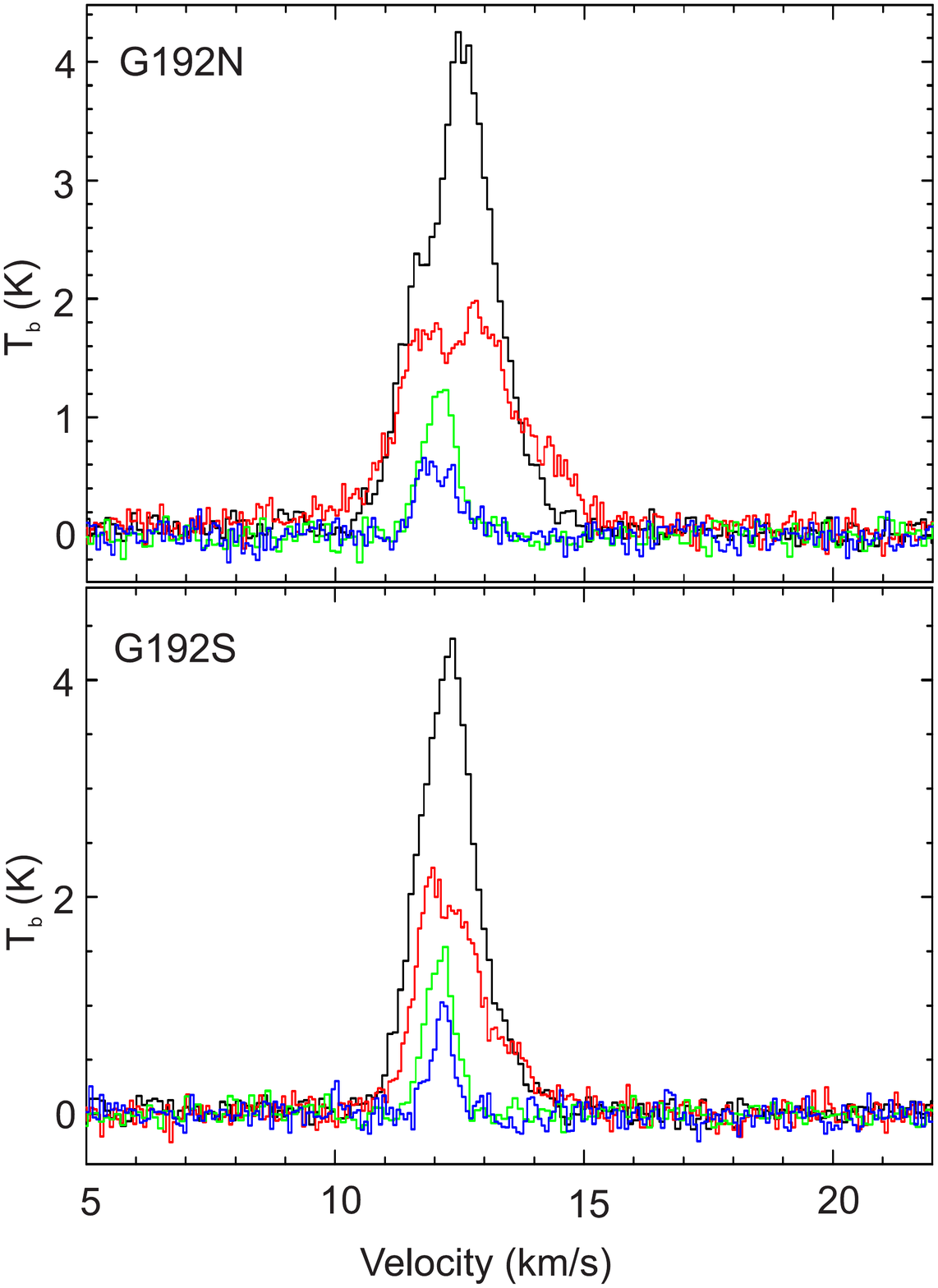}
\caption{KVN spectra toward G192N (upper) and G192S (lower). Spectra of HCO$^{+}$ (1-0), H$^{13}$CO$^{+}$ (1-0), o-H$_{2}$CO ($2_{1,2}-1_{1,1}$) and HDCO ($2_{0,2}-1_{0,1}$) are shown in black, green, red and blue, respectively.}
\end{figure}

Figure 11 shows the spectra of N$_{2}$H$^{+}$ (1-0) toward G192N and G192S obtained with the KVN Yonsei 21 m telescope. We fitted the hyperfine structure of N$_{2}$H$^{+}$ (1-0) with the pyspeckit package\footnote{PySpecKit is an extensible spectroscopic analysis toolkit for astronomy. http://pyspeckit.bitbucket.org/html/sphinx/index.html\#} assuming LTE. The excitation temperature T$_{ex}$(0), optical depth $\tau$(0), systemic velocity v(0) and velocity dispersion of the central component are shown in the upper-right corners in each panel. G192N has a lower excitation temperature and a higher optical depth than G192S, indicating that G192N might be colder and denser than G192S.

\begin{deluxetable}{ccc}
\tabletypesize{\scriptsize} \tablecolumns{3} \tablewidth{0pc}
\tablecaption{Column densities and abundances} \tablehead{
\colhead{Species}& \colhead{Column density} & \colhead{Abundance\tablenotemark{a}}   \\
\colhead{} &\colhead{(cm$^{-2}$)} &} \startdata
&G192N &\\
\cline{1-3}
H$_{2}$\tablenotemark{b} &          (1.1$\pm0.1)\times10^{22}$     & ---  \\
C$^{18}$O&         (2.5$\pm0.2)\times10^{15}$     & $(1.4\pm0.2)\times10^{-6}$    \\
H$^{13}$CO$^{+}$&  (1.1$\pm0.1)\times10^{12}$     & $(5.2\pm0.5)\times10^{-10}$    \\
N$_{2}$H$^{+}$&    (4.4$\pm0.6)\times10^{13}$     & $(2.1\pm0.4)\times10^{-8}$      \\
o-H$_{2}$CO &              $\sim$1.5$\times10^{14}$     & $(4.1\pm0.3)\times10^{-8}$     \\
HDCO               &  $(1.9\pm0.1)\times10^{12}$  & $(5.0\pm0.6)\times10^{-10}$     \\
\cline{1-3}
&G192S   & \\
\cline{1-3}\\
H$_{2}$\tablenotemark{b} &          (6.8$\pm0.2)\times10^{21}$ & ---\\
C$^{18}$O &        (2.8$\pm0.2)\times10^{15}$      & $(2.4\pm0.2)\times10^{-6}$  \\
H$^{13}$CO$^{+}$&  (1.1$\pm0.1)\times10^{12}$      & $(8.4\pm1.0)\times10^{-10}$   \\
N$_{2}$H$^{+}$&    (2.3$\pm0.5)\times10^{13}$      & $(1.8\pm0.4)\times10^{-8}$       \\
o-H$_{2}$CO &              $\sim$1.1$\times10^{14}$      & $(2.9\pm0.3)\times10^{-8}$      \\
HDCO               &  $(1.7\pm0.1)\times10^{12}$   &  $(7.3\pm0.5)\times10^{-10}$
\enddata
\tablenotetext{a}{corrected with filling factor as described in section 3.3.2}\\
\tablenotetext{b}{derived from SCUBA-2 850 $\micron$ continuum as described in section 3.2.}
\end{deluxetable}

Figure 12 shows spectra of other molecular transitions from the KVN single-dish observations. For G192N, both HCO$^+$ (1-0) and o-H$_{2}$CO ($2_{1,2}-1_{1,1}$) lines show clear line wing emission. The o-H$_{2}$CO ($2_{1,2}-1_{1,1}$) line has a double peak profile and larger line width than HCO$^+$ (1-0), indicating that the profile of the o-H$_{2}$CO ($2_{1,2}-1_{1,1}$) line is greatly affected by outflows. In contrast, the optically thin lines H$^{13}$CO$^+$ (1-0) and HDCO ($2_{0,2}-1_{0,1}$) have much smaller line widths and seem to be unaffected by outflow activities. Interestingly, HDCO ($2_{0,2}-1_{0,1}$) also has a double peak profile, which might be a hint for rotation. The separation of the two peaks is $\sim$0.6 km~s$^{-1}$, about 8 times larger than the spectral resolution (0.073 km~s$^{-1}$). We will discuss the profile of HDCO ($2_{0,2}-1_{0,1}$) in section 5.2. For G192S, both HCO$^+$ (1-0) and o-H$_{2}$CO ($2_{1,2}-1_{1,1}$) lines show weak line wings. H$^{13}$CO$^+$ (1-0) and HDCO ($2_{0,2}-1_{0,1}$) are single-peaked and have very narrow line widths. We fitted all the spectra with Gaussian profiles and listed the results in Table 2. In general, molecular lines of G192N have larger line widths than those of G192S. The systemic velocity for both G192N and G192S is 12 km~s$^{-1}$, which is determined from optically thin lines (C$^{18}$O (2-1), H$^{13}$CO$^+$ (1-0) and HDCO ($2_{0,2}-1_{0,1}$)). For a kinetic temperature of 18.9 K, the sound speed is $\sim$0.24 km~s$^{-1}$. We found that the non-thermal velocity dispersions (0.33-0.39 km~s$^{-1}$) of all optically thin lines C$^{18}$O (2-1), H$^{13}$CO$^+$ (1-0) and HDCO ($2_{0,2}-1_{0,1}$) toward G192N are larger than the sound speed. The non-thermal velocity dispersions (0.28-0.31 km~s$^{-1}$) of C$^{18}$O (2-1) and H$^{13}$CO$^+$ (1-0) toward G192S are also larger than the sound speed. The non-thermal velocity dispersion (0.20$\pm$0.01 km~s$^{-1}$) of HDCO ($2_{0,2}-1_{0,1}$) toward G192S is smaller than the sound speed, indicating that the linewidth of HDCO ($2_{0,2}-1_{0,1}$) contains a subsonic turbulent component, probably suggesting turbulence dissipation.

We calculated the beam-averaged column densities of different molecular species in Appendix B and present them in Table 3. The abundance ``X$_{m}$" of molecule ``m" can be derived as:
\begin{equation}
X_{m}=\frac{N_{m}\cdot\eta_{m}}{N_{H_{2}}\cdot\eta_{c}}
\end{equation}
where $N_{m}$, $\eta_{m}$, $N_{H_{2}}$ and $\eta_{c}$ are the beam-averaged column density of molecule ``m", filling factor of molecular line emission, beam-averaged column density of H$_{2}$ estimated from SCUBA-2 850 $\micron$ continuum and filling factor of continuum emission. We remind that our single-dish observations could not resolve the sources. In another Class 0 source HH 212 located in Orion complex, the extents of continuum and dense molecular line (e.g. HCO$^{+}$) emissions are about 2$\arcsec$-5$\arcsec$ (Lee et al. 2014), which are much smaller than the beam sizes of JCMT and KVN. We assume that continuum and molecular line emissions have similar source sizes that are much smaller than the beam sizes of JCMT and KVN in G192N and G192S, $\frac{\eta_{m}}{\eta_{c}}\approx\frac{\theta_{KVN}^{2}}{\theta_{JCMT}^{2}}$, where $\theta_{KVN}$ and $\theta_{JCMT}$ are the beam sizes of KVN and JCMT, respectively. Then the abundances were derived from equation (1) and presented in the last column of Table 3. It should be noted that further higher angular resolution observations are needed to resolve the line and continuum emission to determine their source filling factors and to study the chemistry in detail.

G192N and G192S have a similar N$_{2}$H$^{+}$ abundance. The H$^{13}$CO$^+$ abundance of G192S is about 1.6 times larger than that of G192N, indicating that H$^{13}$CO$^+$ in G192N might be as depleted as CO. The HDCO column densities of G192N and G192S are $(1.9\pm0.1)\times10^{12}$ and $(1.7\pm0.1)\times10^{12}$ cm$^{-2}$, respectively. The o-H$_{2}$CO column densities of G192N and G192S are $\sim$1.5$\times10^{14}$ and $\sim$1.1$\times10^{14}$ , respectively. Assuming that the ortho/para ratio for H$_{2}$CO is the statistical value of 3 (Roberts et al. 2012), the [HDCO]/[H$_{2}$CO] ratios of G192N and G192S are $(1.0\pm0.1)\times10^{-2}$ and $(1.1\pm0.1)\times10^{-2}$, respectively.

\section{Results of SMA observations}

\subsection{SMA 1.3 mm continuum sources and their associations with infrared sources}

\begin{figure*}
\centering
\includegraphics[angle=90,scale=0.5]{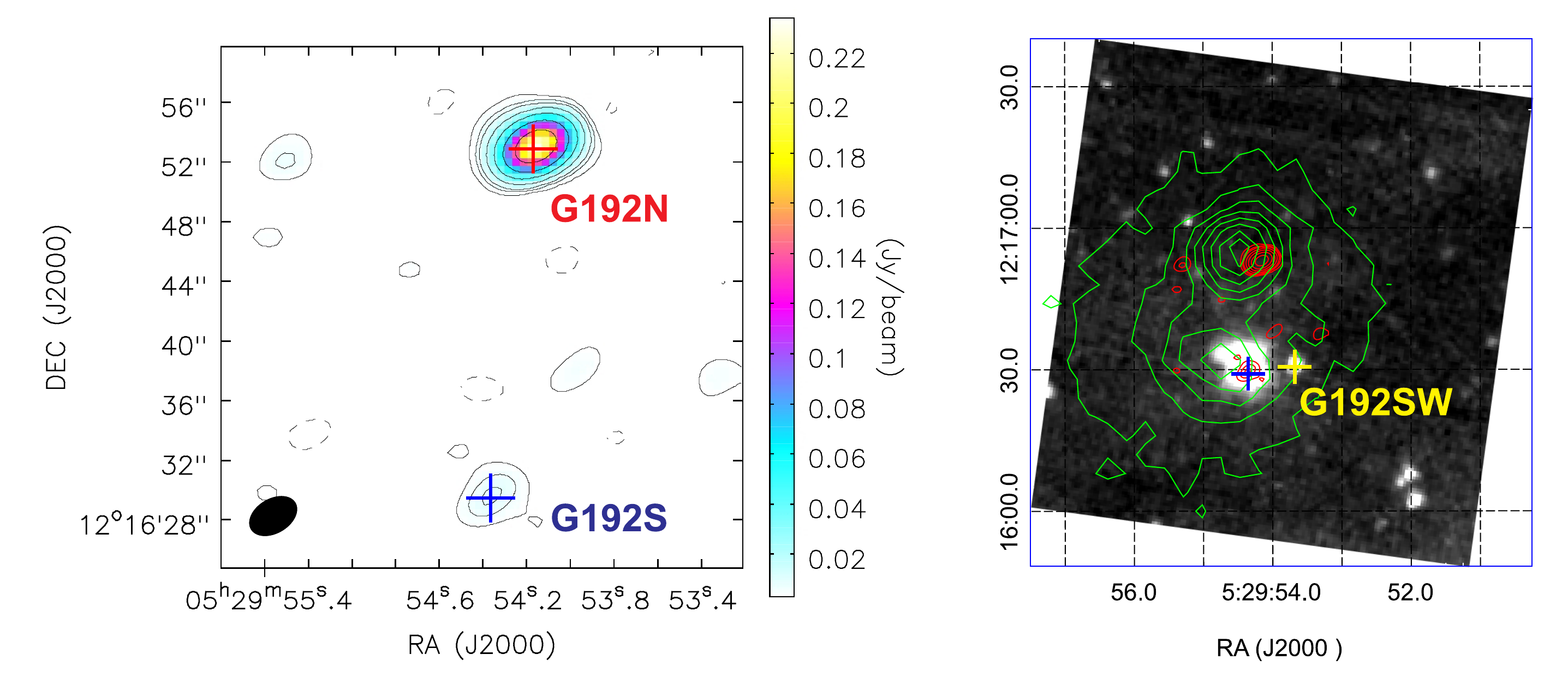}
\caption{Left: The SMA 1.3 mm continuum is shown in black contours and color scale. The contours are (-3, 3, 6, 9, 18, 36, 72, 144)$\times$1 $\sigma$ (1 $\sigma$=1 mJy~beam$^{-1}$). The red and blue crosses mark G192N and G192S, respectively. Right: The SMA 1.3 mm continuum is shown in red contours overlayed on Spitzer/IRAC 4.5 $\micron$ emission. JCMT/SCUBA-2 850 $\micron$ continuum emission is shown in green contours. The contours are from 10\% to 90\% in steps of 10\% of the peak value 0.63 Jy~beam$^{-1}$. The yellow cross marks another YSO G192SW.  }
\end{figure*}

As shown in the left panel of Figure 13, G192N contains a compact condensation in SMA 1.3 mm continuum, which is marked with a red cross. Its peak position is R.A.(J2000)~=~05$^{\rm h}$29$^{\rm m}$54.16$^{\rm s}$ and
DEC.(J2000)~=~$12\arcdeg16\arcmin53.08\arcsec$. Its deconvolved FWHM angular size is $1\arcsec.26\times0\arcsec.36$ with P.A. of 76$\arcdeg$.3, corresponding to a radius of 255 AU at a distance of 380 pc. As marked with a blue cross, G192S contains a relatively weak source, which was detected at a level of 9 $\sigma$.  Its peak position is R.A.(J2000)~=~05$^{\rm h}$29$^{\rm m}$54.35$^{\rm s}$ and
DEC.(J2000)~=~$12\arcdeg16\arcmin29.68\arcsec$. That weak source has a deconvolved FWHM angular size of $2\arcsec.45\times1\arcsec.43$ with P.A. of -42$\arcdeg$.1, corresponding to a radius of 710 AU. The total fluxes of G192N and G192S in the 1.3 mm continuum are 0.223 and 0.013 Jy, respectively. The condensations detected in SMA observations are much more compact than the cold envelope traced by 850 $\micron$ continuum in JCMT/SCUBA-2 observations. The condensations G192N and G192S may be more affected by heating from protostars or outflows and should have higher dust temperatures. In section 4.3, we will show evidence that the central region may be heated by outflows up to $\sim$30 K.  Assuming a dust temperature of 30 K, the masses of G192N and G192S are 0.38 and 0.02 M$_{\sun}$, respectively. The volume densities of G192N and G192S are $7.2\times10^{8}$ and $1.9\times10^{6}$ cm$^{-3}$. As shown in the right panel of Figure 13, we found that the emission peaks of SMA 1.3 mm continuum for both G192N and G192S are significantly offset from the emission peaks of the JCMT/SCUBA-2 850 $\micron$ continuum. The offsets for G192N and G192S are $\sim5\arcsec$ and $\sim7\arcsec$, respectively. Such offsets are much larger than the pointing error ($\sim2\arcsec$) of the JCMT.

\begin{figure*}
\centering
\includegraphics[angle=-90,scale=0.5]{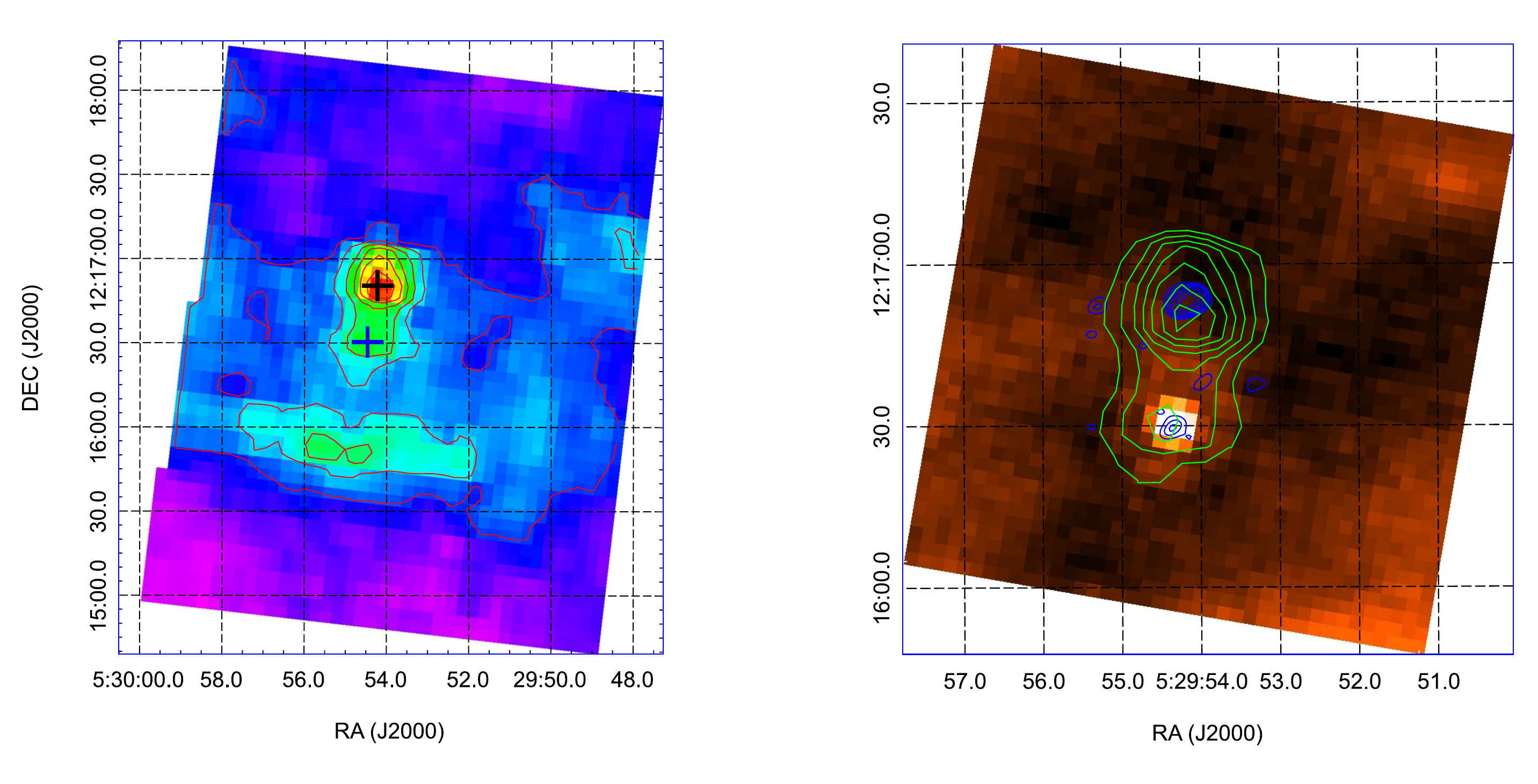}
\caption{Left: Spitzer/MIPS 70 $\micron$ emission is shown in contours and color scale. The contours are from 40\% to 90\% in steps of 10\% of the peak value 102 MJy/sr. The black and blue crosses mark G192N and G192S, respectively. Right: Spitzer/MIPS 24 $\micron$ emission is shown in color scale. Spitzer/MIPS 70 $\micron$ emission is shown in green contours. The contours are from 50\% to 95\% in steps of 7.5\% of the peak value 102 MJy/sr. The SMA 1.3 mm continuum is shown in blue contours. The contours are (-3, 3, 6, 9, 18, 36, 72, 144)$\times$1 $\sigma$ (1 $\sigma$=1 mJy~beam$^{-1}$).  }
\end{figure*}

As shown in the right panel of Figure 13, G192N is undetected in the Spitzer/IRAC 4.5 $\micron$ band. G192S shows extended 4.5 $\micron$ emission. The SMA 1.3 mm continuum emission of G192S coincides with 4.5 $\micron$ emission very well. To the west of G192S, another point source G192SW is seen in 4.5 $\micron$ emission but not in SMA 1.3 mm continuum emission, indicating that it might be a foreground or background star. In Figure 14, we present Spitzer/MIPS 70 and 24 $\mu$m emission as color images. G192N is not detected at Spitzer/MIPS 24 $\mu$m or IRAC bands but has a large envelope seen in Spitzer/MIPS 70 $\mu$m and 1.3 mm continuum from SMA observations. After subtracting a local background flux density of $\sim$44.9 MJy/sr, the total flux of G192N at 70 $\micron$ is $475\pm15$ mJy, corresponding to an internal luminosity of $0.21\pm0.01$ L$_{\sun}$ derived by using an empirical correlation between fluxes at 70 $\micron$ and internal luminosities (Dunham et al. 2008). The southern object (G192S) is a point source at Spitzer/MIPS 24 $\mu$m and has weak emission in Spitzer/MIPS 70 $\mu$m and 1.3 mm continuum. It has a flux of $184\pm10$ mJy at 70 $\micron$, corresponding to an internal luminosity of $0.08\pm0.01$ L$_{\odot}$. Due to its low luminosity, we classify it as a very low-luminosity object, i.e. a VeLLo (Di Francesco et al. 2007).

It should be noted that we did not consider systematic effects (e.g., distance, calibration) on the internal luminosities. The uncertainties of internal luminosities caused by calibration uncertainties of fluxes at MIPS 70 $\micron$ are $\sim$20\% (Evans et al. 2007; Dunham et al. 2008). Additionally, if the distance was actually 10\% higher, the internal luminosities can increase by $\sim$20\%. Therefore, the internal luminosities estimated above have uncertainties of at least 20\%.

\subsection{Compact CO outflows}

\begin{figure*}
\centering
\includegraphics[angle=-90,scale=0.5]{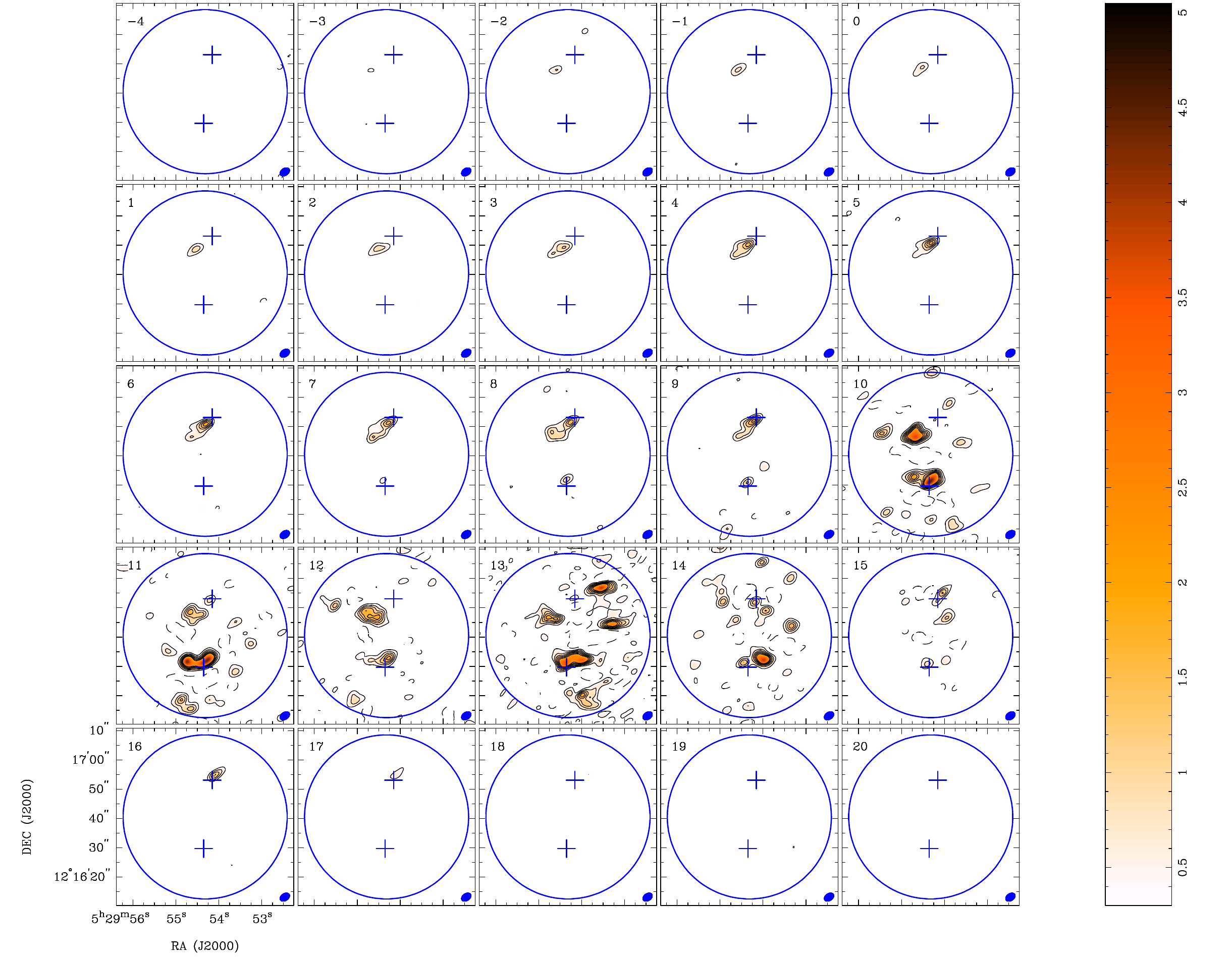}
\caption{SMA $^{12}$CO (2-1) channel map. The contours are -5 and from 5 to 30 in steps of 5 $\sigma$ (1 $\sigma$=0.07 Jy~beam$^{-1}$). The primary beam of the SMA is shown in blue circles. The synthesised beam is shown as blue filled ellipses in the lower right corner of each panel. The two continuum sources G192N and G192S are marked with blue crosses. The channel velocity is labeled in the upper-left corner of each panel.  }
\end{figure*}

Figure 15 shows the channel maps of $^{12}$CO (2-1) emission from the SMA observations. The $^{12}$CO (2-1) emission in the central channels [10,14] km~s$^{-1}$ is very clumpy. Blueshifted high-velocity emission associated with G192N is clearly seen from -2 to 9 km~s$^{-1}$, indicating that the maximum outflow velocity can be up to $\sim$14 km~s$^{-1}$ without correction of the inclination angle. In contrast, the redshifted high-velocity emission is much weaker, which can only be seen in channels from 15 to 17 km~s$^{-1}$. From channel maps, one can also identify weak outflows associated with G192S. The blueshifted high-velocity emission of G192S can be seen from 7 to 9 km~s$^{-1}$, while redshifted high-velocity emission is only apparent in the 15 km~s$^{-1}$ channel. The integrated intensity of the high-velocity $^{12}$CO (2-1) emission is shown in Figure 16. The outflow associated with G192N is well collimated. As shown in the right panel of Figure 16, the outflow associated with G192S is very compact and its redshifted and blueshifted emission overlap. We listed the radius and maximum outflow velocity of each lobe in the second and third columns of Table 4, respectively. The outflow dynamical age ($t_{dyn}$) is estimated as $t_{dyn}=2R_{lobe}/V_{max}$ and listed in the fourth column of Table 4. $R_{lobe}$ is the radius of each outflow lobe.

\begin{figure*}
\centering
\includegraphics[angle=90,scale=0.5]{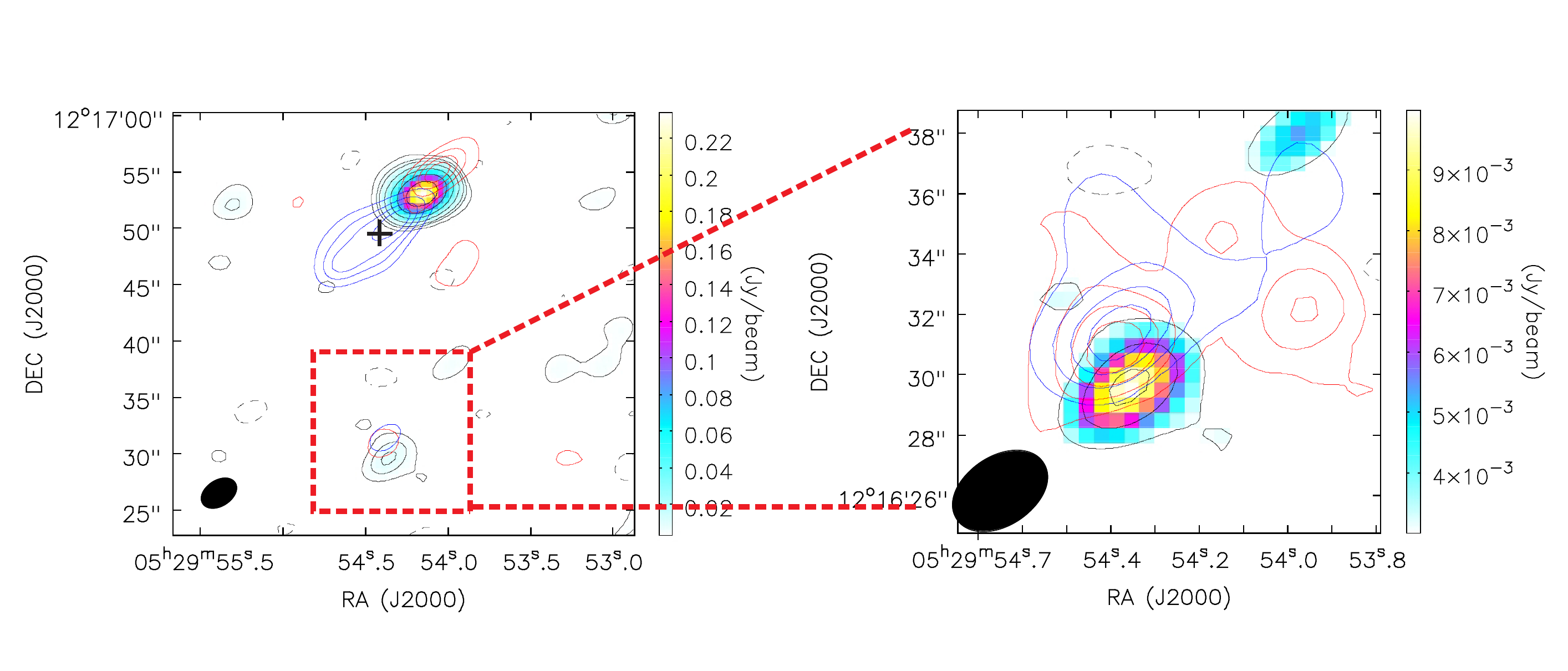}
\caption{The SMA 1.3 mm continuum is shown in black contours and color scale. The contours are (-3, 3, 6, 9, 18, 36, 72, 144)$\times$1 $\sigma$ (1 $\sigma$=1 mJy~beam$^{-1}$). The integrated intensity of high-velocity blueshifted (-2 to 9 km~s$^{-1}$) and redshifted (15 to 17 km~s$^{-1}$) $^{12}$CO (2-1) emission are shown in blue and red contours respectively. The contours for $^{12}$CO (2-1) emission are from 20\% to 80\% of peak values. The peak values for blueshifted and redshifted emission of $^{12}$CO (2-1) in the left panel are 10.82 and 3.51 Jy~beam$^{-1}$~km~s$^{-1}$, respectively. The peak values for blueshifted and redshifted emission of $^{12}$CO (2-1) in the right panel are 3.1 and 1.1 Jy~beam$^{-1}$~km~s$^{-1}$, respectively. The emission peak of blueshifted outflow emission is marked with a black cross in the left panel. The spectra in Figure 17 were taken at this position.}
\end{figure*}

The gas mass of the outflow can be estimated following Qiu et al. (2009):
\begin{equation}
M_{out}=1.39\times10^{-6}exp(\frac{16.59}{T_{ex}})(T_{ex}+0.92)D^{2}\int\frac{\tau_{12}}{1-e^{-\tau_{12}}}S_{v}dv,
\end{equation}
where $M_{out}$, $T_{ex}$, D, $\tau_{12}$ and $S_{v}$ are the outflow gas mass in $M_{\sun}$, excitation temperature of $^{12}$CO (2-1), source distance in kpc, optical depth of $^{12}$CO (2-1) and line flux in Jy, respectively. The mass of each outflow lobe is presented in the fifth column of Table 4. We sum both redshifted and blueshifted outflow masses to derive total outflow mass. The total outflow masses for G192N and G192S are $\sim5.3\times10^{-4}$ and $\sim6.7\times10^{-5}$ M$_{\sun}$, respectively.

\begin{deluxetable*}{ccccccccc}
\tabletypesize{\scriptsize} \tablecolumns{9} \tablewidth{0pc}
\tablecaption{Parameters of CO outflow in the SMA observations} \tablehead{
\colhead{Lobes\tablenotemark{a}}  & \colhead{Radius\tablenotemark{b}} & \colhead{V$_{max}$} & \colhead{Age} & \colhead{Mass} & \colhead{$\dot{M}_{loss}$} & \colhead{F$_{flow}$}  &
\colhead{$\dot{M}_{acc}$}  & \colhead{L$_{acc}$} \\
\colhead{} &\colhead{(10$^{3}$ AU)} & \colhead{(km~s$^{-1}$)}    &\colhead{($10^{3}$ yr)} &
\colhead{($10^{-5}$ M$_{\sun}$)} &
\colhead{($10^{-8}$ M$_{\sun}$~yr$^{-1}$)} &
\colhead{($10^{-7}$ M$_{\sun}$~km~s$^{-1}$~yr$^{-1}$)} &
\colhead{($10^{-8}$ M$_{\sun}$~yr$^{-1}$)} &
\colhead{($10^{-5}$ L$_{\sun}$)} }
\startdata
G192N-Red	  &  0.9	  &   5	  &  1.7	 &   7.4	&  4.5	&  1.8   &  4.8   &4.0 \\
G192N-blue	&  1.8        &  14	  &  1.2	 &   45	    &  36	&  22    &  58    &430 \\
G192S-red	  &  0.8	  &   3	  &  2.6	 &   1.6	&  0.6	&  0.2   &  0.5   &0.1 \\
G192S-blue	&  1.0	      &   4	  &  2.3	 &   5.1	&  2.2	&  0.9   &  2.3   &1.3
\enddata
\tablenotetext{a}{Outflow lobes associated with G192N and G192S as shown in the left panel of Figure 16.}
\tablenotetext{b}{The radius of each outflow lobe is defined as $R=\sqrt{a\cdot b}$, where a and b are the full width at half-maximum (FWHM) deconvolved sizes of the major and minor axes, respectively.}
\end{deluxetable*}

The mass loss rate ($\dot{M}_{loss}$) of the outflow can be derived as $\dot{M}_{loss}=M_{out}/t_{dyn}$. The total mass loss rates of the outflows associated with G192N and G192S are 4.1$\times10^{-7}$ and 2.8$\times10^{-8}$ M$_{\sun}$~yr$^{-1}$. Under the assumption that jet energy and wind energy are due to the gravitational energy released by mass accretion
onto the protostar, as stated in Bontemps et al. (1996), the
outflow force ($F_{out}$) is related to the mass accretion rate
($\dot{M}_{acc}$) as given in the following equation obtained from the
principle of momentum conservation and some manipulation of
the equation:

\begin{equation}\label{eq_finalmdot}
\dot{M}_{\rm acc} = \frac{1}{f_{\rm ent}} \, \frac{\dot{M}_{\rm acc}}{\dot{M}_w} \frac {1}{V_w} \, F_{flow}
\end{equation}

We assume a typical jet/wind velocity of $V_w \sim 150$ km~s$^{-1}$ (Bontemps et al. 1996). $\dot{M}_w$ is the
wind/jet mass loss rate. Models of jet/wind formation predict, on average, $\dot{M}_w / \dot{M}_{acc}\sim$ 0.1 (Shu et al. 1994; Pelletier
\& Pudritz 1992; Wardle \& K\"{}nigl 1993; Bontemps et
al. 1996).  The entrainment efficiency is typically $f_{\rm ent} \sim 0.1-0.25$. Here we take 0.25. The outflow force $F_{flow}$ is calculated as:

\begin{equation}
F_{flow}=\frac{P_{flow}}{t_{dyn}}=\frac{M_{out}V_{char}}{t_{dyn}}=\dot{M}_{out}V_{char}
\end{equation}
where V$_{char}$ is the characteristic outflow velocity, which is estimated from the intensity weighted velocity. The calculated outflow forces and mass accretion rates are listed in the seventh and eighth columns, respectively. The total mass accretion rate for G192N is 6.3$\times10^{-7}$ M$_{\sun}$~yr$^{-1}$. While the total mass accretion rate for G192S is 2.8$\times10^{-8}$ M$_{\sun}$~yr$^{-1}$.

The accretion luminosity was calculated as:

\begin{equation}\label{eq_lacc}
L_{acc} = \frac{G M_{\rm acc} \dot{M}_{acc}}{R}~~,
\end{equation}
where $M_{acc}=\dot{M}_{acc}\times t_{dyn}$ is the mass accreted during dynamical time. We present the accretion luminosity in the last column of Table 4. The total accretion luminosities for G192N and G192S are 4.3$\times10^{-3}$ and 1.4$\times10^{-5}$ L$_{\sun}$, respectively.

\begin{figure}
\centering
\includegraphics[angle=-90,scale=0.3]{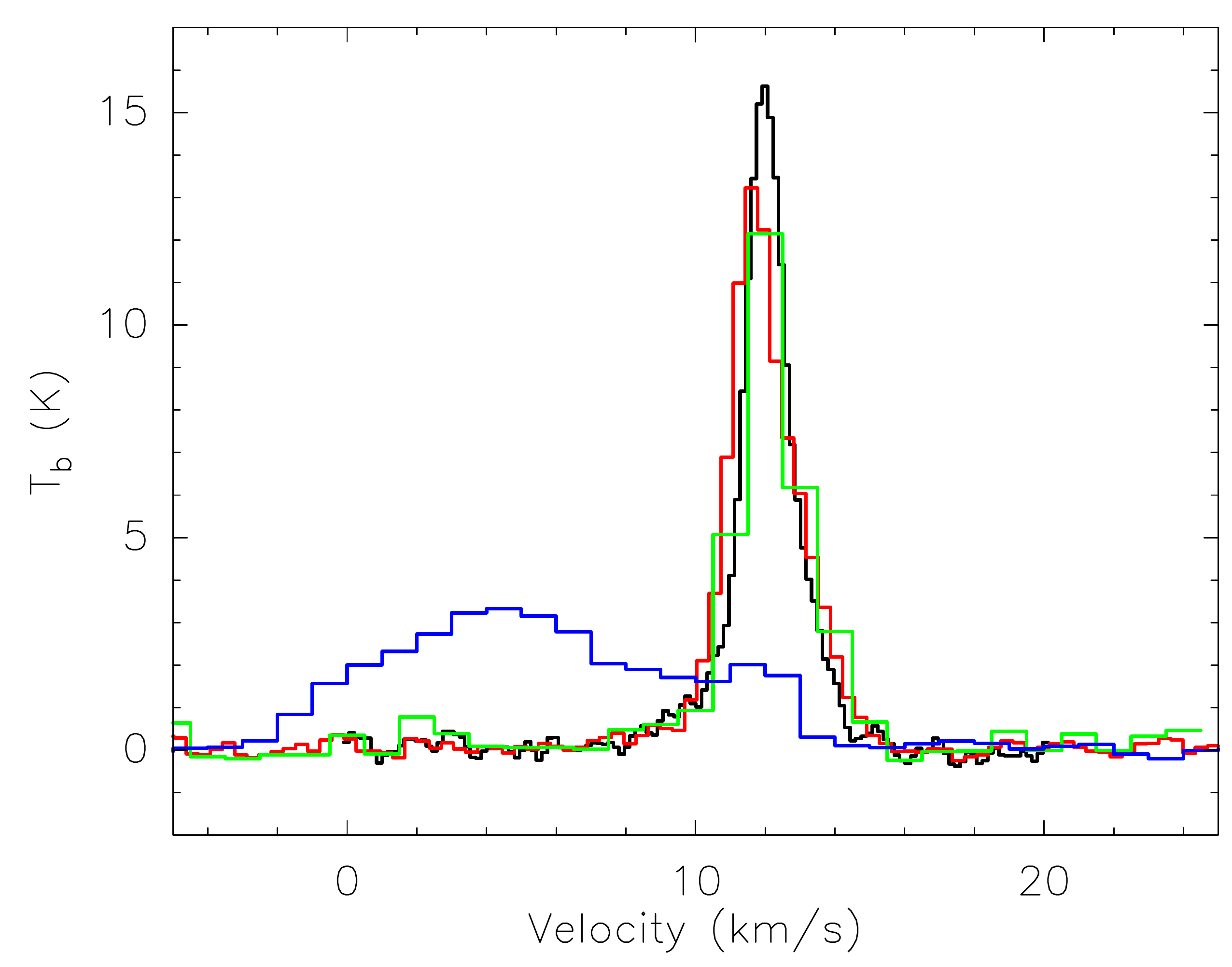}
\caption{Spectra at the emission peak of blueshifted outflow lobe of G192N. The beam averaged PMO 13.7-m $^{12}$CO (1-0), CSO $^{12}$CO (2-1), SMA $^{12}$CO (2-1) are shown in black, red and blue, respectively. The SMA+CSO $^{12}$CO (2-1) spectrum convolved with CSO beam is shown in green.}
\end{figure}

In Figure 17, we present the spectra of CO lines at the emission peak of the blueshifted lobe of G192N. The PMO 13.7-m CO (1-0) and CSO CO (2-1) lines show weak line wings with V$_{lsr}$ up to 8 km~s$^{-1}$ at the blueshifted side and 16 km~s$^{-1}$ at the redshifted side, respectively. However, the blueshifted high-velocity emission in SMA CO (2-1) covers a much larger velocity range from -2 to 10 km~s$^{-1}$. For comparison, we convolved the SMA data with the CSO beam and found that even for the low-velocity part (8-10 km~s$^{-1}$), the CSO intensity is only 3-5 times larger than the SMA emission, while the blueshifted emission in SMA CO (2-1) is mainly dominated by higher velocity ($V_{lsr}<$8 km~s$^{-1}$) gas. The SMA+CSO CO (2-1) emission recovers most of the missing flux at low velocities. However, the high velocity outflow emission is not clearly seen in either CSO or SMA+CSO CO (2-1) emission, indicating that the outflow is very compact and its emission is totally diluted by single-dish observations. Therefore it seems that the missing flux of the SMA is not significant for high velocity outflow gas. The outflow mass or mass loss rate should only be underestimated by a factor of $<$3-5 due to missing flux.

\subsection{Gas fragmentation}

\begin{figure*}
\centering
\includegraphics[angle=90,scale=0.5]{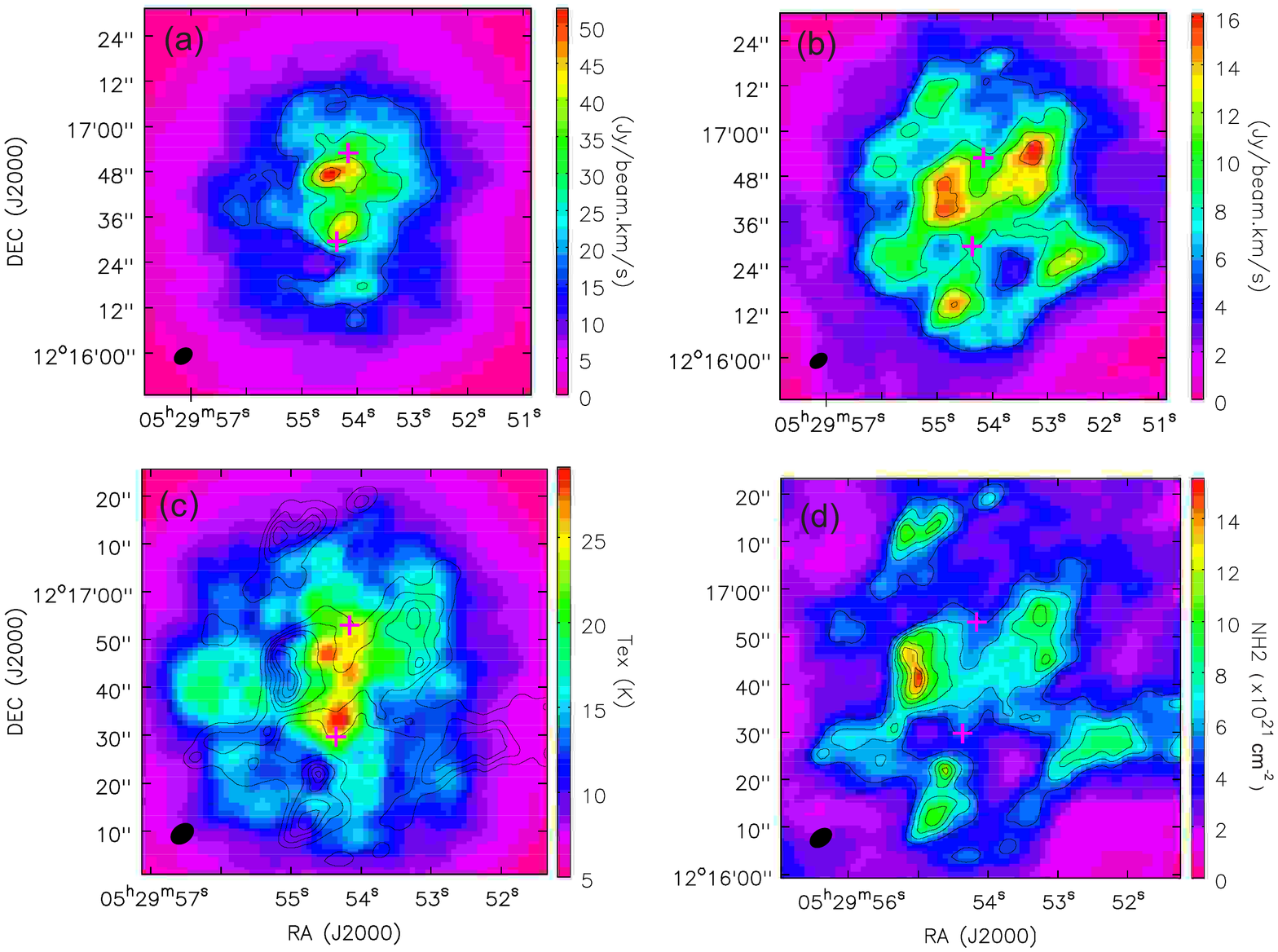}
\caption{(a). Integrated intensity of SMA+CSO $^{12}$CO (2-1) emission from 10 to 14 km~s$^{-1}$ is shown in contours overlayed on its color image. The contours are from 30\% to 90\% in steps of 20\% of the peak value 52.5 Jy~beam$^{-1}$~km~s$^{-1}$. (b). Integrated intensity of SMA+CSO $^{13}$CO (2-1) emission from 10 to 14 km~s$^{-1}$ is shown in contours overlayed on its color image. The contours are from 30\% to 90\% in steps of 20\% of the peak value 16.2 Jy~beam$^{-1}$~km~s$^{-1}$. (c). H$_{2}$ column density estimated from $^{13}$CO (2-1) emission is shown in contours overlayed on color image of excitation temperature of $^{12}$CO (2-1). The contours are from 30\% to 90\% in steps of 10\% of the peak value 1.56$\times10^{22}$ cm$^{-2}$. (d). H$_{2}$ column density in contours overlayed on its color image. The two continuum sources G192N and G192S are marked with pink crosses. }
\end{figure*}

\begin{figure}
\centering
\includegraphics[angle=90,scale=0.35]{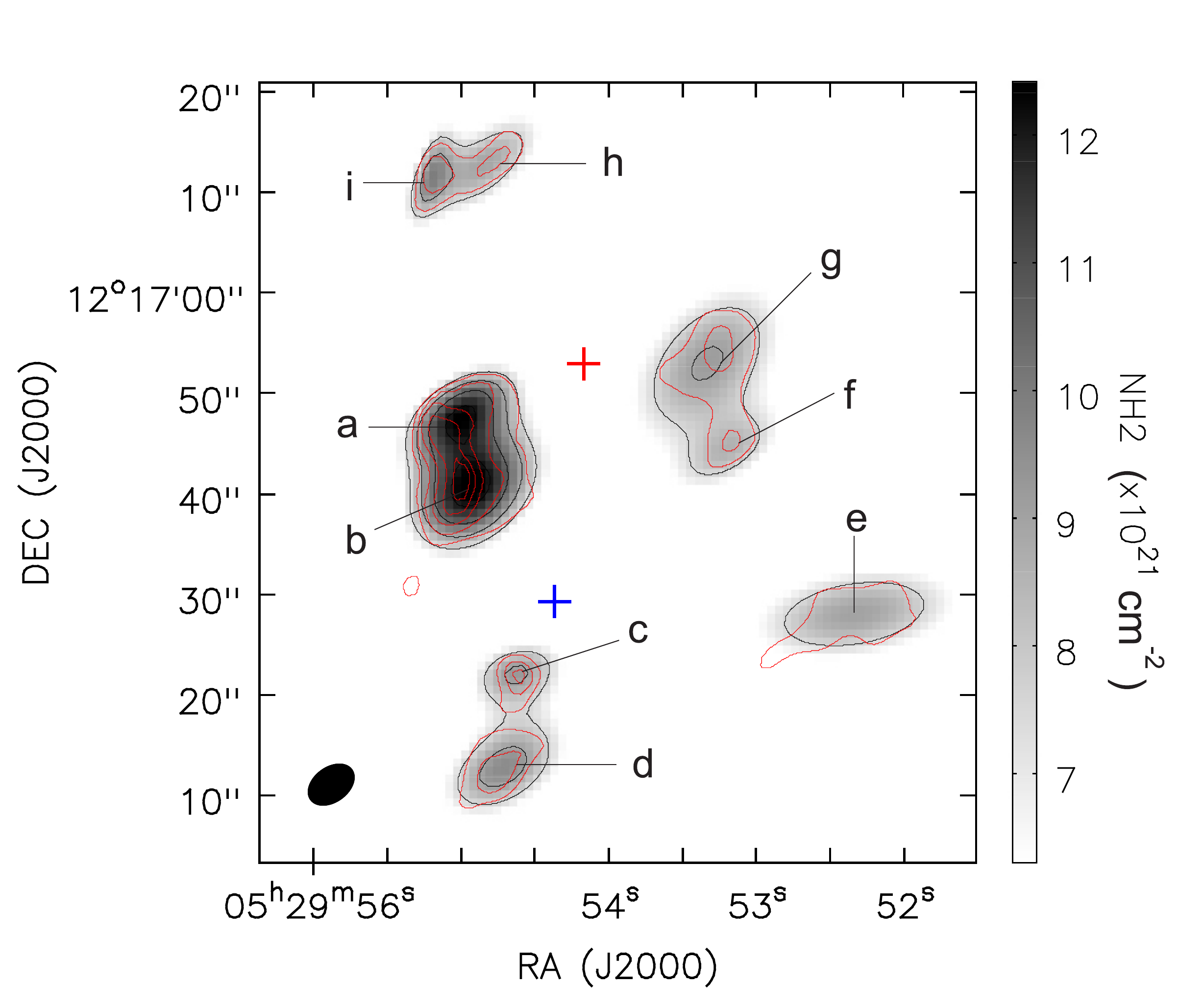}
\caption{H$_{2}$ column density estimated from SMA+CSO $^{13}$CO (2-1) emission data is shown in red contours. Gaussian fits of the nine dense fragments are shown in grey image and black contours. The contours are from 50\% to 90\% in steps of 10\% of the peak value 1.56$\times10^{22}$ cm$^{-2}$. The names of fragments are labeled from a to i. G192N and G192S are marked with red and blue crosses, respectively. }
\end{figure}

\begin{figure}
\centering
\includegraphics[angle=0,scale=0.45]{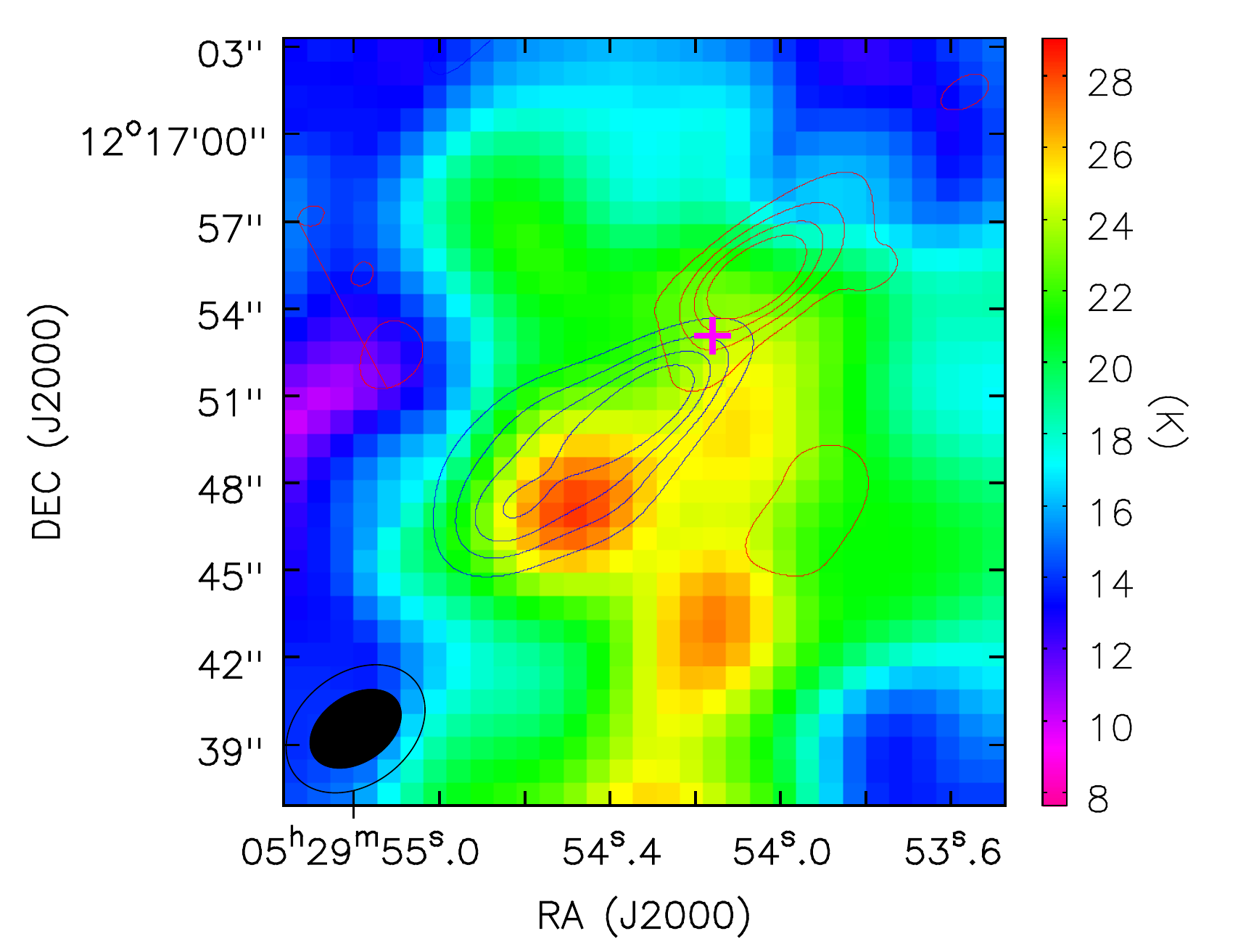}
\caption{The integrated intensity of high-velocity blueshifted and redshifted $^{12}$CO (2-1) emission are shown in blue and red contours, respectively. The contour levels are the same as in the left panel of Figure 16. The excitation temperature of $^{12}$CO (2-1) is shown as color image. G192N is marked with a pink cross.  }
\end{figure}

The integrated intensity maps of SMA+CSO $^{12}$CO (2-1) and $^{13}$CO (2-1) emissions are shown in the panels (a) and (b) of Figure 18, respectively. The $^{12}$CO (2-1) emission has two emission peaks, which are offset from the 1.3 mm continuum emission. The $^{13}$CO gas is highly fragmented. All the $^{13}$CO emission peaks are also away from the 1.3 mm continuum peaks. We derived the excitation temperature of $^{12}$CO (2-1) emission from equation (A2) assuming that its emission is optically thick. The excitation temperature is shown as color image in the panel (c) of Figure 18. Assuming that $^{13}$CO (2-1) emission has the same excitation temperature as $^{12}$CO (2-1) emission under LTE condition, we derived the column density of H$_{2}$ from equation (A1) by adopting typical abundance ratios [H$_{2}$]/[$^{12}$CO]=$10^{4}$ and [$^{12}$CO]/[$^{13}$CO]=60. The H$_{2}$ column density map revealed from SMA+CSO $^{13}$CO (2-1) emission data is shown in the panel (d) of Figure 18. We identified 9 fragments from the column density map within its 50\% contour as shown in Figure 19. From Gaussian fits, their peak positions, radii (R) and peak column densities (N$_{H_{2}}$) are derived and shown in columns 2-4 in Table 5. Their volume densities (n$_{H_{2}}$) and masses (M$_{LTE}$) are calculated as $n_{H_{2}}=\frac{N_{H_{2}}}{2R}$ and $M_{LTE}=\frac{4}{3}\pi R^{3}\cdot n_{H_{2}}\cdot m_{H}\cdot\mu_{g}$, respectively. The Jeans masses of the fragments are derived following Wang et
al. (2014):
\begin{equation}\label{eq_lacc}
M_{J}=\frac{\pi^{5/2}c_{s}^{3}}{6\sqrt{G^{3}\rho}}=0.877M_{\sun}(\frac{T}{10 K})^{3/2}(\frac{n}{10^{5} cm^{-3}})^{-1/2}.
\end{equation}
We use the mean excitation temperature of $^{12}$CO (2-1) emission to derive the Jeans mass for each fragment. The M$_{LTE}$ and M$_{J}$ are shown in the last two columns of Table 5. We find that the gas fragments have Jeans masses significantly larger than their LTE masses, indicating that the gas fragments will not collapse but will be dispersed eventually.

\begin{deluxetable*}{cccccccc}
\tabletypesize{\scriptsize} \tablecolumns{8} \tablewidth{0pc}
\tablecaption{Parameters of $^{13}$CO fragments in the SMA+CSO observations} \tablehead{
\colhead{Fragment}& \colhead{offset\tablenotemark{a}} & \colhead{Radius} & \colhead{N$_{H_{2}}$} & \colhead{n$_{H_{2}}$} & \colhead{T$_{ex}$ } & \colhead{M$_{LTE}$ }  &
\colhead{M$_{J}$}   \\
\colhead{} &\colhead{(arcsec, arcsec)} & \colhead{(10$^{-2}$ pc)}    &\colhead{(10$^{21}$ cm$^{-2}$)} &\colhead{(10$^{4}$ cm$^{-3}$)} &
\colhead{(K)} &
\colhead{(10$^{-1}$ M$_{\sun}$)} &
\colhead{(M$_{\sun}$)}
  } \startdata
a       &  (10.13, 8.53)                           &  1.6                     &  8.4      &   8.5        &  15           &  1.0   &   1.7      \\
b       &  (9.08, -0.10)                           &  2.3                     &  12.4     &   8.7        &  17           &  3.1   &   2.1      \\
c       &  (4.68, -17.82)                          &  1.4                     &  7.8      &   9.0        &  12           &  0.7   &   1.2     \\
d       &  (5.80, -27.82)                          &  2.2                     &  9.6      &   7.1        &  12           &  2.2   &   1.4       \\
e       &  (-29.19, -12.43)                        &  3.2                     &  9.0      &   4.6        &  10           &  4.3   &   1.3     \\
f       &  (-17.28, 3.14)                          &  1.5                     &  4.9      &   5.3        &  17           &  0.5   &   2.7     \\
g       &  (-14.55, 12.62)                         &  3.3                     &  9.2      &   4.5        &  16           &  4.7   &   2.6     \\
h       &  (7.60, 32.31)                           &  2.2                     &  8.9      &   6.6        &  10           &  2.0   &   1.1     \\
i       &  (12.96, 32.92)                          &  $<$0.8                  &  4.6      &   9.3        &  10           &  0.1   &   0.9
\enddata
\tablenotetext{a}{offset from phase center R.A.(J2000)~=~05$^{\rm h}$29$^{\rm m}$54.32$^{\rm s}$ and
DEC.(J2000)~=~$12\arcdeg16\arcmin40.50\arcsec$.}
\end{deluxetable*}

As shown in Figure 18, the two continuum sources G192N and G192S are located in the cavities enclosed by the gas fragments, indicating that the gas distribution may be affected by ongoing star formation. In Figure 20, we plot the outflow emission contours on the excitation temperature of $^{12}$CO (2-1) emission. Interestingly, we find that the CO gas at the blueshifted outflow lobe has the highest excitation temperature ($\sim$30 K), hinting for outflow heating.

\begin{figure}
\centering
\includegraphics[angle=-90,scale=0.35]{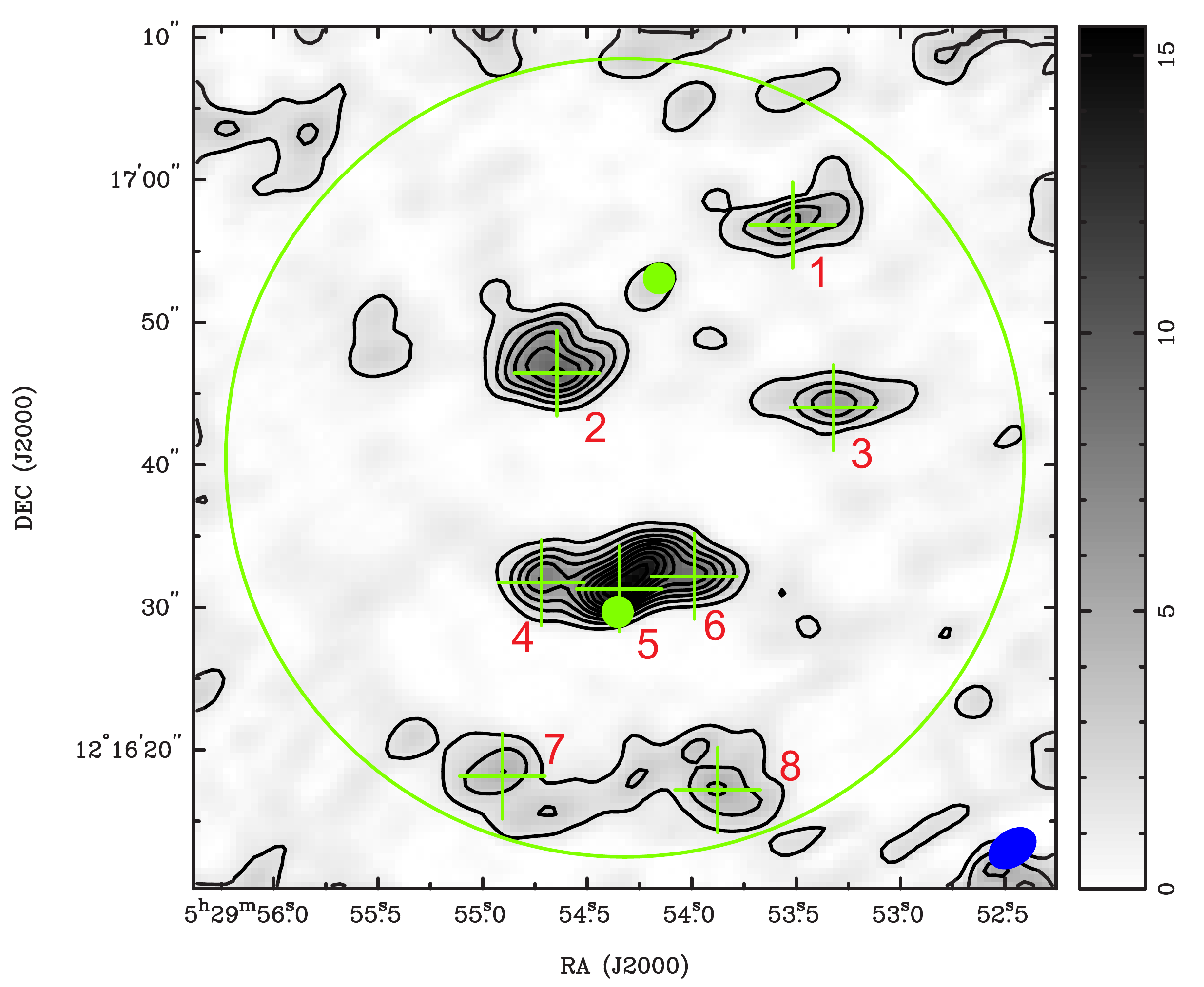}
\caption{Integrated intensity of $^{12}$CO (2-1) from 10 to 14 km~s$^{-1}$ is shown in contours. The contours are from 10\% to 90\% in steps of 10\% of the peak value (15.52 Jy~beam$^{-1}$~km~s$^{-1}$). The primary beam of SMA is shown in green circle. The synthesised beam is shown in blue filled ellipse. The two continuum sources G192N and G192S are marked with green filled circles. CO fragments with integrated intensity larger than 30\% of peak value are marked with green crosses.}
\end{figure}

\begin{deluxetable*}{ccccccc}
\tabletypesize{\scriptsize} \tablecolumns{7} \tablewidth{0pc}
\tablecaption{Parameters of CO fragments in the SMA observations} \tablehead{
\colhead{Fragment}& \colhead{offset\tablenotemark{a}} & \colhead{Area} & \colhead{V$_{lsr}$} & \colhead{FWHM } & \colhead{I$_{peak}$ }  &
\colhead{rms}   \\
\colhead{} &\colhead{(arcsec, arcsec)} & \colhead{(Jy~beam$^{-1}$~km~s$^{-1}$)}    &\colhead{(km~s$^{-1}$)} &
\colhead{(km~s$^{-1}$)} &
\colhead{(Jy~beam$^{-1}$)} &
\colhead{(mJy~beam$^{-1}$)}
  } \startdata
G192N   &  (-2.38, 12.58)                          &  5.06$\pm$0.17            &   9.07$\pm$0.04    &  4.74$\pm$0.19      & 1.00    &   42.9      \\
        &  (-2.38, 12.58)                          &  2.99$\pm$0.14            &  15.16$\pm$0.07    &  3.39$\pm$0.17      & 0.83    &   42.9      \\
G192S   &  (0.51, -10.82)                          &  6.77$\pm$0.11            &  11.55$\pm$0.04    &  5.02$\pm$0.09      & 1.26    &   42.1      \\
1       &  (-11.76, 16.33)                         &  6.32$\pm$0.12            &  13.51$\pm$0.02    &  2.52$\pm$0.05      & 2.35    &   50.7     \\
2       &  (4.78, 5.93)                            &  2.74$\pm$0.08            &   0.24$\pm$0.09    &  5.21$\pm$0.10      & 0.49    &   24.4       \\
        &  (4.78, 5.93)                            &  3.00$\pm$0.01            &   5.14$\pm$0.08    &  4.30$\pm$0.12      & 0.65    &   24.4      \\
        &  (4.78, 5.93)                            & 12.59$\pm$0.08            &  11.26$\pm$0.02    &  5.70$\pm$0.03      & 2.07    &   24.4      \\
3       &  (-14.61, 3.52)                          &  5.45$\pm$0.05            &  13.84$\pm$0.01    &  2.88$\pm$0.03      & 1.77    &   39.3     \\
4       &  (5.87, -8.75)                           &  7.85$\pm$0.10            &  11.36$\pm$0.01    &  2.85$\pm$0.04      & 2.58    &   40.8     \\
5       &  (0.40, -9.19)                           & 19.14$\pm$0.10            &  11.61$\pm$0.01    &  5.14$\pm$0.03      & 3.49    &   46.3     \\
6       &  (-4.86, -8.31)                          &  9.16$\pm$0.07            &  14.08$\pm$0.01    &  3.06$\pm$0.03      & 2.81    &   24.7     \\
7       &  (8.61, -22.33)                          &  4.22$\pm$0.19            &  11.73$\pm$0.06    &  3.15$\pm$0.17      & 1.25    &   72.4   \\
8       &  (-6.50, -23.31)                         &  5.45$\pm$0.17            &  12.23$\pm$0.08    &  4.98$\pm$0.18      & 1.02    &   50.9
\enddata
\tablenotetext{a}{offset from phase center R.A.(J2000)~=~05$^{\rm h}$29$^{\rm m}$54.32$^{\rm s}$ and
DEC.(J2000)~=~$12\arcdeg16\arcmin40.50\arcsec$.}
\end{deluxetable*}

We integrated SMA $^{12}$CO (2-1) emission between 10 and 14 km~s$^{-1}$ and present the integrated intensity in Figure 21. We identified 8 dense and compact fragments with integrated intensity larger than 30\% of the peak value 15.52 Jy~beam$^{-1}$~km~s$^{-1}$. Figure 22 shows the beam-averaged spectra of $^{12}$CO (2-1) toward each fragment. We fitted the spectra with Gaussian profiles and presented the fitted results in Table 6.

\begin{figure}
\centering
\includegraphics[angle=0,scale=0.4]{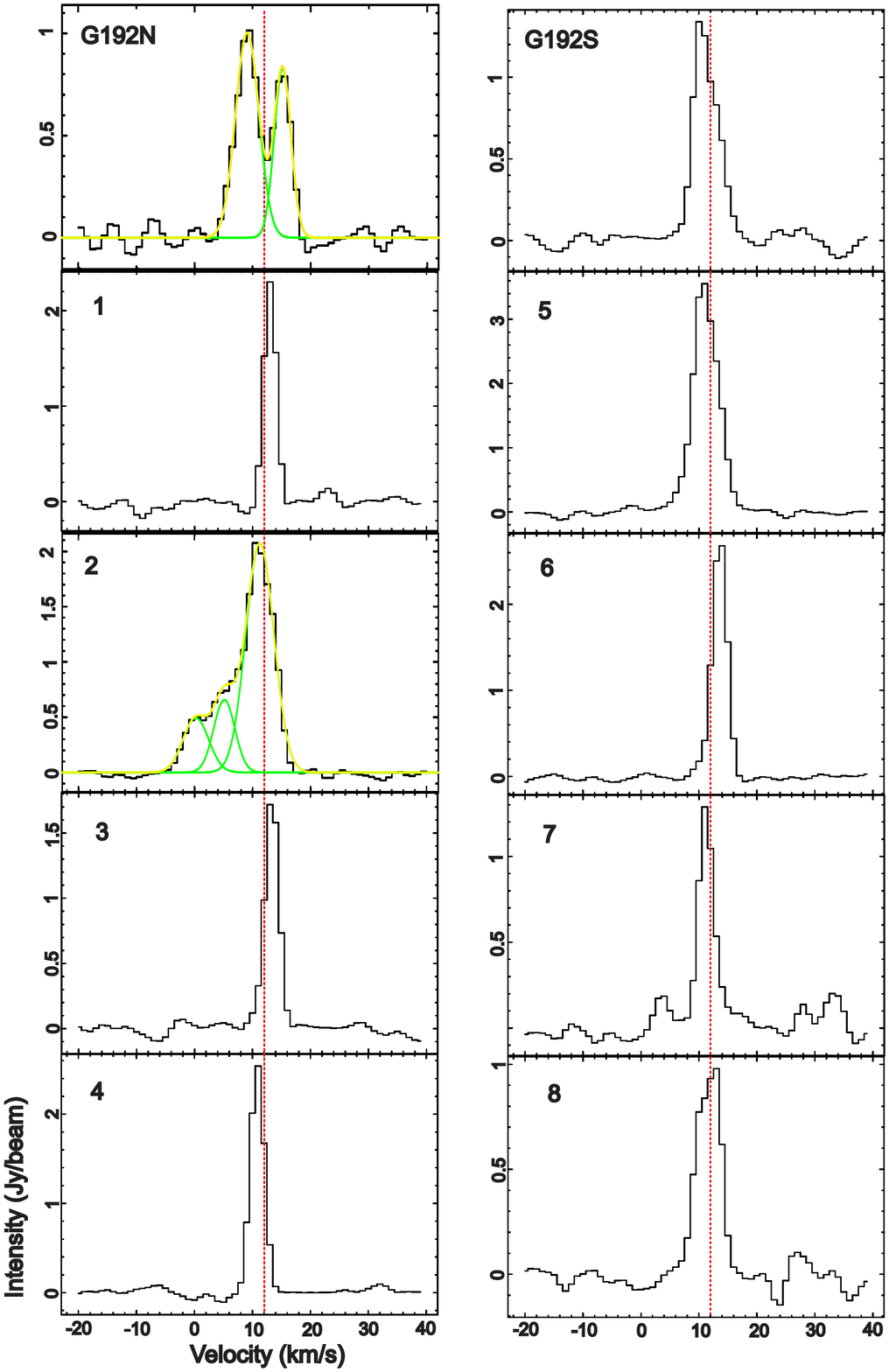}
\caption{SMA spectra of $^{12}$CO (2-1) at continuum sources and CO fragments. The vertical red lines mark the systemic velocity of 12 km~s$^{-1}$. The green and yellow lines toward G192N and fragment 2 represent multiple Gaussian fits and their sum. }
\end{figure}

Except for G192N, the spectra of $^{12}$CO (2-1) emission at other positions are all single-peaked. The SMA $^{12}$CO (2-1) spectrum toward G192N shows a typical ``blue profile" with the blueshifted emission peak stronger than the redshifted peak, a signature for infall motions (Zhou et al. 1993). The nondimensional parameter $\delta V=(V_{thick}-V_{thin})/\Delta V_{thin}$ is usually used to quantify the spectral line asymmetries (Mardones et al. 1997).. Here $V_{thick}$ is the peak velocity of the optically thick line; $V_{thin}$ and $\Delta V_{thin}$ are the peak velocity and line width of the optically thin line. Infall candidates will have $\delta V<-0.25$. In the case of G192N, the $\delta V$ derived from SMA $^{12}$CO (2-1) line and CSO C$^{18}$O (2-1) line is -3.1, much smaller than -0.25, indicating that SMA $^{12}$CO (2-1) line toward G192N might trace an infalling envelope. However, such double peak profile might also be caused by outflows as shown by o-H$_{2}$CO ($2_{1,2}-1_{1,1}$) line (Fig. 12) or missing flux. We do not see such double peak profile in the SMA+CSO combined spectra because the combined data are dominated by large-scale extended emission. Higher angular resolution and higher sensitivity observations of dense molecular tracers (e.g., HCN, HCO$^{+}$) are needed to resolve the envelope to test the infall scenario.

Fragments 1 and 3 have velocities $\sim$1.5 and $\sim$1.8 km~s$^{-1}$ blueshifted with respect to the systemic velocity (12 km~s$^{-1}$). Fragment 2, which is close to the blueshifted outflow lobe, is greatly affected by the outflow. The $^{12}$CO (2-1) line of Fragment 2 can be fitted with three Gaussian components with velocities at $\sim$0.2, $\sim$5.1 and $\sim$12.6 km~s$^{-1}$, respectively. The components at $\sim$0.2 and $\sim$5.1 km~s$^{-1}$ trace outflow gas with different velocities, hinting that the outflow jet might be episodically changing. The linewith of the $\sim$12.6 km~s$^{-1}$ component toward Fragment 2 is more than two times larger than that of Fragments 1 and 3, indicating that the $^{12}$CO (2-1) emission at Fragment 2 may be broadened by the outflow shocks. Fragments 4, 5, and 6 are connected and located close to G192S. Fragment 6 has a velocity about 2.5 km~s$^{-1}$ redshifted to that of Fragment 5, indicating that Fragment 6 is impelled by the outflows. G192S itself is about 1$\arcsec$.6 south of Fragment 5. The peak intensity of $^{12}$CO (2-1) line toward G192S is only about one third of that of Fragment 5. It seems that G192S has moved out of its natal core. Fragments 7 and 8 are located in a filament.

In conclusion, we think that the feedback of star forming activities (e.g. outflow shocks) in this region can propel, heat, unbind and disperse the surrounding gas.

\section{Discussion}

\subsection{Chemical evolutionary stages of G192N and G192S}

\begin{figure}
\centering
\includegraphics[angle=0,scale=0.4]{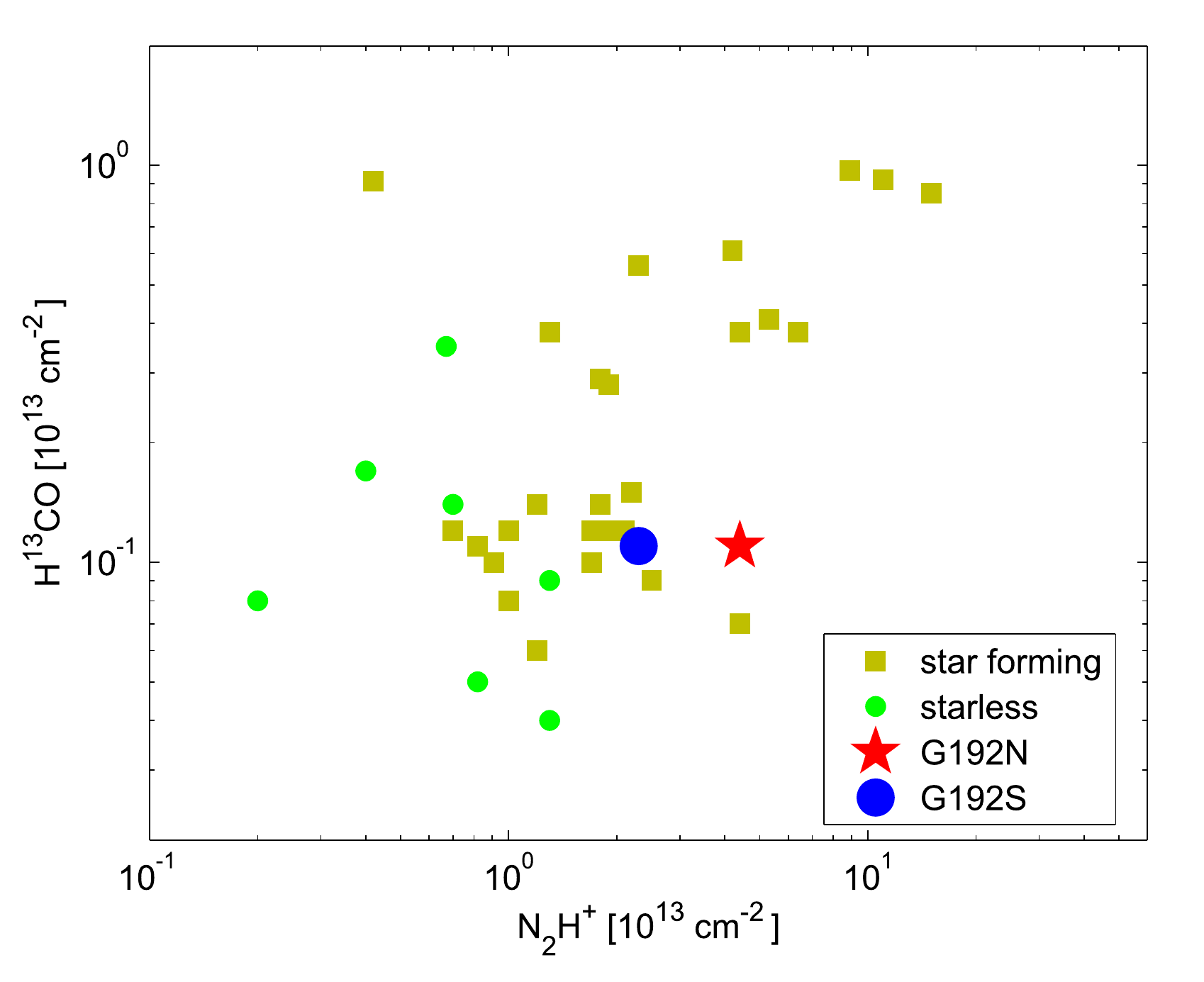}
\caption{The column density of N$_{2}$H$^{+}$ is plotted against the column density of H$^{13}$CO$^{+}$. The data of protostellar or starless cores in the Orion A cloud are from Table 3 of Ohashi et al.\ (2014).}
\end{figure}

As shown in Table 3, the C$^{18}$O and H$^{13}$CO$^{+}$ abundances in G192N are smaller than those in G192S, indicating that a larger amount of CO is frozen out in icy mantles which results in a depletion of HCO$^{+}$ in G192N. As a major destroyer of N$_{2}$H$^{+}$, the absence of CO in the gas phase could enhance the abundance of N$_{2}$H$^{+}$. Indeed, the N$_{2}$H$^{+}$ abundance in G192N is larger than that in G192S. In Figure 23, we compared the N$_{2}$H$^{+}$ and H$^{13}$CO$^{+}$ abundances in G192N and G192S with that of other dense cores in the Orion A cloud (Ohashi et al.
2014). In protostellar cores, the N$_{2}$H$^{+}$ and H$^{13}$CO$^{+}$ column densities are correlated with each other (Ohashi et al.
2014). The N$_{2}$H$^{+}$ and H$^{13}$CO$^{+}$ column densities in G192S follow this correlation well. This behavior has also been observed in a large sample of IRDC clumps (Sanhueza et al. 2012). While G192N seems to have relatively larger N$_{2}$H$^{+}$ to H$^{13}$CO$^{+}$ abundance ratio than other protostellar cores. In warm regions, CO evaporation could destroy N$_{2}$H$^{+}$ and enhance the abundance of HCO$^{+}$ by the following reaction (Lee, Bergin \& Evans 2004; Busquet et al. 2011):
\begin{equation}\label{eq_lacc}
N_{2}H^{+}+CO \rightarrow HCO^{+}+N_{2}
\end{equation}
The larger N$_{2}$H$^{+}$ to H$^{13}$CO$^{+}$ abundance ratio in G192N indicates that G192N is much colder and is thus chemically younger than other protostellar cores including G192S.

In a survey of low-mass protostellar cores, Roberts et al. (2002) found that the observed [HDCO]/[H$_{2}$CO] abundance ratios are $\sim$0.05-0.07. These ratios are larger than those of G192N and G192S. Bell et al. (2011) found that turbulent mixing in clouds could reduce the efficiency of deuteration by increasing the ionization fraction and reducing freeze-out of heavy molecules. The low [HDCO]/[H$_{2}$CO] ratios ($\sim1\times10^{-2}$) in G192N and G192S could be produced in their chemical models at very early evolutionary stages, i.e., with a chemical age less than $<1\times10^{4}$ yr (Bell et al. 2011). As shown in the left panel of Figure 13, we notice that G192N and G192S are surrounded by extended warm dust emission as traced by 70 $\micron$ continuum. The 70 $\micron$ continuum even shows a bar-like structure close to the southern edge of the cloud, probably indicating the existence of an ionization front. The ionization front from the H{\sc ii} region can heat the cloud and increase its ionization fraction, leading to lower abundances of deuterated species in prestellar phase as described in Bell et al.
(2011). Therefore, we argue that the chemistry in PGCC G192.32-11.88 might be greatly affected by stellar feedback of the H{\sc ii} region. However, we notice that the small value of [HDCO]/[H$_{2}$CO] ratio could also be partly attributed to the very different regions traced by these two isotopologues. This can be deduced from the very different line widths of the two lines (Fig. 12). o-H$_{2}$CO ($2_{1,2}-1_{1,1}$) could have significant contribution from either outflow regions or a large volume of turbulent cloud, while HDCO ($2_{0,2}-1_{0,1}$) may come from a compact region, which needs to be confirmed by further higher angular resolution observations in future.

Anyway, G192N and G192S are more chemically evolved than starless cores but might be still at earlier evolutionary stages than most protostellar objects so far observed.

\subsection{Is G192N the youngest Class 0 source?}

\begin{figure*}
\centering
\includegraphics[angle=90,scale=0.5]{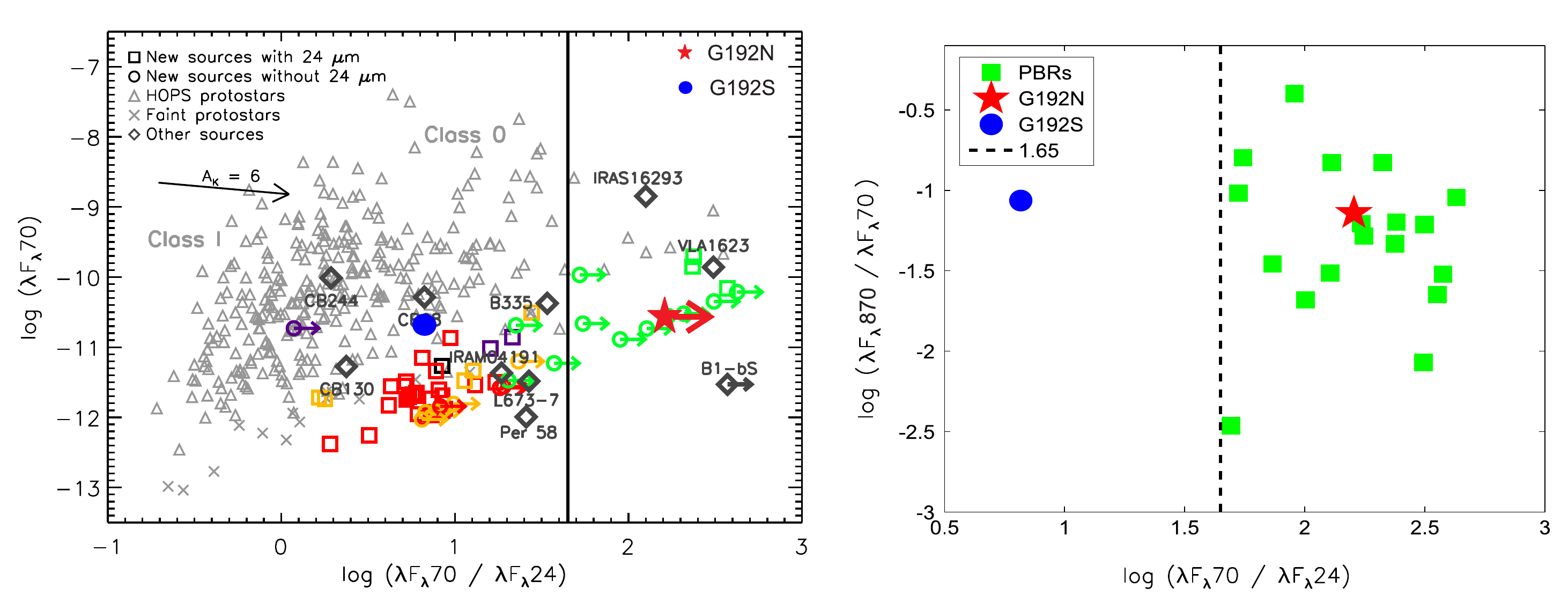}
\caption{Left: 70 $\micron$ flux vs. 70 $\micron$ to 24 $\micron$ flux ratio for HOPS detected Orion protostars (from (Stutz, Tobin, \& Stanke 2013). G192S and G192N are marked with a blue filled circle and a red star, respectively. Right: 70 $\micron$ to 24 $\micron$ flux ratio vs. 870 $\micron$ to 70 $\micron$ flux ratio for PBRs. G192S and G192N are marked with a blue filled circle and a red star, respectively. The vertical lines in both panels mark 70 $\micron$ to 24 $\micron$ flux ratio of 1.65, the PBRs selection criterion (Stutz, Tobin, \& Stanke 2013). }
\end{figure*}

\begin{figure*}
\centering
\includegraphics[angle=0,scale=0.6]{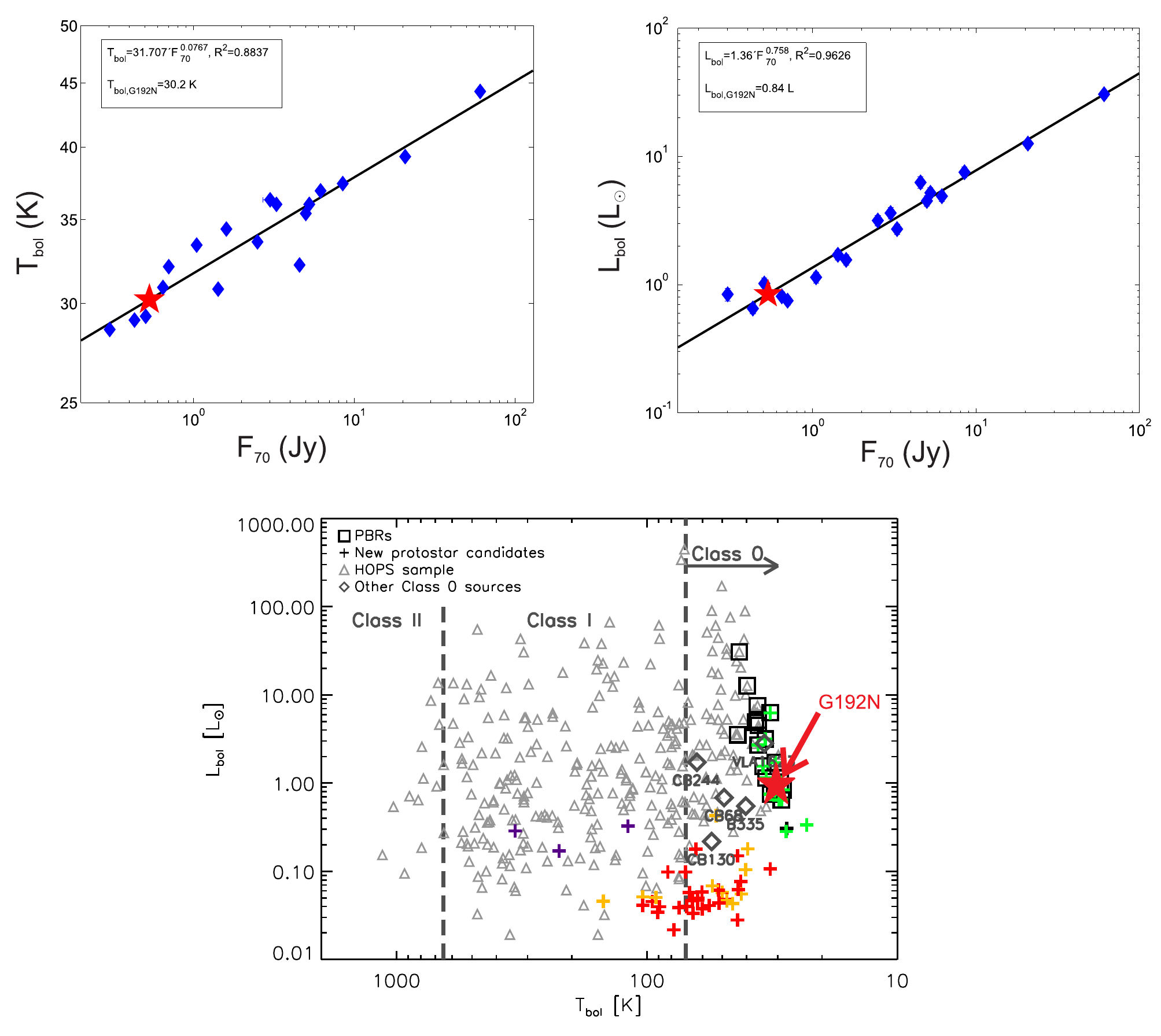}
\caption{Upper-left: 70 $\micron$ flux vs. T$_{bol}$ for PBRs. Upper-right: 70 $\micron$ flux vs. L$_{bol}$ for PBRs. Lower: T$_{bol}$ vs. L$_{bol}$ for HOPS protostars (from Stutz, Tobin, \& Stanke 2013). G192N is marked with red star. }
\end{figure*}

In Figure 24, we compare the infrared properties of G192N with other candidates of young Class 0 protostars (e.g. PACS Bright Red Sources (PBRs): (Stutz,
Tobin, \& Stanke 2013; Tobin, Stutz \& Megeath 2015), which are deeply embedded in dense envelopes, and are also either too faint (m$_{24}~>$ 7 mag) or undetected in the Spitzer/MIPS 24 $\mu$m band. Those PBRs are very rare and only 18 were identified in the whole Orion molecular cloud (Stutz, Tobin, \& Stanke 2013). G192N is located in the same region as PBRs in both 70 $\micron$ flux vs. 70 $\micron$ to 24 $\micron$ flux ratio plot and 70 $\micron$ to 24 $\micron$ flux ratio vs. 870 $\micron$ to 70 $\micron$ flux ratio plot, indicating that G192N should have similar properties as PBRs. As shown in the upper panels of Figure 25, the bolometric temperature and luminosity of PBRs in the Orion complex are strongly correlated to the flux of 70 $\micron$ emission (Stutz, Tobin, \& Stanke 2013). Assuming that G192N also follows these relationships, the bolometric temperature and luminosity of G192N are derived as $\sim$30 K and $\sim$0.8 L$_{\sun}$. As shown in the lower panel of Figure 25, G192N is also located in the same region as PBRs in T$_{bol}$ vs. L$_{bol}$ plot. When compared with other class 0 protostars, PBRs (including G192N) have relatively low bolometric temperatures. G192N itself has a relatively smaller bolometric luminosity when compared with other PBRs.

In a survey of 15 class 0 protostars, Yildiz et al. (2015) found that the median values of outflow mass and outflow mass loss rate are $1.7\times10^{-2}$ M$_{\sun}$ and $3.3\times10^{-5}$ M$_{\sun}$~yr$^{-1}$, which are about two orders of magnitude larger than those values of the outflow associated with G192N. But we should notify that the outflow parameters in Table 4 were not corrected with missing flux and inclination angle. As mentioned in section 4.2, the outflow mass or mass loss rate could be underestimated by a factor of $<$3-5 due to missing flux. Since the blue and red components of the elongated outflow are well separated, neither the very low (pole-on) inclination nor the very high (edge-on) inclination is likely. The outflow mass loss rate and accretion rate can increase by a factor of $\sim$1.6 and 2.9, respectively, at the mean angle of 57$\arcdeg$.3 of random outflow inclination. Even considering these effects, the outflow mass and outflow mass loss rate of G192N are still about one orders of magnitude smaller than those of other protostars. Therefore, we argue that G192S probably has the outflow with the lowest mass and mass loss rate among known Class 0 sources.

As mentioned in section 3.3.2, HDCO ($2_{0,2}-1_{0,1}$) line of G192N has a double peak profile. The linewidth of HDCO ($2_{0,2}-1_{0,1}$) is much smaller than those of o-H$_{2}$CO ($2_{1,2}-1_{1,1}$) line, indicating that its line emission is not affected by outflow activities. The critical density of HDCO ($2_{0,2}-1_{0,1}$) is much larger than those of other lines (e.g. J=1-0 of N$_{2}$H$^{+}$ and H$^{13}$CO) and therefore could trace the inner envelope or disk. Therefore the double peak profile of HDCO ($2_{0,2}-1_{0,1}$) might be produced by rotation. If that is true, we can estimate a dynamical mass M$_{dyn}$ by:
\begin{equation}\label{eq_lacc}
M_{dyn}=\frac{Rv_{rot}^{2}}{G}
\end{equation}
where R is the distance to the object of mass M creating the gravitational
field, G is the gravitational constant, and v$_{rot}$ is the rotational velocity. Taking R of 255 AU as estimated from SMA 1.3 mm continuum and v$_{rot}$ of 0.3 km~s$^{-1}$, which is half of the velocity separation of the two emission peaks, the dynamical mass is inferred to be $\sim$0.026 M$_{\sun}$, much smaller than that of other Class 0 sources (Yen et al. 2015).

In general, when compared with other young Class 0 sources in the Orion complex, G192N has potentially the smallest stellar mass, lowest bolometric luminosity and accretion rate, indicating that G192N might be at an earlier evolutionary phase than other Class 0 sources. The formation of the
central protostar must be preceded by the formation of a larger, less dense, first hydrostatic core (hereafter FHSC) (Larson 1969). The FHSC
thus represents an intermediate evolutionary stage between the
prestellar and protostellar phases. Theoretical works indicate that FHSCs might have a maximum mass of 0.01-0.1 M$_{\sun}$, lifetime of 0.5-50 kyr and internal luminosity of 10$^{-4}$-10$^{-1}$ L$_{\sun}$ (Dunham et al. 2014). Their SEDs are characterized by emission from 10-30 K dust and no observable emission below $\sim$20-50 $\mu$m (Dunham et al. 2014). As mentioned before, G192N is not visible at 24 $\mu$m or shorter wavelength bands. The stellar mass ($\sim$0.026 M$_{\sun}$) and lifetime (1.2-1.7 kyr) estimated from outflows of G192N are consistent with the predictions of FHSCs. However, the internal luminosity (0.21$\pm$0.01 L$_{\sun}$) of G192N is slightly larger than the upper limit of that of FHSCs. Machida, Inutsuka \& Matsumoto (2008) showed that FHSCs drive slow ($\sim$5 km~s$^{-1}$) outflows with wide opening-angles, while Price, Tricco \& Bate (2012) showed that FHSCs can indeed produce tightly collimated outflows with speeds of $\sim$2-7 km~s$^{-1}$. The maximum outflow speed ($>$14 km~s$^{-1}$) of G192N runs against the predictions for FHSCs. However, it should be noted that velocities $>$9 km~s$^{-1}$ would not be reached in the simulations of Price, Tricco \& Bate (2012) due to the 5 AU size of the central sink particle. When compared with other FHSC candidates (Dunham et al. 2014 and references therein), G192N is more luminous and has a more energetic outflow. Therefore, although G192N shares some similar properties as predicted FHSCs, it might be more evolved than FHSCs. G192N might be among the youngest Class 0 sources, which are slightly evolved than a FHSC.

To summarize, we could not exclude the possibility that G192N is a relatively more evolved Class 0 object in the low mass end of the Class 0 type distribution. Considering its low envelope mass, which is 0.35-0.43 M$_{\sun}$ from SMA and JCMT/SCUBA-2 observations, G192N may only form a very low mass hydrogen-burning star (M$\leq$0.2 M$_{\sun}$) assuming a typical core-to-star formation efficiency of 30\% (Alves, Lombardi \& Lada 2007; Andr\'{e} et al.
2014). However, the dynamical age of the outflow is (1.2-1.7)$\times10^{3}$ yr, which decreases to (0.8-1.1)$\times10^{3}$ yr at an outflow inclination angle of 57$\arcdeg$.3, indicating that G192N should be at a very early evolutionary phase.

\subsection{Is G192S a proto-brown dwarf candidate?}

\begin{figure}
\centering
\includegraphics[angle=-90,scale=0.35]{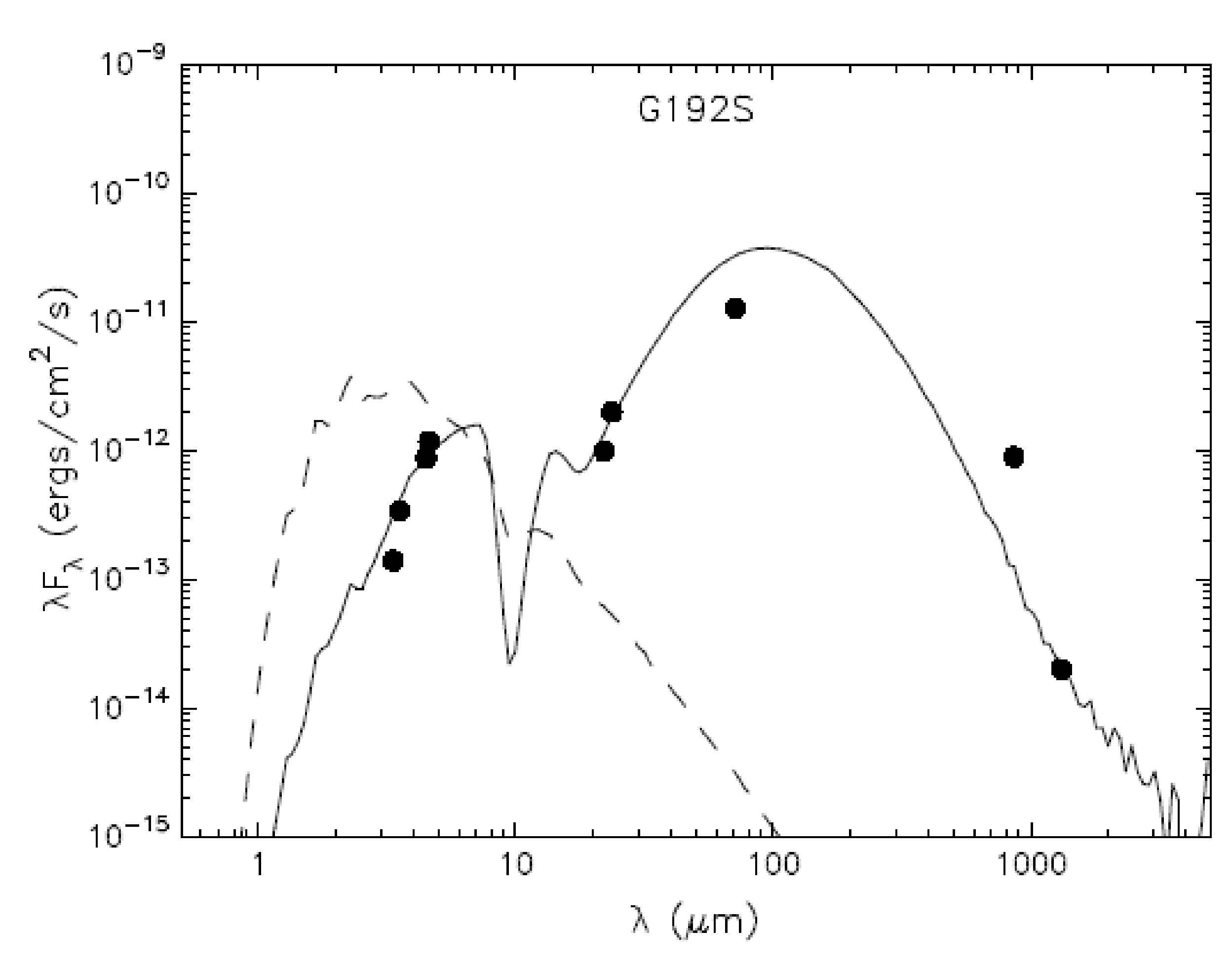}
\caption{Spectral energy distribution (SED) of G192S. The solid line shows the best modeled SED. The dashed line represents
best stellar fit (for comparison), including the effect of foreground extinction. }
\end{figure}

As shown in Figure 26, we fitted the SED of G192S with a YSO SED fitting tool (Robitaille et al. 2006,
2007). The stellar mass, disk mass and envelope mass of the best fit model are 0.12 M$_{\sun}$, 7.43$\times10^{-3}$ M$_{\sun}$, 0.27 M$_{\sun}$, respectively. The envelope mass is consistent with that estimated from JCMT/SCUBA-2 850 $\micron$ continuum emission. The stellar age is 8.2$\times10^{4}$ yr. The disk and envelope accretion rates are 1.5$\times10^{-8}$ M$_{\sun}$~yr$^{-1}$ and 1.8$\times10^{-5}$ M$_{\sun}$~yr$^{-1}$, respectively, indicating that G192S is still at the evolutionary stage 0/I (Robitaille et al. 2006,
2007). G192S also probably has a much smaller stellar mass and envelope mass than the values estimated from SED fitting because: (a) the minimum stellar mass in model SEDs of Robitaille et al. 2006,
2007 is 0.1 M$_{\sun}$ and thus those model SEDs could not fit stellar masses below 0.1 M$_{\sun}$; (b) the modeled 70 $\micron$ flux is more than 2 times larger than the observed value, indicating that the stellar mass should be overestimated; (c) as shown in the left panel of Figure 13, the 4.5 $\micron$ emission of G192S is offset from the core center in JCMT/SCUBA-2 850 $\micron$ emission, indicating that G192S might have a much smaller envelope mass than that estimated from SED modeling or JCMT/SCUBA-2 850 $\micron$ continuum emission. (d). Its vega magnitudes at Spitzer/IRAC 3.4 (ch1) and 4.5 (ch2) $\micron$ bands are 14.613$\pm$0.009 and 12.836$\pm$0.001 mag, which are compliant with the criteria (ch1-ch2$\geq$1.5, ch2$\leq$17.0 mag) for brown dwarf candidates (Griffith et
al. 2012). Therefore, we suggest that G192S is more likely forming a brown dwarf (M$\leq0.075$M$_{\sun}$).

The accretion rate and accretion luminosity for G192S are
comparable to or much smaller than that of other proto-brown
dwarf candidates (Lee et al. 2009; Palau et al. 2014; Morata et al. 2015). For example, the typical proto-brown dwarf L328-IRS has an accretion rate and accretion luminosity of 3.6$\times10^{-7}$ M$_{\sun}$~yr$^{-1}$ and 0.02 L$_{\sun}$(Lee et al. 2009), which are about one orders of magnitude larger than the ones of G192S.
Its extremely low accretion rate (2.8$\times10^{-8}$ M$_{\sun}$~yr$^{-1}$) also indicates that G192S is indeed likely to form a brown dwarf. Additionally, the 4.5 $\micron$ emission and SMA 1.3 mm continuum emission of G192S are clearly offset from SCUBA-2 850 $\micron$ continuum emission and CO gas fragments, indicating that G192S might have moved out of its natal envelope. The total mass estimated from SMA 1.3 mm continuum emission is only 0.02 M$_{\sun}$, which is insufficient for accretion to form a low or very low mass star. In conclusion, we argue that G192S is an ideal proto-brown dwarf candidate.

\subsection{Suppressed star formation in PGCC G192.32-11.88?}

PGCC G192.32-11.88 is a typical bright-rimmed cloud. As shown in Figure 1, it is surrounded by H$_{\alpha}$ emission from the H{\sc ii} region. The Spitzer 70 $\micron$ continuum emission reveals hot dust inside PGCC G192.32-11.88 and possibly also traces the ionization front in the south. The velocity and temperature gradients revealed in line emissions of CO isotopologues also indicate that the cloud might be compressed by the H{\sc ii} region. All this evidence suggests that star formation in PGCC G192.32-11.88 should be affected by the feedback of the H{\sc ii} region.

The total mass of the dense cores G192N and G192S in the Clump-S is 0.63$\pm$0.06 M$_{\sun}$. Assuming a typical core-to-star formation efficiency of 30\%, the total eventual stellar mass of Clump-S should be about 0.2 M$_{\sun}$. By comparing with the total clump mass (49$\pm$3 or 59$\pm$10 M$_{\sun}$) derived from CO isotopologues or virial analysis, we infer a SFE of $\sim$0.3\%-0.4\% in the southern clump of PGCC G192.32-11.88. Such low SFE is about one order of magnitude lower than that (3\%-6\%) in nearby giant molecular clouds (Evans et al. 2009). We can also calculate a core formation efficiency (CFE), $M_{core}/(M_{cloud}+M_{core})$. The CFE of the southern clump in PGCC G192.32-11.88 is $\sim$1\%. When compared with the CFE in the whole L1641 cloud, which is $\sim$4\% and increases to the dense filaments, which is 12\% (Polychroni
et al. 2013), the CFE in PGCC G192.32-11.88 is much lower. It seems that star formation in PGCC G192.32-11.88 is greatly suppressed.

Additionally, as mentioned in section 4.1, we found that G192N and G192S significantly depart from their parent cores along the direction of velocity/temperature gradient. This offset is more evident for G192S. G192S departs about 7$\arcsec$ from the SCUBA-2 850 $\micron$ emission and 1$\arcsec$.6 from the nearest CO clump in SMA observations. It seems that G192N and G192S will move out of their parent cores or their envelopes will be blown away due to photo-erosion during their formation. Since G192S is close to the ionization front, its formation is very similar to the brown dwarf formation model of Whitworth
\& Zinnecker (2004), in which a prestellar core's growth to a normal star is limited by photo-erosion of the ionizing radiation from a nearby OB star. We need further observations to test this scenario. Additionally, we notice that a large population of brown dwarfs have been discovered in the $\lambda$ Orionis complex (Barrado y Navascu{\'e}s et al. 2004, 2007; Bouy et al. 2009; Bayo et al. 2011, 2012). Those brown dwarfs are not located at the edges of the star cluster as the ejection mechanism would predict (Bayo et al. 2011), indicating that other mechanisms (like photo-erosion) may be responsible for the brown dwarf formation in the $\lambda$ Orionis complex.

At any rate, the stellar feedback seems to have a very negative effect on the star formation in PGCC G192.32-11.88.

\section{Summary}

To investigate the effect of stellar feedback on star formation, we are performing a series of observations with ground-based telescopes (PMO 13.7-m, KVN, CSO, JCMT and SMA) toward PGCCs in the $\lambda$ Orionis complex. In this work, we report the first results toward a particular source PGCC G192.32-11.88. Our main results are summarized below.

1. Mapping observations in $^{13}$CO (1-0) emission reveals two clumps in PGCC G192.32-11.88. We see a clear velocity gradient of 1.6-1.8 km~s$^{-1}$~pc$^{-1}$ across PGCC G192.32-11.88. The excitation temperature of $^{12}$CO (1-0) also reveals a temperature gradient. Through LTE and non-LTE RADEX analysis of CO isotopologues, we found that PGCC G192.32-11.88 may be externally heated by the H{\sc ii} region. The LTE masses of the southern and northern clumps are 49$\pm$3 and 99$\pm$3 M$_{\sun}$, respectively. Both clumps are gravitationally bound.

2. JCMT/SCUBA-2 observations of dust continuum emission at 850 $\micron$ detected two dense cores (G192N and G192S) in ``Clump-S". G192N is undetected at Spitzer/MIPS 24 $\mu$m or IRAC bands but has a large envelope seen in Spitzer/MIPS 70 $\mu$m. The southern object (G192S) is a point source at Spitzer/MIPS 24 $\mu$m, it has weaker 70 $\mu$m emission than G192N. G192S also shows extended 4.5 $\micron$ emission. G192N with an internal luminosity of $0.21\pm0.01$ L$_{\sun}$ shares similar infrared properties as other young Class 0 sources (e.g. PBRs). G192S is a very low-luminosity object with an internal luminosity of $0.08\pm0.01$ L$_{\odot}$. The core masses of G192N and G192S derived from 850 $\micron$ continuum are 0.43$\pm$0.03 and 0.23$\pm$0.03 M$_{\sun}$, respectively.

3. G192N has smaller CO and HCO$^{+}$ abundances than G192S, indicating that CO and HCO$^{+}$ gas might be more depleted toward G192N. The depletion of CO enhanced the abundance of N$_{2}$H$^{+}$ in G192N. The N$_{2}$H$^{+}$ to H$^{13}$CO$^{+}$ abundance ratio in G192N is larger than that of other protostellar cores in the Orion A cloud as well as G192S, indicating that G192N is colder and thus chemically younger. G192N and G192S have very low [HDCO]/[H$_{2}$CO] ratios ($<2\times10^{-3}$), which is much lower than that in other protostellar cores (Roberts et
al. 2002). The H{\sc ii} region could heat the cloud and increase its ionization fraction, leading to lower abundances of deuterated species. The chemistry in PGCC G192.32-11.88 might be greatly affected by stellar feedback (e.g. photoionizing radiation, heating) from the H{\sc ii} region, which needs to be tested by further chemical studies.

4. The SMA 1.3 mm continuum may be tracing the disks or inner envelopes of G192N and G192S. The masses of G192N and G192S derived from SMA 1.3 mm continuum are 0.38 and 0.02 M$_{\sun}$, respectively. The volume densities of G192N and G192S are $7.2\times10^{8}$ and $1.9\times10^{6}$ cm$^{-3}$. The extremely high volume density of G192N indicates that it is at an very early evolutionary stage. The 1.3 mm continuum peaks are consistent with Spitzer observations but significantly offset from JCMT/SCUBA-2 850 $\micron$ continuum, indicating that G192N and G192S might have moved out of their natal envelopes.

5. The $^{12}$CO (2-1) emission from the SMA observations reveals a collimated outflow associated with G192N and a weak compact outflow associated with G192S. The outflows have a dynamic timescale of $\sim1\times10^{3}$ yr. The total mass accretion rates for G192N and G192S are 6.3$\times10^{-7}$ M$_{\sun}$~yr$^{-1}$ and 2.8$\times10^{-8}$ M$_{\sun}$~yr$^{-1}$, respectively.

6. The $^{13}$CO (2-1) emission from the SMA+CSO combined data is highly fragmented and the fragments have masses much smaller than their Jeans masses, indicating that the gas fragments will be dispersed. The $^{12}$CO fragment close to the blueshifted outflow lobe has largest excitation temperature and broadened line width, hinting for outflow feedback in the form of heating and shocks.

7. G192N probably has the smallest stellar mass, lowest bolometric luminosity and accretion rate when compared with other young Class 0 sources (e.g. PBRs), indicating that G192N might be at an earlier evolutionary phase than other Class 0 sources. G192N has slightly larger internal luminosity and outflow velocity than the predictions of FHSCs. Therefore G192N might be among the youngest Class 0 sources, probably at a transitional phase from a FHSC to a Class 0 protostellar object. Considering its low internal luminosity, accretion rate and envelope mass, G192S is an ideal proto-brown dwarf candidate. Since G192S is close to the ionization front, its formation might be greatly affected by photo-erosion of the H{\sc ii} region as described in the model of Whitworth
\& Zinnecker (2004), which needs to be confirmed by further observations.

8. The star formation efficiency ($\sim$0.3\%-0.4\%) and core formation efficiency ($\sim$1\%) in PGCC G192.32-11.88 are significantly smaller than in other giant molecular clouds and dense filaments, indicating that the star formation therein is greatly suppressed due to stellar feedback.

\section*{Acknowledgment}
\begin{acknowledgements}

We are grateful to the SMA, PMO, CSO, JCMT and KVN staff. Tie Liu is supported by KASI fellowship. Y. Wu is partly supported by the China Ministry of Science and
Technology under State Key Development Program for Basic Research (No.2012CB821800), the grants of NSFC No.11373009 and No.11433008. Ke Wang acknowledge the support from ESO fellowship and DFG Priority Program 1573 (``Physics of the Interstellar Medium'') grant WA3628-1/1. C.W. Lee was supported by Basic Science Research Program though the National Research Foundation of Korea (NRF) funded by the Ministry of Education, Science, and Technology (NRF-2013R1A1A2A10005125) and also by the global research collaboration of Korea Research Council of Fundamental Science \& Technology (KRCF). This work was carried out in part at the Jet Propulsion Laboratory, operated for NASA by the California Institute of Technology. J.-E. Lee was supported by the Basic Science Research Program through the National Research Foundation of Korea (NRF) (grant No. NRF-2015R1A2A2A01004769) and the Korea
Astronomy and Space Science Institute under the R\&D
program (Project No. 2015-1-320-18) supervised by the
Ministry of Science, ICT and Future Planning. The KVN is a facility operated by the Korea Astronomy and Space Science Institute. S.P.L. thanks the support of the Ministry of Science and Technology of Taiwan
with Grant MoST 102-2119-M-007-004-MY3. Figures 24 and 25 are used with permission from Dr. A. Stutz.

\end{acknowledgements}

\section*{APPENDIX A\\
LTE analysis of J=1-0 transitions of CO isotopologues}

The column density N of a linear, rigid rotor molecule can be obtained with the
theory of radiation transfer and molecular excitation as following (Garden et al. 1991):
\begin{equation}
\begin {split}
N=\frac{3k}{8\pi^{3}B\mu^{2}}\frac{exp[hBJ(J+1)/kT_{ex}]}{(J+1)}\\ \times\frac{(T_{ex}+hB/3k)}{[1-exp(-h\nu/kT_{ex})]}\int\tau_{v}dv
\end {split}
\tag{A1}
\end{equation}
where h is the Planck constant, k is the Boltzmann constant. B, $\mu$, J, $\nu$, $\tau_{v}$ and T$_{ex}$ are the rotational constant, permanent dipole
moment, the rotational quantum number of the lower state of the
molecular transition, the frequency, optical depth and the excitation temperature of the observed transition, respectively.

The brightness temperature of a molecular transition can be expressed as following (Garden et al. 1991):
\begin{equation}
\begin {split}
T_{b}=\frac{T_{a}^{*}}{\eta_{b}}=\frac{h\nu}{k}\left[\frac{1}{exp(h\nu/kT_{ex})-1}-\frac{1}{exp(h\nu/kT_{bg})-1}\right]\\ \times\left[1-exp(-\tau)\right]f
\end {split}
\tag{A2}
\end{equation}
Here $T_{a}^{*}$ is the antenna temperature, and $T_{b}$ is the brightness temperature corrected with beam
efficiency $\eta_{b}$. The optical depth of $^{12}$CO (1-0) can be estimated by comparing the $^{13}$CO (1-0) emission with the $^{12}$CO (1-0) emission (Liu, Wu \& Zhang 2013; Sanhueza et al. 2012):

\begin{equation}
\frac{T_{b} (^{12}CO)}{T_{b} (^{13}CO)}\approx\frac{1-exp(-\tau_{12})}{1-exp(-\tau_{13})}=\frac{1-exp(-\tau_{12})}{1-exp(-\tau_{12}/60)}
\tag{A3}
\end{equation}
where 60 is the [$^{12}$CO]/[$^{13}$CO] abundance ratio.
Assuming the $^{12}$CO (1-0) emission is optically thick and its filling factor of f=1, the excitation temperature $T_{ex}$ can be
obtained from Equation (A2).

The integral in Equation (1) can be written as (Buckle
et al. 2010):
\begin{equation}
\begin{split}
  \int\tau(v)\mathrm{d}v  =  \frac{1}{J(T_{\rm ex})-J(T_{\rm bg})}\int\frac{\tau(v)}{1-e^{-\tau(v)}}T_{\textsc{b}}\mathrm{d}v \\
    \approx \frac{1}{J(T_{\rm ex})-J(T_{\rm bg})}\frac{\tau(v_{0})}{1-e^{-\tau(v_{0})}}\int T_{\textsc{b}}\mathrm{d}v
\end{split}
\tag{A4}
\end{equation}
where $v_{0}$ is the central velocity of the line, $T_{\textsc{b}}$ is the observed main beam temperature and $J(T)$ is the source function,
\begin{equation}
  J(T)=\frac{T_{0}}{e^{T_{0}/T}-1}
  \tag{A5}
\end{equation}
with $T_{0}=h\nu/k_{\textsc{b}}$, $T_{\rm bg}$ is the cosmic microwave background temperature, 2.73 K.

Assuming $^{13}$CO (1-0) has the same excitation temperature as $^{12}$CO (1-0), the peak optical depth $\tau(v_{0})$ can be derived from Equation (A2) and the column density of $^{13}$CO can be calculated with Equation (A1).

\section*{APPENDIX B\\
Calculation of column density}

\renewcommand{\thefigure}{B\arabic{figure}}

\setcounter{figure}{0}

\subsection*{B1. N$_{2}$H$^{+}$}

The column density of N$_{2}$H$^{+}$ is calculated as Furuya,
Kitamura, \& Shinnaga (2006):

\begin{equation}
N_{N_{2}H^{+}}=3.30\times10^{11}\frac{T_{ex}+0.75}{1-e^{-4.47/T_{ex}}}(\frac{\tau_{tot}}{1.0})(\frac{\Delta v}{1.0~km~s^{-1}})~cm^{-2}
\tag{B1}
\end{equation}
where $\tau_{tot}$ and $\Delta v$ are the total optical depth and linewidth obtained from
the N$_{2}$H$^{+}$ spectra through the HFS analysis.

\subsection*{B2. H$^{13}$CO$^+$}

\begin{figure}[tbh!]
\centering
\includegraphics[angle=0,scale=0.35]{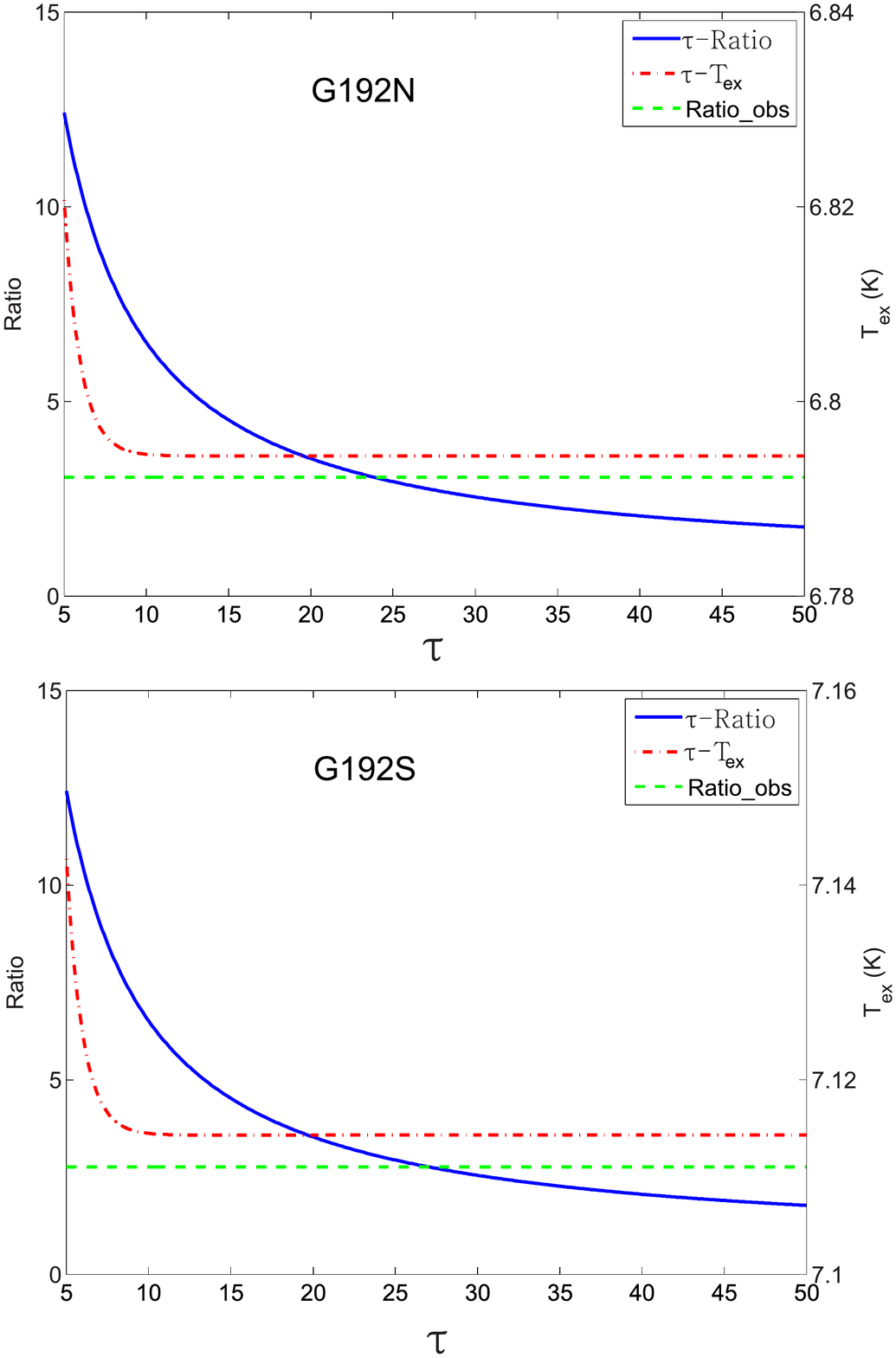}
\caption{ The intensity ratio of HCO$^{+}$ (1-0) to H$^{13}$CO$^{+}$ (1-0) and excitation temperature of HCO$^{+}$ (1-0)  vs. optical depth of HCO$^{+}$ (1-0).}
\end{figure}

The optical depth of HCO$^+$ (1-0) ($\tau_{12}$) of HCO$^+$ (1-0) can be derived from Equation (A3) from the intensity ratio of HCO$^+$ (1-0) to H$^{13}$CO$^+$ (1-0). After determining the optical depth of HCO$^+$ (1-0), its excitation temperature can be derived from Equation (A2) assuming that the filling factor is 1. In Figure B1, the intensity ratio of HCO$^{+}$ (1-0) to H$^{13}$CO$^{+}$ (1-0) and excitation temperature of HCO$^{+}$ (1-0) calculated with equations (A2) and (A3) are plotted as a function of optical depth of HCO$^{+}$ (1-0). It can be seen that the excitation temperature of HCO$^+$ (1-0) stays constant when optical depths are larger than $\sim$10. The excitation temperatures of HCO$^+$ (1-0) toward G192N and G192S are 6.79 and 7.11 K, respectively. Assuming H$^{13}$CO$^+$ (1-0) has the same excitation temperature as HCO$^+$ (1-0), the column density of H$^{13}$CO$^+$ can be derived from Furuya,
Kitamura, \& Shinnaga (2006):

\begin{equation}
\begin{split}
N_{H^{13}CO^{+}}=2.32\times10^{11}\frac{T_{ex}+0.69}{1-e^{-4.16/T_{ex}}}(\frac{1}{J(T_{\rm ex})-J(T_{\rm bg})})\\ \times(\frac{\int T_bdv}{1.0~K~km~s^{-1}})~cm^{-2}
\end{split}
\tag{B2}
\end{equation}

\subsection*{B3. o-H$_{2}$CO and HDCO}

\begin{figure}[tbh!]
\centering
\begin{minipage}[c]{0.5\textwidth}
  \centering
  \includegraphics[angle=0,scale=.4]{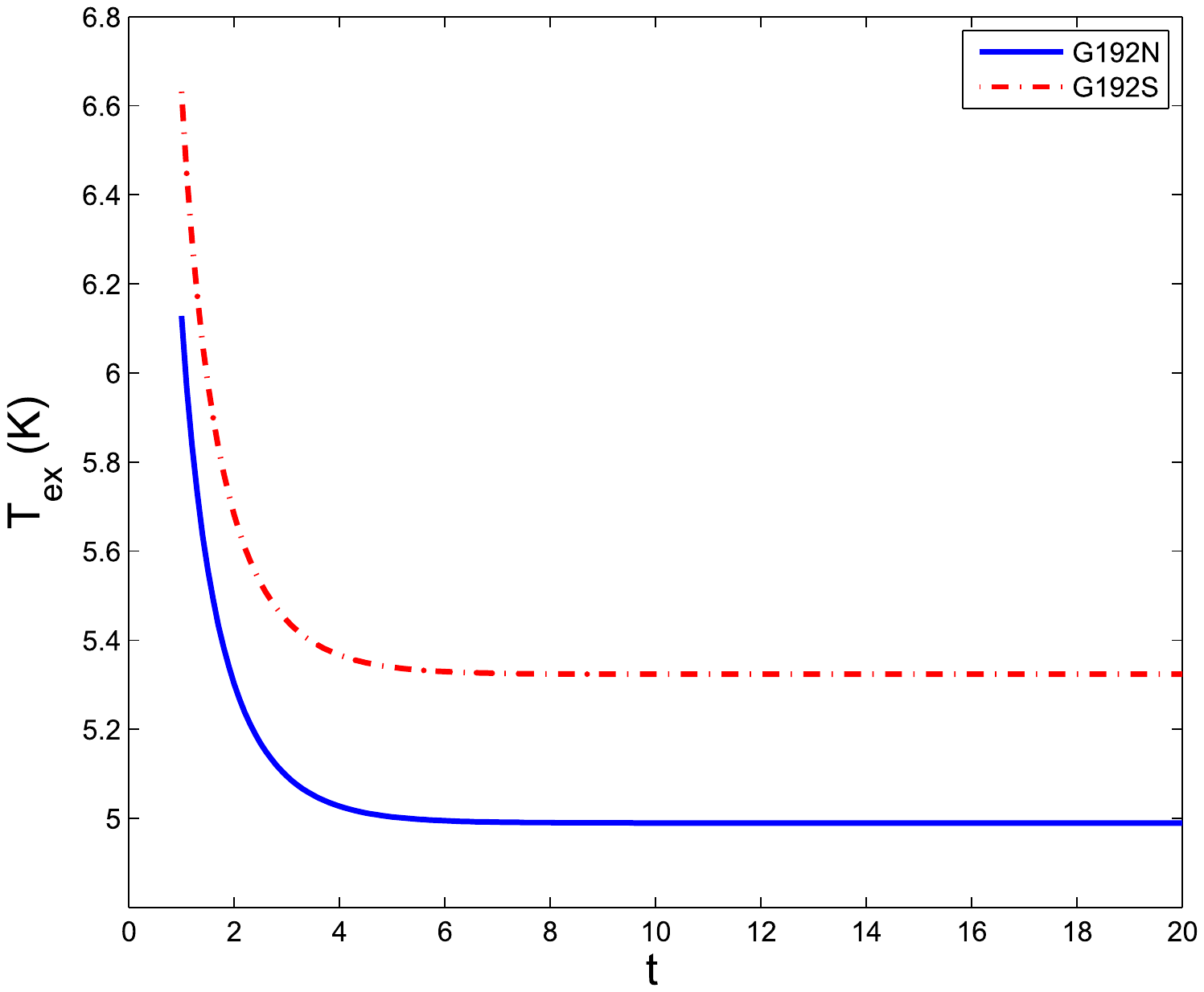}
\end{minipage}
\begin{minipage}[c]{0.5\textwidth}
  \centering
  \includegraphics[angle=0,scale=.4]{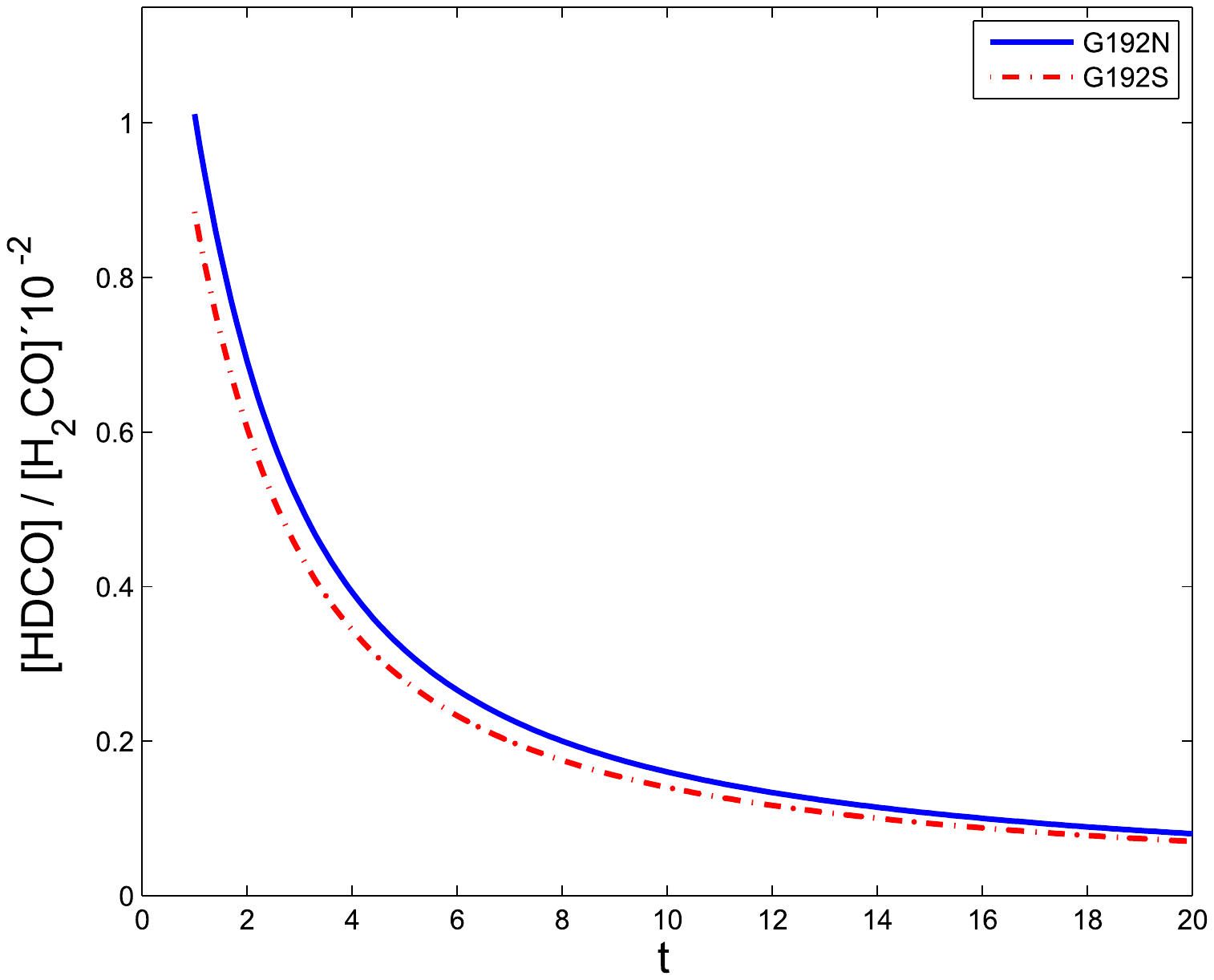}
\end{minipage}
\caption{The excitation temperature of H$_{2}$CO ($2_{1,2}-1_{1,1}$) (upper) and  [HDCO]/[H$_{2}$CO] ratio (lower) vs. optical depth of H$_{2}$CO ($2_{1,2}-1_{1,1}$). }
\end{figure}

We estimated the column density of o-H$_{2}$CO with RADEX. We adopted a kinetic temperature of 18.9 K, as determined from CO isotopologues. We convolved the SCUBA-2 850 $\mu$m data with KVN beam (24$\arcsec$). Then we derived a mean volume density of 2.1$\times10^{4}$ cm$^{-3}$ from Gaussian fit toward the continuum source. Since o-H$_{2}$CO ($2_{1,2}-1_{1,1}$) has much higher critical density than J=1-0 and 2-1 transitions of CO isotopologues (Shirley et al. 2015), we adopt 2.1$\times10^{4}$ cm$^{-3}$ in RADEX calculations. From RADEX calculations, we found the column density and excitation temperature of o-H$_{2}$CO of G192N are $\sim$1.1$\times10^{15}$ cm$^{-2}$ and $\sim$5.0 K, respectively. For G192S, the column density and excitation temperature of o-H$_{2}$CO are $\sim$7.8$\times10^{14}$ cm$^{-2}$ and $\sim$5.3 K, respectively.

Assuming HDCO ($2_{0,2}-1_{0,1}$) is optically thin and has the same excitation temperature as o-H$_{2}$CO ($2_{1,2}-1_{1,1}$), the column density of HDCO can be calculated following Roberts et al. (2002):

\begin{equation}
N_{TOT}=N_{u}Q(T_{ex})e^{-E_{u}/kT_{ex}}/g_{u}
\tag{B3}
\end{equation}
where $N_{TOT}$ is the total column density, $Q(Tex)$ is the partition fraction and $g_{u}$ is the degeneracy factor in the upper energy level. $N_{u}$ is the column density in the upper level of the transition:

\begin{equation}
N_{u}=\frac{8\pi k\nu^2}{A_{ul}hc^{3}}\int T_{b}dv
\tag{B4}
\end{equation}

The HDCO column densities of G192N and G192S are $(1.9\pm0.1)\times10^{12}$ and $(1.7\pm0.1)\times10^{12}$ cm$^{-2}$, respectively. Assuming that the ortho/para ratio for H$_{2}$CO is the statistical value of 3 (Roberts et al. 2002), the [HDCO]/[H$_{2}$CO] ratios of G192N and G192S are $(1.0\pm0.1)\times10^{-2}$ and $(1.1\pm0.1)\times10^{-2}$, respectively.

We can also calculate the o-H$_{2}$CO column density with Equation (B3). In the optically thin case, the o-H$_{2}$CO column densities of G192N and G192S are $\sim$1.2$\times10^{14}$ and $\sim$7.1$\times10^{13}$ cm$^{-2}$, respectively. The corresponding [HDCO]/[H$_{2}$CO] ratios of G192N and G192S are $\sim1.6\times10^{-2}$ and $\sim1.4\times10^{-2}$, respectively, which are larger than the values obtained from RADEX calculations and should be taken as upper limits. In the optically thick case, the column density of o-H$_{2}$CO should be corrected with a factor of $\frac{\tau}{1-e^{-\tau}}$, where $\tau$ is the optical depth of o-H$_{2}$CO ($2_{1,2}-1_{1,1}$). The excitation temperatures of o-H$_{2}$CO ($2_{1,2}-1_{1,1}$) toward G192N and G192S derived from RADEX calculations are consistent with the values derived by assuming LTE and that o-H$_{2}$CO ($2_{1,2}-1_{1,1}$) emission is optically thick ($\tau>5$) as shown in the upper panel of Figure B2. In the lower panel of Figure B2, we shown the [HDCO]/[H$_{2}$CO] ratio as function of the optical depth of o-H$_{2}$CO ($2_{1,2}-1_{1,1}$). When compared with the value derived in the optically thin case, the [HDCO]/[H$_{2}$CO] ratio decreases by a factor of larger than five if the optical depth of o-H$_{2}$CO ($2_{1,2}-1_{1,1}$) is larger than 6. Therefore, we argue that [HDCO]/[H$_{2}$CO] ratios of G192N and G192S might be even lower than 0.01, which needs to be confirmed by further observations.

\end{document}